\documentclass{article}

\usepackage{arxiv}
\usepackage[utf8]{inputenc} % allow utf-8 input
\usepackage[T1]{fontenc}    % use 8-bit T1 fonts
\usepackage{hyperref}       % hyperlinks
\usepackage{booktabs}       % professional-quality tables
\usepackage{amsfonts}       % blackboard math symbols
\usepackage{nicefrac}       % compact symbols for 1/2, etc.
\usepackage{microtype}      % microtypography
\usepackage{graphicx,rotating,url,soul,multirow}
\usepackage{amsthm,amsmath,amsfonts,amssymb}
\usepackage[english]{babel}
\usepackage{natbib}

\title{Modeling Heterogeneity and Missing Data of Multiple Longitudinal Outcomes in Electronic Health Records}

\author{
  Rebecca~Anthopolos \\
  Division of Biostatistics, Department of Population Health\\
  New York University Grossman School of Medicine\\
  New York, NY \\
  \texttt{rebecca.anthopolos@nyulangone.org} \\
   \And
 Ying~Wei \\
  Department of Biostatistics\\
  Columbia University Mailman School of Public Health \\
  New York, NY \\
  \texttt{yw2148@cumc.columbia.edu} \\
     \And
 Qixuan~Chen \\
  Department of Biostatistics\\
  Columbia University Mailman School of Public Health \\
  New York, NY \\
  \texttt{qc2138@cumc.columbia.edu} \\
}

\begin{document}
\maketitle

\begin{abstract}
In  electronic health records (EHRs), latent subgroups of patients may exhibit distinctive patterning in their longitudinal health trajectories. For such data, growth mixture models (GMMs) enable classifying patients into different latent classes based on individual trajectories and hypothesized risk factors. However, the application of GMMs is hindered by the special missing data problem in EHRs, which manifests two patient-led missing data processes: the visit process and the response process for an EHR variable conditional on a patient visiting the clinic. If either process is associated with the process generating the longitudinal outcomes, then valid inferences require accounting for a nonignorable missing data mechanism. We propose a Bayesian shared parameter model that links GMMs of multiple longitudinal health outcomes, the visit process, and the response process of each outcome given a visit using a discrete latent class variable. Our focus is on multiple longitudinal health outcomes for which there can be a clinically prescribed visit schedule. We demonstrate our model in EHR measurements on early childhood weight and height z-scores. Using data simulations, we illustrate the statistical properties of our method with respect to subgroup-specific or marginal inferences. We built the \texttt{R} package \texttt{EHRMiss} for model fitting, selection, and checking.
\end{abstract}

% keywords can be removed
\keywords{Electronic health records \and Gibbs sampling \and Latent class modeling \and Missing not at random \and Multiple longitudinal health outcomes \and Shared parameter model}

\section{Introduction}

\label{s:intro}

As electronic health records (EHRs) are increasingly adopted in US health systems, an estimated one billion patient visits may be documented per year \citep{Hripcsak2013}. Thanks to the rapid advancement of big data management and processing, EHRs are often computable, representing an exceptional observational data resource for new discoveries in science and medicine. A natural feature of such big data may be unobserved, or ``latent'' heterogeneity, whereby latent subgroups of patients are characterized by distinctive patterning in their longitudinal health trajectories. Researchers from diverse biomedical fields, such as psychology \citep{Elliott2005} and maternal and infant health \citep{Neelon2011}, have used growth mixture models (GMMs) \citep{Muthen2002,Verbeke1996} to analyze latent heterogeneity in longitudinal data from diverse data sources other than EHRs. GMMs enable classifying subjects into different subgroups, often called latent classes, according to individual longitudinal trajectories and risk factors hypothesized to be associated with class membership.

Despite the utility of GMMs for EHR-based research, their application is hindered by the special missing data problem in EHRs. In the prototypical mixed model for longitudinal data analysis \citep{Laird1982}, missed measurements are assumed to be missing at random (MAR). However, this assumption may not be valid in EHRs due to the presence of two patient-led missing data processes. First, a patient's \textit{visit process}, defined as the probability of observing a clinic visit at a given time, is driven by some combination of a patient's own prerogative and physician recommendation. The second missing data process is the response process given a clinic visit, defined by the conditional probability of observing a response on an EHR variable given a clinic visit. In EHRs, the likelihood that a variable gets investigated -- and in turn, recorded -- may depend on a patient's stated medical reasons for the visit, in addition to clinical judgment. When either missing data process is associated with the underlying process generating the longitudinal health outcomes, then valid inferences for any models require accounting for a missing not at random (MNAR) mechanism. To our knowledge, no methods have been developed to fully accommodate a two-fold MNAR mechanism in EHRs.

In the missing data literature, a visit process that is associated with the underlying process generating the longitudinal outcomes has been  characterized as a special case of MNAR known as  ``informative'' \citep{Wu1988,Follmann1995}. A common approach to handling an informative visit process is a shared parameter framework \citep{Wu1988,Follmann1995}. It assumes that the distributions of the longitudinal outcomes and visit process \textit{share} a continuous or discrete latent variable, which drives the correlation between missed visits and longitudinal outcomes. Once conditioning on the latent variable, the longitudinal outcomes and visit process are assumed to be independent \citep{Liang2009,Sun2007,McCulloch2016,Lin2004}. However, existing shared parameter models are insufficient to describe the complexity of EHRs where the response process of individual health outcomes given a clinic visit may also exhibit an MNAR mechanism. 

To apply growth mixture modeling to longitudinal health outcomes collected in EHRs, we propose a Bayesian shared parameter model, which integrates GMMs of the longitudinal health outcomes, the visit process, and the response process of individual health outcomes given a clinic visit, using a discrete latent variable to indicate the latent class to which each patient belongs. Our focused applications are preventive care (e.g., screenings for cholesterol) and chronic disease management (e.g., HbA1c \% among patients with type 2 diabetes) in which certain health outcomes are routinely collected with a clinically prescribed visit schedule. We demonstrate our proposed model with early childhood weight and height measurements, which should be collected according to the well-child check schedule \citep{AAP2018}. Using the prescribed visit schedule, we construct time windows of observation to measure each patient's visit process, and in turn, response process for each health outcome. We developed an efficient Markov chain Monte Carlo (MCMC) algorithm based on easily sampled closed-form full conditional distributions. To conduct model fitting, selection, and checking, we built the user-friendly \texttt{R} package \texttt{EHRMiss} available at \url{https://github.com/anthopolos/EHRMiss}.

\section{Statistical Method}

We formulate our proposed Bayesian shared parameter model of longitudinal health outcomes collected in EHRs. First, in Section \ref{s:CompleteDataModel}, we present the complete-data model. In Section \ref{s:MissingDataProcesses}, we extend the complete-data model to account for a nonignorable visit process and response process given a clinic visit. Lastly, in Section \ref{s:BayesianComputation}, we explicate our Bayesian computation. 

\subsection{Complete-data model} 
\label{s:CompleteDataModel}

Suppose there are $K$ latent classes of patients with distinctive patterning in their trajectories of $R$ health outcomes collected over $J$ prescribed clinical time windows. The complete-data model is a  Bayesian multivariate GMM with submodels for latent class membership and the longitudinal health outcomes. We begin with latent class membership. 
Let $c_i$ be a discrete latent variable taking values $k = 1,\dots,K$ to indicate the latent class membership of patient $i$ for $i = 1,\dots,n$. We assume that
\vspace{-0.1in}
\begin{eqnarray}
c_i  \sim  Multinomial \begin{pmatrix}
1; \, \pi_{i1},\dots,\pi_{iK} 
\end{pmatrix} \label{e:multinomial},
\end{eqnarray} 
where $\pi_{ik}$ are patient-specific latent class membership probabilities. To connect $\pi_{ik}$ with covariates of interest, we introduce $K$ latent random variables $\xi^{*}_{ik}$ ($k = 1,\dots,K$) such that $\pi_{ik} = Pr(\xi_{ik}^{*} > \xi_{il}^{*} \text{ for all } l \neq k)$. Upon defining latent class $K$ as the reference level by setting $\xi_{ik} = \xi^{*}_{ik} - \xi^{*}_{iK}$ for $k = 1,\dots,K-1$, we specify the following model: \begin{eqnarray} \label{eqn:c}
\xi_{ik} = \mathbf{w}_i \mathbf{\delta}_k^T + \epsilon_{ik}
\quad\text{and}\quad
c_i  =  \begin{cases} K & \mbox{if } max(\xi_{i1}, \dots, \xi_{iK-1}) < 0\\ 
k & \mbox{if } max(\xi_{i1}, \dots, \xi_{iK-1}) = \xi_{ik} \geq 0.
\end{cases} \label{e:clatentvariable}
\end{eqnarray} In (\ref{e:clatentvariable}), $\mathbf{w}_i$ is a $1 \times s$ row vector containing a one for an intercept and $s-1$ covariates, such as patient-level risk factors, with corresponding regression coefficients in $\mathbf{\delta}_k$. The $\epsilon_{ik}$ are normal random errors with mean zero and identity variance-covariance matrix \citep{StataCorp2019}. For $K=2$, this set-up corresponds to the standard Bayesian probit model for a binary outcome \citep{Albert1993}.

The multivariate model of longitudinal health outcomes is then specified conditional on $c_i$. Let $y_{1ij},\dots,y_{Rij}$ be longitudinal measurements on $R$ health outcomes for patient $i$, $i=1,\dots,n$, in time window $j$, $j=1,\dots,J$. Then, \begin{gather}
\left[
\begin{array}{c|c}
y_{1ij} &  \\
\vdots &  c_i = k \\
y_{Rij} &  \\
\end{array}
\right]  \sim  MVN_{R} \begin{pmatrix}
\begin{bmatrix}
\mathbf{\beta}_{1k} \mathbf{x}_{ij}^T + \mathbf{b}_{1i} \mathbf{z}_{ij}^T \\
\vdots \\
\mathbf{\beta}_{Rk} \mathbf{x}_{ij}^T + \mathbf{b}_{Ri} \mathbf{z}_{ij}^T  
\end{bmatrix}, \,
\mathbf{\Sigma}_k
\end{pmatrix} \label{e:modely} \\[3\jot] 
\left[
\begin{array}{c|c}
\mathbf{b}_{1i} &  \\
\vdots &  c_i = k \\
\mathbf{b}_{Ri} &  \\
\end{array}
\right]  \sim  MVN_{Rq} \begin{pmatrix}
\begin{bmatrix}
\mathbf{0} \\
\vdots \\
\mathbf{0}  
\end{bmatrix}, \,
\mathbf{\Psi}_k
\end{pmatrix}. \label{e:modelb} 
\end{gather} In (\ref{e:modely}), conditional on $c_i$, $y_{rij}$ ($r = 1,\dots,R$) are modeled as a smooth function of time in window $j$, with $\mathbf{x}_{ij}^T$ being a $p$-length column vector containing a one and $(p-1)$ polynomial terms for time. The corresponding regression coefficients in $\mathbf{\beta}_{rk}$ capture the average trajectory for health outcome $r$ in latent class $k$. Covariates other than time may be included in $\mathbf{x}_{ij}$. The $\mathbf{\Sigma}_k$ is an $R \times R$ latent class-specific residual variance-covariance among $y_{rij}$ ($r = 1,\dots,R$). 
% , and $\mathbf{x}_{ij}$ and $\mathbf{w}_{ij}$ in (\ref{e:clatentvariable}) may be overlapping as long as careful attention\footnote{Reviewers may raise questions here on "what careful attention means". \textcolor{red}{Rebecca says: OK.}} is paid to issues of identifiability \citep{Jiang1999}

For each longitudinal health outcome $r$, $\mathbf{b}_{ri}$ is a $1 \times q$ row vector of patient-specific random effects associated with $\mathbf{z}_{ij}^T$, the columns of which are a subset of $\mathbf{x}_{ij}^T$. As shown in (\ref{e:modelb}), $\mathbf{b}_{ri}$ are modeled given $c_i$, thus reflecting patient-specific variability around the average health trajectory in latent class $k$. The latent class-specific variance-covariance $\mathbf{\Psi}_{k}$ in (\ref{e:modelb}) is an $Rq \times Rq$ block diagonal matrix with entries $\mathbf{\Psi}_{kr}$ ($q \times q$), the elements of which compose a variance-covariance for $\mathbf{b}_{ri}$ ($i = 1,\dots,n$). For simplicity, we have used the same $\mathbf{x}_{ij}$ and $\mathbf{z}_{ij}$ for each longitudinal health outcome $y_{rij}$ ($r = 1,\dots,R$), but this is not required.

% = (b_{ri1},\dots,b_{riq})^T$
%Fruhrwirth page 275 about identifiabiltiy

\subsection{Nonignorable missing data processes in EHRs}
\label{s:MissingDataProcesses}
We extend the complete-data model in (\ref{e:multinomial}) - (\ref{e:modelb}) to account for nonignorable missing data mechanisms for the visit process and the response process given a clinic visit in EHRs. 

To specify the full data, for health outcome $r$, consider the elements $y_{ri1},\dots,y_{riJ}$ for patient $i$ over $J$ time windows. Let $d_{ij}$ ($j=1,\dots,J$) be an indicator for the visit process such that $d_{ij} = 1$ if patient $i$ has a clinic visit during time window $j$, and 0 otherwise. The response process for the $r^{th}$ health outcome given a clinic visit is defined for the subset of time windows when patient $i$ visits the clinic. Let $A = \{j: \, \, d_{ij} = 1 \text{ for } j = 1,\dots,J \}$, and let the total number of clinic visits for patient $i$ be $n_i = \sum_{j =1}^J d_{ij}$. Then, for $l = 1,\dots,n_i$, define $m_{riA(l)} = 1$ if a response is observed for health outcome $r$ at window $A(l)$, and 0 otherwise. The full data are given by $y_{rij}$, $d_{ij}$, and $m_{riA(l)}$. 

To ease computational burden in MCMC estimation, we use a probit link function in modeling the probability of a clinic visit. For patient $i$ in time window $j$, 
\begin{gather}
\left[
\begin{array}{c|c}
d_{ij} &  c_i = k \\
\end{array}
\right]  \sim  Bernoulli \begin{pmatrix}
\Phi\{\mathbf{x}_{ij} \mathbf{\phi}_{k}^T + \mathbf{z}_{ij} \mathbf{\tau}_{i}^T\}
\end{pmatrix} \label{e:modeld} \\[3\jot] 
\left[
\begin{array}{c|c}
\mathbf{\tau}_{i} &  c_i = k
\end{array}
\right]  \sim  MVN_{q} \begin{pmatrix}
\mathbf{0}, \,
\mathbf{\Omega}_{k}
\end{pmatrix}, \label{e:modeltau}
\end{gather} where $\Phi\{.\}$ is the cumulative distribution function of the standard normal distribution. Analogous to (\ref{e:modely}) and (\ref{e:modelb}), in (\ref{e:modeld}) and (\ref{e:modeltau}), the regression coefficients in the $p \times 1$ column vector $\mathbf{\phi}_k^T$ reflect the average visit process trajectory in latent class $k$, with $q$ patient-specific random effects in $\mathbf{\tau}_i$ that capture individual-level variations within each latent class. The $\mathbf{\Omega}_k$ is a latent class-specific variance-covariance.
%($q \times q$)

Correspondingly, the probability of response for health outcome $r$ in $A(l)$ is specified as
\begin{gather}
\left[
\begin{array}{c|c}
m_{riA(l)} &  c_i = k \\
\end{array}
\right]  \sim  Bernoulli \begin{pmatrix} 
\Phi\{\mathbf{x}_{iA(l)} \mathbf{\lambda}_{rk}^T + \mathbf{z}_{iA(l)} \mathbf{\kappa}_{ri}^T\}
\end{pmatrix} \label{e:modelm} \\[3\jot] 
\left[
\begin{array}{c|c}
\mathbf{\kappa}_{ri} &  c_i = k
\end{array}
\right]  \sim  MVN_{q} \begin{pmatrix}
\mathbf{0}, \,
\mathbf{\Theta}_{rk}
\end{pmatrix}, \label{e:modelkappa}
\end{gather} where $\mathbf{\lambda}_{rk}^T$ is a $p \times 1$ column vector that represents the latent class-specific average response process for health outcome $r$, and the $q$ patient-specific random effects in $\mathbf{\kappa}_{ri}$ are modeled with a latent class-specific variance-covariance $\mathbf{\Theta}_{rk}$. For simplicity, in (\ref{e:modeld}) and (\ref{e:modelm}), we have assumed that $\mathbf{x}_{iA(l)}$ and $\mathbf{z}_{iA(l)}$ are the same as in the longitudinal health outcome model (\ref{e:modely}). 

Conditional on $c_i$, the longitudinal health outcomes, visit process, and response process given a clinic visit are assumed to be independent. The MNAR mechanism is evident because the visit and response processes depend on missing longitudinal health outcomes indirectly through latent class membership. The proposed shared parameter model can be easily altered to an MAR mechanism for one or both of the visit process and response process given a clinic visit. For example, the visit process is MAR if $f(d_{ij}, \mathbf{\tau}_i \, | \, c_i, rest) = f(d_{ij}, \mathbf{\tau}_i \, | \, rest)$. Then, assuming separable parameter spaces, the visit process can be ignored in statistical analysis.

\subsection{Bayesian computation}
\label{s:BayesianComputation}

To complete the Bayesian model specification, we assign prior distributions to all of the parameters. For each parameter, we use the same prior distribution across mixture components. In the latent class membership model, we assign the probit regression coefficients $\mathbf{\delta}_k$ in (\ref{e:clatentvariable}) the prior distribution $MVN_s(\mathbf{0},\mathbf{I})$ such that on the probability scale, the mode of the prior probability of latent class membership is approximately $\frac{1}{K}$ \citep{Garrett2000,Elliott2005}. In the models for the longitudinal health outcomes, visit process, and response process given a clinic visit, we assign diffuse multivariate normal prior distributions for the latent class-specific regression coefficients $\mathbf{\beta}_{rk}$ in (\ref{e:modely}), $\mathbf{\phi}_{k}$ in (\ref{e:modeld}), and $\mathbf{\lambda}_{rk}$ in (\ref{e:modelm}), and inverse-Wishart prior distributions for the hierarchical variance-covariances $\mathbf{\Psi}_{kr}$ in (\ref{e:modelb}), $\mathbf{\Omega}_k$ in (\ref{e:modeltau}), and $\mathbf{\Theta}_{rk}$ in (\ref{e:modelkappa}), respectively. In the longitudinal health outcome model (\ref{e:modely}), we also assign the observation-level variance-covariance $\mathbf{\Sigma}_k$ an inverse-Wishart prior distribution. 

% \mathbf{y}_{iA(l)} 1 by R
% \mathbf{\beta}_k R by p
% \mathbf{b}_{i} R by q
 Let $\mathbf{y}_{iA(l)} = (y_{1iA(l)},\dots,y_{RiA(l)})^T$, $\mathbf{\beta}_k =  (\mathbf{\beta}_{1k}^T,\dots,\mathbf{\beta}_{Rk}^T)^T$, and $\mathbf{b}_i = (\mathbf{b}_{1i}^T,\dots,\mathbf{b}_{Ri}^T)^T$. Assuming prior independence, we specify the joint posterior distribution as  \begin{eqnarray}
\lefteqn{p(\mathbf{c};\, \mathbf{\beta}, \mathbf{b}, \mathbf{\Sigma}, \mathbf{\Psi}; \, \mathbf{\phi}, \mathbf{\tau}, \mathbf{\Omega}; \, \mathbf{\lambda}, \mathbf{\kappa}, \mathbf{\Theta} \, | \, \mathbf{y}, \mathbf{d}, \mathbf{m}; \mathbf{x}, \mathbf{z}, \mathbf{w})}  \nonumber \\
& = &  \prod_{k = 1}^K \, \Bigg\{ \prod_{i = 1}^n \pi_{ik} \Bigg[ \left( \prod_{j = 1}^J  f(d_{ij} \, | \, \mathbf{\tau}_{i}, \mathbf{\phi}_k) \, f(\mathbf{\tau}_{i} \, | \, \mathbf{\Omega}_k) \right) \, \nonumber \\
& \times & \prod_{l=1}^{n_i} \left( f(\mathbf{y}_{iA(l)} \, | \, \mathbf{b}_i, \mathbf{\beta}_{k}, \mathbf{\Sigma}_k) \, f(\mathbf{b}_{i} \, | \, \mathbf{\Psi}_{k}) \prod_{r = 1}^R  f(m_{riA(l)} \, | \, \mathbf{\kappa}_{ri}, \mathbf{\lambda}_{rk}) \, f(\mathbf{\kappa}_{ri} \, | \, \mathbf{\Theta}_{rk}) \right) \Bigg]^{\mathbf{1}_{c_i = k}}  \, \nonumber \\
& \times & p(\mathbf{\beta}_{k}) \, p(\mathbf{\Sigma}_k) \, p(\mathbf{\Psi}_k) \, p(\mathbf{\phi}_k) \, p(\mathbf{\Omega}_k) \, \prod_{r = 1}^R  p(\mathbf{\lambda}_{rk}) \, p(\mathbf{\Theta}_{rk}) \Bigg\} \prod_{k = 1}^{K-1} p(\mathbf{\delta}_k), \nonumber
\end{eqnarray} where $p(.)$ indicates a prior distribution, and to simplify notation, the design matrices for $d_{ij}$, $\mathbf{y}_{iA(l)}$, and $m_{riA(l)}$ are suppressed.

We propose an MCMC algorithm that uses easily sampled closed-form full conditionals. Upon initialization, the algorithm iterates among the following steps: \begin{enumerate} 
	\item For $k = 1,\dots,K-1$, update $\mathbf{\delta}_k$ and $\xi_{ik}$ in (\ref{e:clatentvariable}). Compute $\pi_{ik}$ for $k = 1,\dots,K$ in (\ref{e:multinomial}).
	\item For $k = 1,\dots,K$, update $\mathbf{\beta}_{rk}$, $\mathbf{b}_{ri}$, $\mathbf{\Sigma}_{k}$, and $\mathbf{\Psi}_{k}$ in (\ref{e:modely}) and (\ref{e:modelb}).
	\item For $k = 1,\dots,K$, update $\mathbf{\phi}_{k}$, $\mathbf{\tau}_{i}$, and $\mathbf{\Omega}_{k}$ in (\ref{e:modeld}) and (\ref{e:modeltau}).
	\item For $k = 1,\dots,K$, update  $\mathbf{\lambda}_{rk}$, $\mathbf{\kappa}_{ri}$, and $\mathbf{\Theta}_{rk}$ in (\ref{e:modelm}) and (\ref{e:modelkappa}).
	\item  Sample latent class indicators $c_i$ for $i = 1,\dots,n$ from $Multinomial(1; \, p_{i1},\dots,p_{iK})$, where $p_{i1},\dots,p_{iK}$ are the posterior probabilities of latent class assignment given by 	\begin{eqnarray}
	p_{ik} 
	& = & Pr(c_i = k \, | \, \pi_{ik}; \, \mathbf{y}^{*}_{i}, \mathbf{b}_i; \, \mathbf{d}_{i}, \mathbf{\tau}_{i}; \, \mathbf{m}_{1i},\dots, \mathbf{m}_{Ri}, \mathbf{\kappa}_{1i},\dots,\mathbf{\kappa}_{Ri}; \, rest) \nonumber \\
	& \propto & \pi_{ik} \, f(\mathbf{y}^{*}_{i} \, | \, \mathbf{b}_i, \mathbf{\beta}_k, \mathbf{\Sigma}_k^{*}) \, f(\mathbf{b}_{i} \, | \, \mathbf{\Psi}_k) \, f(\mathbf{d}_{i} \, | \, \mathbf{\tau}_{i}, \mathbf{\phi}_k) \, f(\mathbf{\tau}_{i} \, | \, \mathbf{\Omega}_k) \nonumber \\
	& \times & \prod_{r = 1}^R f(\mathbf{m}_{ri} \, | \, \mathbf{\kappa}_{ri}, \mathbf{\lambda}_{rk}) \, f(\mathbf{\kappa}_{ri} \, | \, \mathbf{\Theta}_{rk}), \nonumber
	\end{eqnarray} where $\mathbf{y}^{*}_{i} = (\mathbf{y}_{iA(1)}^T,\dots, \mathbf{y}_{iA(n_{i})}^T)$, $\mathbf{d}_i = (d_{i1},\dots,d_{iJ})^T$, and $\mathbf{m}_{ri} = (m_{riA(1)},\dots,m_{riA(n_i)})^T$. $\mathbf{\Sigma}_k^{*}$ is an $n_i R \times n_i R$ block diagonal matrix with elements $\mathbf{\Sigma}_k$ ($R \times R$) for each $\mathbf{y}_{iA(l)}$ ($l=1,\dots,n_i$). 
\end{enumerate} 
The full MCMC algorithm is detailed in Section A of the supplementary material (SM).
%\mathbf{y}^{*}_{i}: R x n_i
%\mathbf{b}_i: R x q
%$\mathbf{d}_i = (d_{i1},\dots,d_{iJ})^T$, and $\mathbf{m}_{ri} = (m_{riA(1)},\dots,m_{riA(n_i)})^T$ for $r = 1,\dots,R$.

\section{Analysis of Early Childhood Weight and Height Measurements}

We apply our proposed model to an illustrative dataset of EHR measurements on weight and height in a sample of US children followed from birth to age 4 years. These EHR measurements were linked to participants in the 1988 National Maternal and Infant Health Survey (NMIHS) and its 1991 Longitudinal Follow-Up, in which low birth weight infants ($<$2,500 g) were oversampled \citep{Sanderson1988}. In this dataset, clinic visit times are available in terms of a child's age in months. Clinical recommendation suggests that in early childhood, weight and height measurements should be collected at clinic visits classified as well-child checks \citep{AAP2018}. The well-child check schedule prescribes clinic visits at age in months 1, 2, 4, 6, 9, 12, 15, 18, 24, 30, 36, and 48. To illustrate our proposed model, we used weight and height measurements from clinic visits classified as check-ups for a random sample of 500 children. We converted weight and height measurements to z-scores using a reference distribution from the Centers for Disease Control and Prevention \citep{CDC2019}. Of the 500 children, we excluded one child whose available measurements were flagged as biologically implausible values. SM Figure B.1 presents the patterns of observed visits and responses for weight and height given a clinic visit. Of 5,988 well-child windows (499 children $\times$ 12 well-child windows), 67\% correspond to missed visits. Among 1,983 observed visits, only 17 weight measurements are missing ($<$ 1\%), whereas 207 height measurements (10\%) are missing.

We analyze early childhood weight and height z-scores using three estimation methods that can be executed via our \texttt{R} package \texttt{EHRMiss}.  First, the \textbf{MNAR} method demonstrates our proposed model: We assume both the visit process and response process for height are MNAR, while since weight z-scores are rarely missing, the response process for weight is MAR. Second, in the \textbf{MAR} method, we assume each of the missing data mechanisms is ignorable. For the \textbf{Na\"ive} method, we fit the complete-data model using only well-child windows in which both weight and height z-scores are observed, herein ``complete pairs''. Whereas the \textbf{MNAR} and \textbf{MAR} methods include all 499 children (1,983 observed visits), the \textbf{Na\"ive} method uses only 471 children who have at least one complete pair, corresponding to 1,759 observed visits. 

We include a child's race, sex, and birth weight in $\mathbf{w}_i$ from the latent class membership submodel in (\ref{e:clatentvariable}). For weight and height z-scores, the visit process, and the response process for height z-scores given a clinic visit, we model longitudinal trajectories as a cubic polynomial function of a child's age in months, and the patient-specific random effects are specified by a random intercept. 

We ran the Gibbs sampler for 20,000 iterations discarding the first 10,000 as burn-in. Using three chains from dispersed initial values, the Gelman-Rubin diagnostic \citep{gelman2014bayesian} indicated model convergence with values near 1 for all parameters. In Bayesian mixture modeling, label switching is a well-known problem for posterior inference \citep{Fruhwirth-Schnatter2006}. We used Stephen’s relabeling method \citep{Stephens2000} to assess the label switching problem via the \texttt{R} package \texttt{label.switching} \citep{Papastamoulis2016}. This method identifies the labeling permutation that minimizes the Kullback-Leibler divergence between the posterior probabilities of latent class assignment averaged over MCMC iterations and the corresponding probabilities at each MCMC iteration. For each GMM in our data application, the original (identity) labeling was returned, which suggests that the label switching problem was not detected.  

We proceed in Section \ref{s:MNARMethod} by demonstrating the \textbf{MNAR} method in analyzing longitudinal trajectories of weight and height z-scores, the visit process, and the response process for height z-scores given a clinic visit, including selecting among models with varying numbers of latent classes and conducting model checking using the posterior predictive distribution. In Section \ref{s:Classification}, we use a 2-latent class model in order to simply explicate the patterns of differential child classification among the \textbf{Na\"ive}, \textbf{MAR}, and \textbf{MNAR} methods.

\subsection{Longitudinal trajectories of weight and height z-scores using the \textbf{MNAR} method}
\label{s:MNARMethod}

A challenge in data applications with GMMs is to select among models that assume a varying number of latent classes $K$. We compared different $K$-class models based on the \textbf{MNAR} method according to model information criteria, including the Bayesian Information Criterion (BIC) \citep{Schwarz1978} and a modified version of the Deviance Information Criterion (DIC) \citep{Spiegelhalter2002} known as the DIC3 recommended for latent variable models \citep{Celeux2006}; the log-pseudo marginal likelihood (LPML) \citep{Geisser1979,Gelfand1994,Ibrahim2001}; a graphical technique known as latent class identifiability displays (LCIDs) \citep{Garrett2000}; and, clinical interpretation. We selected the 3-class model. Details are provided in Section B of the SM.

\begin{figure}[!p]
	\centering\includegraphics[scale = 0.75]{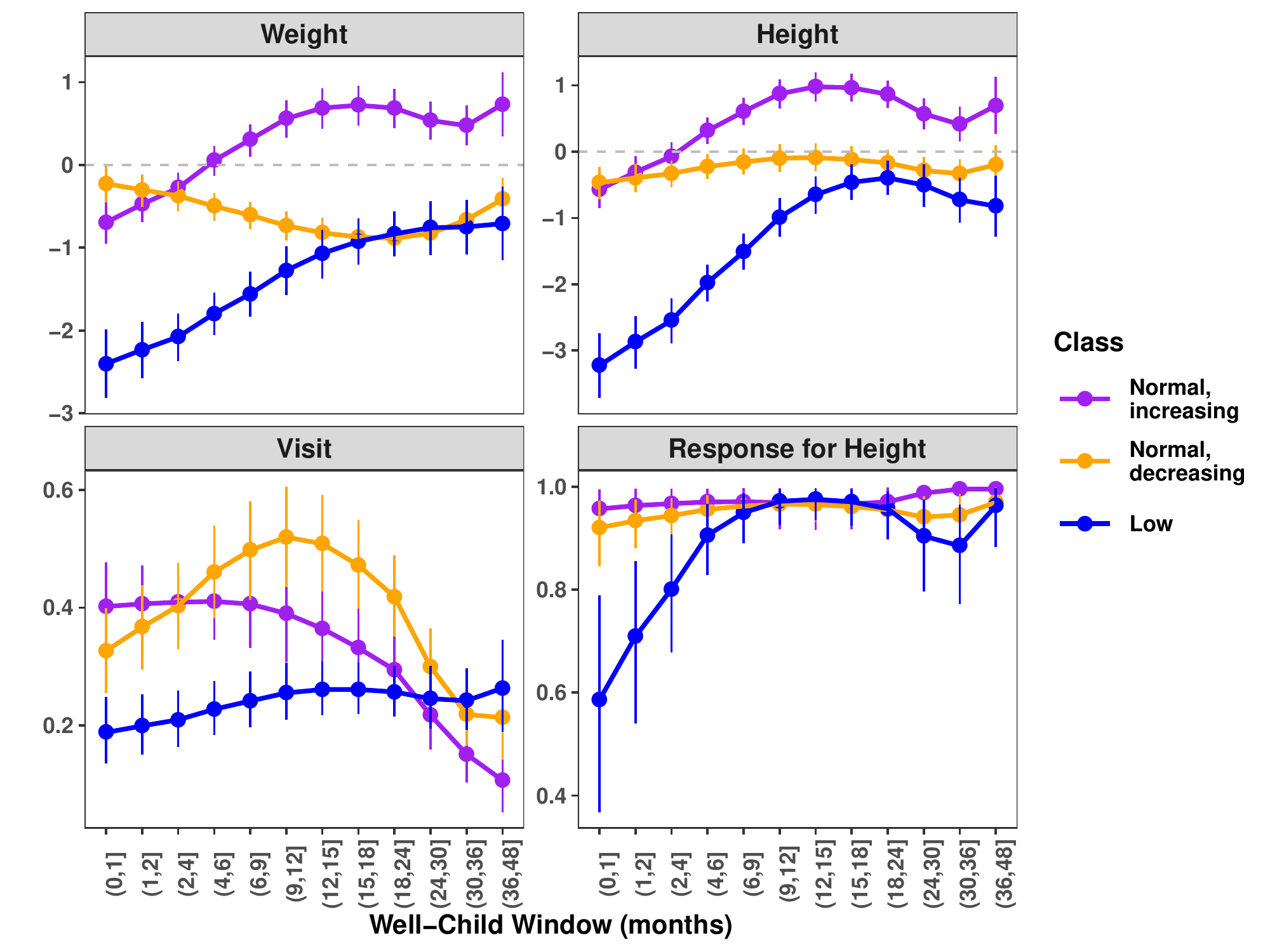}
	\caption{Latent class-specific average trajectories of weight and height z-scores, the probability of a clinic visit, and the probability of a response for height z-scores, estimated by the \textbf{MNAR} method.}
	\label{f:K3_LCSpecific_Y_DM_MNAR}
\end{figure}

Based on the 3-class model, Figure \ref{f:K3_LCSpecific_Y_DM_MNAR} shows the latent class-specific average trajectories of weight and height z-scores, the visit process, and the response process for height z-scores given a clinic visit. The longitudinal trajectories of weight and height z-scores, the visit process, and the response process for height z-scores exhibited latent heterogeneity. Using the weight trajectories to label the latent classes of children, we identified Normal, increasing (purple); Normal, decreasing (orange); and Low (blue) subgroups. The visit process of the Normal, increasing subgroup decreases over follow-up, whereas for the Normal, decreasing subgroup, the probability of a clinic visit rises at the outset before decreasing. The probability of response for height z-scores is indistinguishable for these two subgroups. In the Low subgroup, the probability of clinic visit rises slowly over follow-up, while the response process for height z-scores climbs sharply until about 12 months. Based on the maximum of a child's mean posterior probabilities of belonging to each latent class, we assigned approximately one-third of children to each subgroup, with subgroup mean (median) probability ranging from 0.81 to 0.84 (0.84 to 0.93) (SM Table B.2).

For model checking in the presence of missing data, we used the completed datasets that include observed and imputed weight and height z-scores in each well-child window, and replicates of the completed datasets drawn from the posterior predictive distribution \citep{Gelman2005}. We conducted Bayesian posterior predictive checking using the multivariate mean square error \citep{Daniels2008} as our discrepancy measure, %page 67 in Daniels 20
\begin{eqnarray}
T & = &  \sum_{k=1}^K \sum_{i=1}^n  \sum_{l=1}^{n_{i}} (\mathbf{y}_{iA(l)} - \mathbf{\mu}_{iA(l)}) \mathbf{\Sigma}_k^{-1}(\mathbf{y}_{iA(l)} - \mathbf{\mu}_{iA(l)})^T \times \mathbf{1}_{c_{i} = k}, \label{e:discrepancy} \end{eqnarray}  where $\mathbf{\mu}_{iA(l)} = \mathbf{x}_{iA(l)} \mathbf{\beta}_k^T  +  \mathbf{z}_{iA(l)} \mathbf{b}_i^T$. 
SM Figure B.5 presents a scatter plot of the discrepancy measure $T$ in (\ref{e:discrepancy}) across MCMC samples, with the horizontal and vertical axes being $T$ based on the completed and replicated datasets, respectively. Comparing completed and replicated $T$, the Bayesian predictive p-value of 0.44 suggests adequate overall model fit. In addition, we compared histograms of randomly selected completed and replicated datasets of weight and height z-scores \citep{Gelman2005}. In SM Figures B.6 and B.7, the distribution of z-scores by subgroup and well-child window appears largely consistent between the completed and replicated datasets.

\subsection{Child classification using the different estimation methods in 2-class models}
\label{s:Classification}

Based on the simplifying assumption of two latent classes, we examine the patterns of differential child classification among the \textbf{Na\"ive}, \textbf{MAR}, and \textbf{MNAR} methods. Herein, after briefly describing analysis results under each method, we focus our presentation on classification patterns. See Section B of the SM for details. 

The \textbf{Na\"ive}, \textbf{MAR}, and \textbf{MNAR} methods each detected a Normal trajectory subgroup (purple) and a Low trajectory subgroup (orange) (SM Figure B.8). Despite similar trajectory patterns across methods, the latent classes appear better separated in the \textbf{MNAR} method, particularly for height z-scores for which the response process was modeled. Based on the \textbf{MNAR} method, SM Figure B.9 shows that compared to the Low subgroup, the Normal subgroup generally exhibits a higher probability of a clinic visit. Whereas in the Normal subgroup, the probability of a height response is invariably near 1, in the Low subgroup, the response process climbs sharply at the outset. SM Table B.3 presents a summary of posterior latent class assignment under the three methods. The \textbf{MNAR} method assigned about 8\% fewer children to the Normal subgroup than the other methods. The mean (median) probability of latent class assignment in each subgroup ranged from 0.87 to 0.93 (0.92 to 0.99). 

\begin{sidewaysfigure}[t!]
	\centering
	\begin{minipage}{0.5\textwidth}
		\centerline{\includegraphics[scale = 0.75]{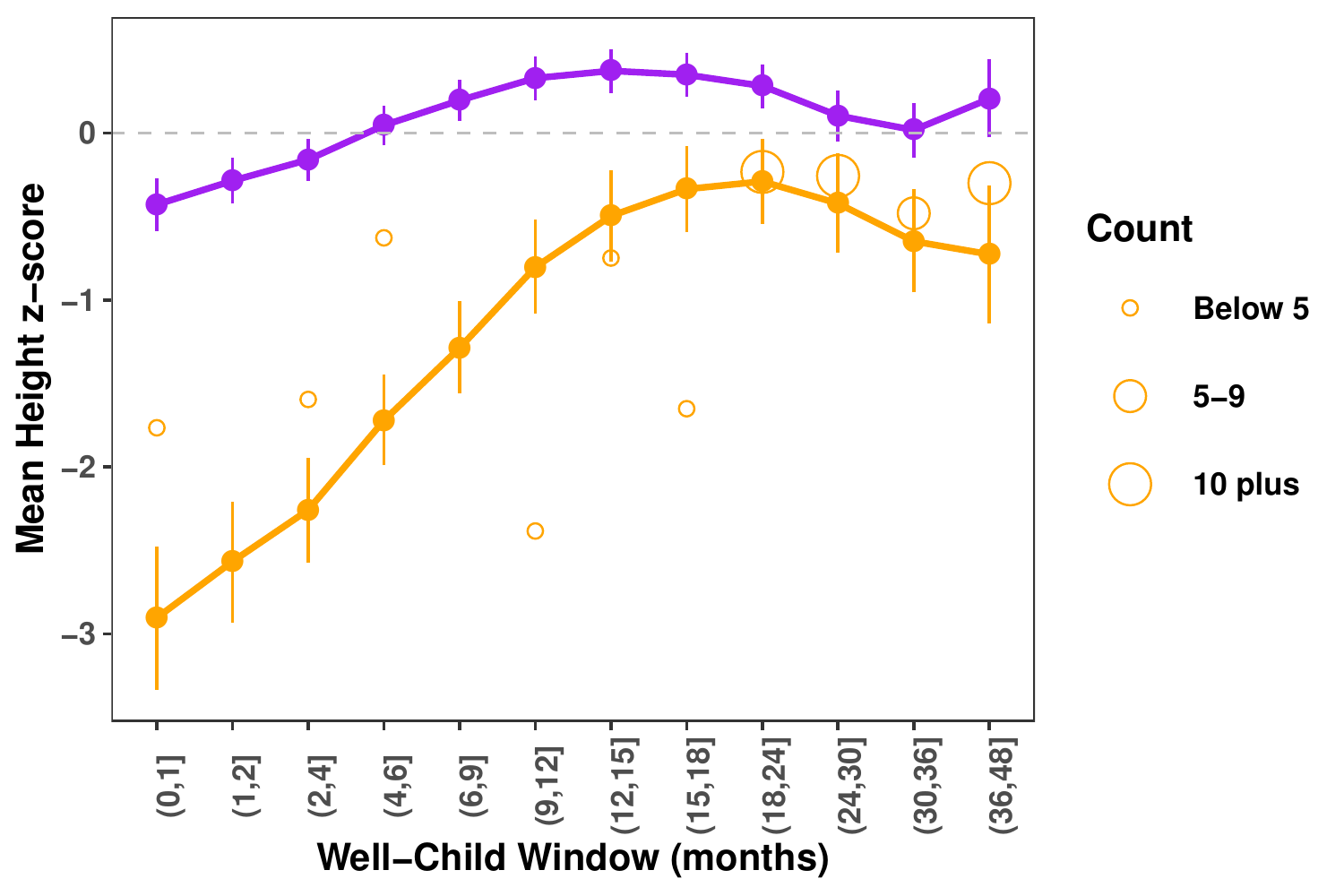}}
	\end{minipage}%
	\begin{minipage}{0.5\textwidth}
		\centerline{\includegraphics[scale = 0.75]{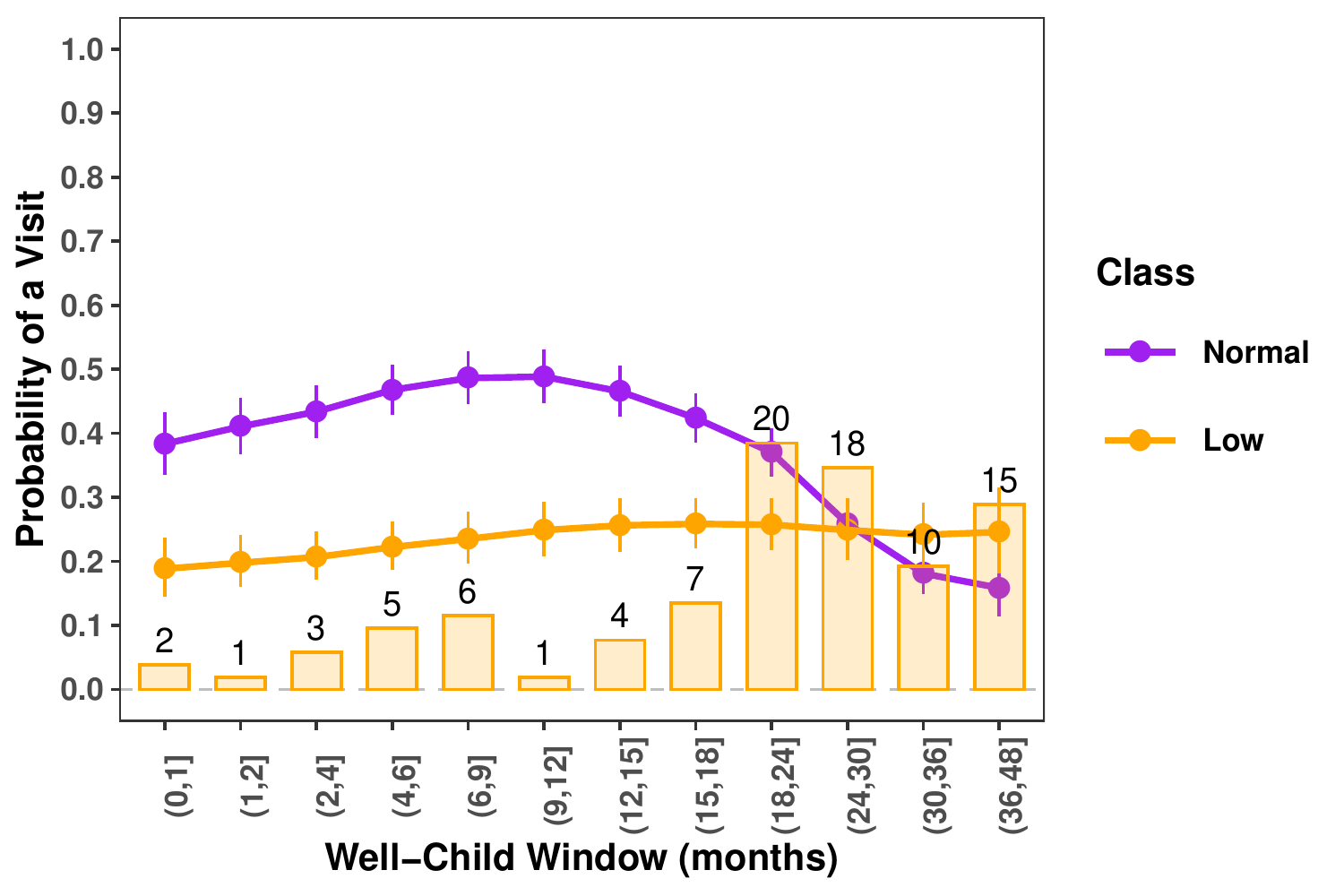}}
	\end{minipage}
	\caption{Characterizing the 52 non-low birth weight children classified into the Low trajectory subgroup in the \textbf{MNAR} method but into the Normal trajectory subgroup in the \textbf{MAR} method, assuming 2 latent classes. The left panel shows the sample means of observed height z-scores (hollow circles) in each well-child window, with the circle size indicating the number of measurements. The right panel shows a bar plot of the observed proportions of children with a clinic visit. The count of children with an observed visit in each window is appended. In each panel, corresponding average latent class-specific trajectories estimated with the \textbf{MNAR} method are overlaid.}
	\label{f:NormaltoLow_nonLBW_K2_MAR_MARMNAR_MNAR}
\end{sidewaysfigure}

To illustrate patterns of differential child classification by estimation method, we compare the \textbf{MAR} versus \textbf{MNAR} methods that used all 499 children. SM Table B.4 cross-classifies the 499 children by their latent class assignment from the \textbf{MAR} and \textbf{MNAR} methods, and the birth weight variable from the latent class membership model. Since few low birth weight (LBW) children were classified differently between the two methods, we focus on the two off-diagonal cells for children born non-LBW. First, 52 non-LBW children were placed in the Normal subgroup by the \textbf{MAR} method but the Low subgroup by the \textbf{MNAR} method. For height z-scores, the left panel in Figure \ref{f:NormaltoLow_nonLBW_K2_MAR_MARMNAR_MNAR} shows the sample means among the 52 children using their observed measurements, overlaid on the average latent class-specific trajectories estimated by the \textbf{MNAR} method. Larger circles indicate sample means with more observed measurements. Sample means with more measurements appear in later follow-up when the latent class-specific trajectories are similar. In fact, the 52 children have few observed measurements in early follow-up when the class trajectories are easily distinguished. In Figure \ref{f:NormaltoLow_nonLBW_K2_MAR_MARMNAR_MNAR}, the right panel shows the pattern of the proportions of observed visits in each well-child window among the 52 children, overlaid by the average latent class-specific visit trajectories. Consistent with the \textbf{MNAR} method classifying the children in the Low subgroup, the observed visit pattern resembles the Low trajectory. 

\begin{sidewaysfigure}[t!]
	\centering
	\begin{minipage}{0.5\textwidth}
		\centerline{\includegraphics[scale = 0.75]{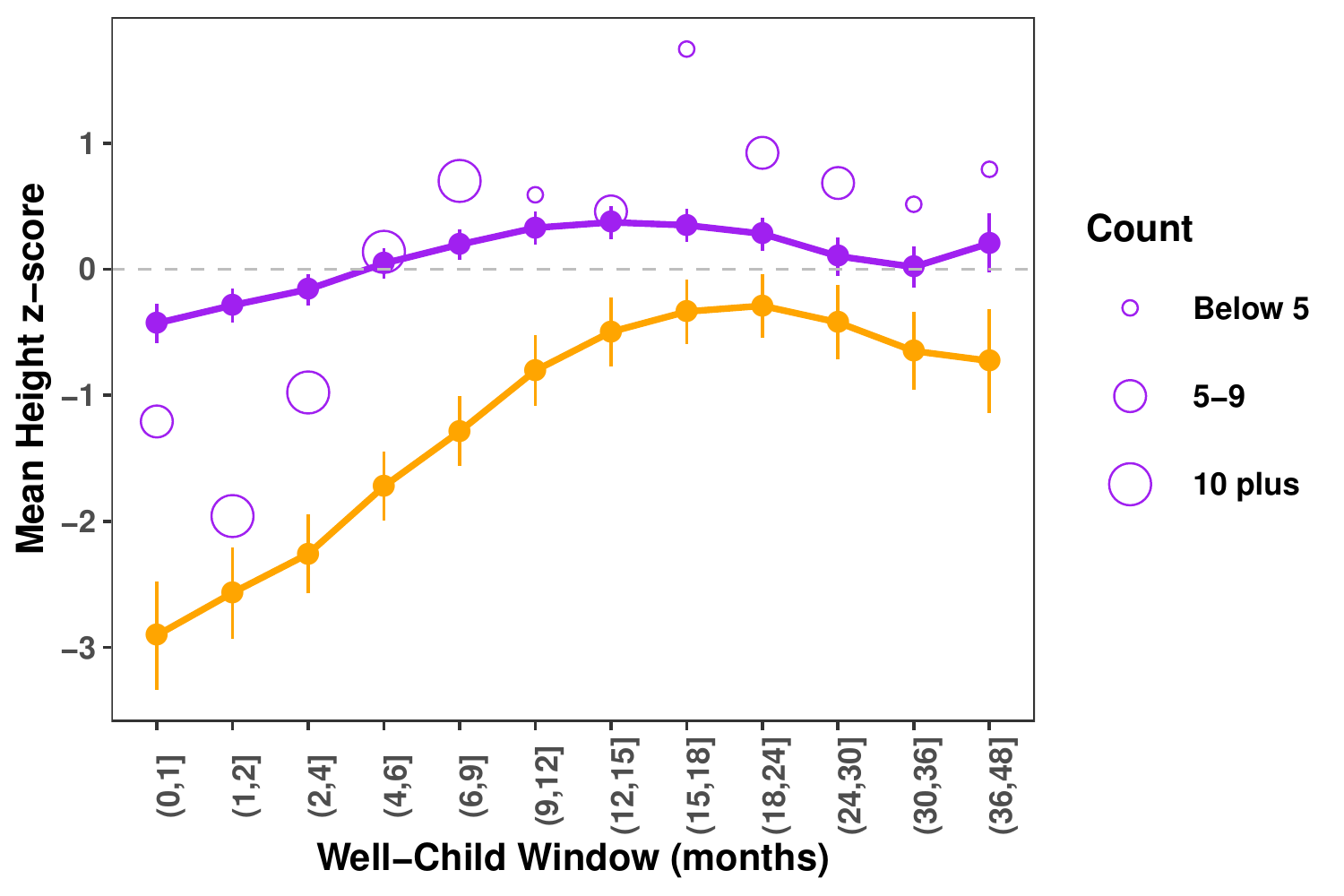}}
	\end{minipage}%
	\begin{minipage}{0.5\textwidth}
		\centerline{\includegraphics[scale = 0.75]{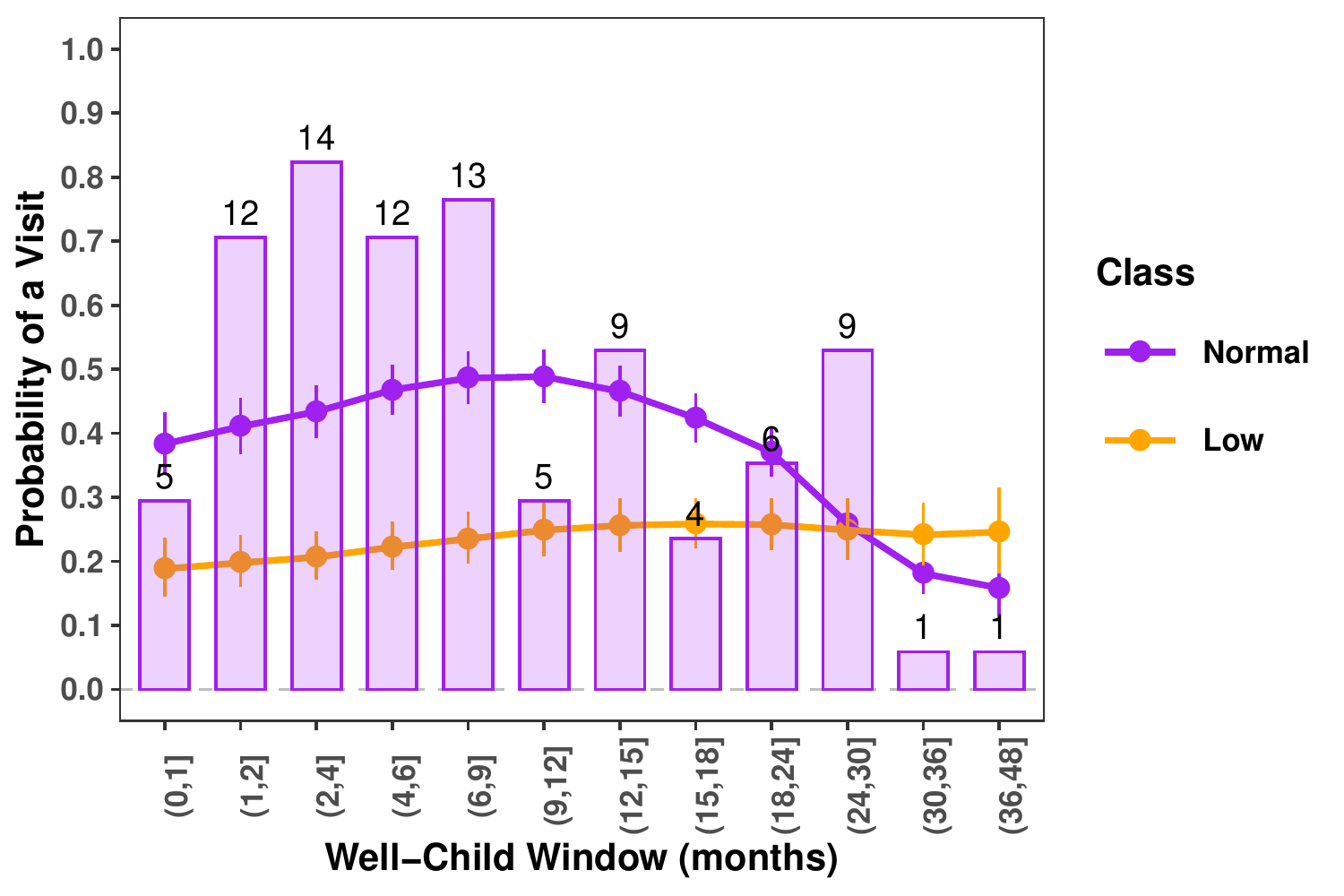}}
	\end{minipage}
	\caption{Characterizing the 17 non-low birth weight children classified into the Normal trajectory subgroup in the \textbf{MNAR} method but into the Low trajectory subgroup in the \textbf{MAR} method, assuming 2 latent classes. The left panel shows sample means of observed height z-scores (hollow circles) in each well-child window, with the circle size indicating the number of measurements. The right panel shows a bar plot of the observed proportions of children with a clinic visit. The count of children with an observed visit in each window is appended. In each panel, corresponding average latent class-specific trajectories estimated with the \textbf{MNAR} method are overlaid.}
	\label{f:LowtoNormal_nonLBW_K2_MAR_MARMNAR_MNAR}
\end{sidewaysfigure}

In the second off-diagonal cell, 17 non-LBW children were placed in the Low subgroup by the \textbf{MAR} method but the Normal subgroup by the \textbf{MNAR} method (Table B.4). In contrast to the 52 children, the 17 children have more observed height z-scores in early follow-up when the Low and Normal trajectories are easily distinguished (Figure \ref{f:LowtoNormal_nonLBW_K2_MAR_MARMNAR_MNAR}, left panel). However, during this period, the observed sample means among the 17 children are located in between the Low and Normal trajectories, rather than showing a clear classification. The \textbf{MNAR} method classified the 17 children in the Normal subgroup because their pattern of proportions of observed visits correspond to the visit process trajectory in the Normal subgroup (Figure \ref{f:LowtoNormal_nonLBW_K2_MAR_MARMNAR_MNAR}, right panel).

The comparison of the \textbf{Na\"ive} and \textbf{MNAR} methods for the 471 common children revealed patterns of classification similar to those heretofore described for the \textbf{MAR} and \textbf{MNAR} methods (data not shown).

\section{Simulation Study}

We conducted a simulation study to examine the effect of estimation method on estimating the latent class-specific average health trajectories $\mathbf{\beta}_{rk}$ in (\ref{e:modely}); and, in predicting a subject's true latent class assignment. In addition, since scientific inquiry may concern the average health trajectory over time, we considered the effect of estimation method on marginal regression coefficients obtained by averaging $\mathbf{\beta}_{rk}$ over the latent class membership probabilities $\pi_{ik}$ in (\ref{e:multinomial}). For example, for longitudinal health outcome $r$, the marginal intercept is given by $\tilde{\beta}_{r1} = \frac{1}{n} \sum_{i=1}^n \sum_{k=1}^K \pi_{ik} \beta_{rk1}$. Here, we summarize the design and results, with details in Section C of the SM.

\subsection{Design}

Based on the real data analysis for $K=2$ using the \textbf{MNAR} method, we generated longitudinal outcomes $y_{1}$ and $y_{2}$ over 12 time windows for 500 subjects, with about 60\% and 40\% of subjects in classes 1 and 2, respectively. We assumed an MNAR visit process and response process for $y_{2}$, while $y_{1}$ is fully observed given a clinic visit. We then considered five scenarios: S0 is the baseline scenario in which we mimic the latent class-specific average trajectories and missingness proportions in the real data analysis. True parameter values for $\mathbf{\beta}_{rk}$ (\ref{e:modely}), $\mathbf{\phi}_{k}$ (\ref{e:modeld}), and $\mathbf{\lambda}_{2k}$ (\ref{e:modelm}) were selected to linearly summarize the estimated trajectories. Latent class 1 is characterized by 55\%  missed clinic visits and 10\% missed $y_{2}$ responses. The corresponding values in class 2 are 70\% and 20\%. SM Figure C.1 depicts S0 for $y_{1}$ and $y_{2}$: Corresponding to Figures \ref{f:NormaltoLow_nonLBW_K2_MAR_MARMNAR_MNAR} and \ref{f:LowtoNormal_nonLBW_K2_MAR_MARMNAR_MNAR} in the real data analysis, in early follow-up when the latent class-specific average trajectories are better separated, missingness in $y_{2}$ is high in class 2, while in later follow-up, missingness in $y_{2}$ is high in class 1. 

S1 -- S4 make selected changes to S0, as shown in SM Figure C.2 for $y_{2}$. S1 and S2 consider whether the effect of estimation method varies by the degree to which the slopes are different for the latent class-specific average trajectories of $y_{2}$. In S1, we made the slopes more different, while in S2, we made them more similar. S3 and S4 examine whether the effect of estimation method varies by the extent of missingness from the visit and response processes whilst maintaining the shapes of their latent class-specific average trajectories. In S3, we reduced the percent of missed clinic visits to 35\% in class 1 and 55\% in class 2. In S4, we increased the percent of missed $y_{2}$ responses to 25\% and 35\% in classes 1 and 2, respectively. 

For S0--S4, we compare estimation using the \textbf{MNAR} method to the \textbf{MAR} and \textbf{Na\"ive} methods, based on $K=2$. For a benchmark, we also include the \textbf{Full} method, in which the complete-data model is fit to the full data before introducing any missed visits or responses. We ran 500 data simulations. For $\mathbf{\beta}_{rk}$ and the marginal effects, we examined bias, mean squared error (MSE), 95\% coverage probability, and the average length of the 95\% credible interval. For subject classification, we considered summary statistics of the proportion of misclassified subjects in each simulation. 

\subsection{Results}

Table \ref{t:param_DG1} shows S0 results. Estimation under the \textbf{Full} method presents the benchmark. For the latent class-specific parameters, compared to the \textbf{Na\"ive} and \textbf{MAR} methods, the \textbf{MNAR} method largely exhibits the smallest bias, the smallest MSE, coverage probability nearest to the nominal level, and the shortest interval length. For example, for $y_{2}$, while the slope in latent class 2, $\beta_{222}$, is estimated with negative bias and poor coverage using the \textbf{Na\"ive} and \textbf{MAR} methods, bias and coverage under the \textbf{MNAR} method are comparable to the \textbf{Full} method. The subpar performance of the \textbf{Na\"ive} and \textbf{MAR} methods appears to be driven by subject misclassification from class 1 to 2. With respect to the marginal effects, the \textbf{MNAR} method again outperforms the \textbf{Na\"ive} and \textbf{MAR} methods, demonstrating the smallest bias and MSE and highest coverage probability. However, coverage falls below the nominal level, ranging from 0.89 to 0.93. Even though the \textbf{Na\"ive} and \textbf{MAR} methods show shorter interval length than the \textbf{MNAR} method, their coverage probabilities are markedly lower.

\begin{table}[ht]
\centering
\caption{Simulation results of S0 for parameter estimation of intercept $\beta_{rk1}$ and slope $\beta_{rk2}$ for longitudinal outcome $r$ in latent class $k$, along with the corresponding marginal intercept and slope, $\tilde{\beta}_{r1}$ and $\tilde{\beta}_{r2}$, respectively, under the \textbf{Full}, \textbf{Na\"ive}, \textbf{MAR}, and \textbf{MNAR} methods. The \textbf{Full} method is the benchmark. The best performing method among the \textbf{Na\"ive}, \textbf{MAR}, and \textbf{MNAR} methods is in bold.} 
\label{t:param_DG1}
\small
 \begin{tabular}{cclcrrrr}
		\hline
		\multicolumn{1}{c}{Outcome} & \multicolumn{1}{c}{Parameter} & \multicolumn{1}{c}{Method} & \multicolumn{1}{c}{Truth} & \multicolumn{1}{c}{Bias} & \multicolumn{1}{c}{MSE} & \multicolumn{1}{c}{Coverage} & \multicolumn{1}{c}{Length} \\ 
		\hline
		\multirow{16}{*}{$y_{1}$} & \multirow{4}{*}{\shortstack{$\beta_{111}$ \\ \vspace{0.1cm} \\ \footnotesize{(Class 1 Intercept)}}} & Full &  \multirow{4}{*}{-0.250}  &  -0.002 & 0.002 & 0.950 & 0.190 \\ 
		& & Na\"ive &  &  0.029 & 0.005 & 0.904 & 0.224 \\ 
		& & MAR &  &  0.019 & 0.004 & 0.908 & 0.219 \\ 
		& & MNAR &  &  \textbf{0.002} & \textbf{0.003} & \textbf{0.942} & \textbf{0.209} \\ \cline{2-8}
		& \multirow{4}{*}{\shortstack{$\beta_{121}$ \\ \vspace{0.1cm} \\ \footnotesize{(Class 2 Intercept)}}} & Full & \multirow{4}{*}{-1.000} &  0.000 & 0.003 & 0.956 & 0.230 \\ 
		& & Na\"ive &  &  0.046 & 0.016 & 0.878 & 0.404 \\ 
		& & MAR &  &  \textbf{0.004} & 0.011 & 0.932 & 0.370 \\ 
		& & MNAR &  & 0.005 & \textbf{0.007} & \textbf{0.936} & \textbf{0.312} \\ \cline{2-8}
		& \multirow{4}{*}{\shortstack{$\beta_{112}$ \\ \vspace{0.1cm} \\ \footnotesize{(Class 1 Slope)}}} & Full & \multirow{4}{*}{0.100} & -0.000 & 0.000 & 0.930 & 0.048 \\ 
		& & Na\"ive &  & -0.011 & \textbf{0.001} & 0.928 & 0.099 \\ 
		& & MAR &   & -0.008 & \textbf{0.001} & 0.926 & 0.094 \\ 
		& & MNAR &  &  \textbf{-0.000} & \textbf{0.001} & \textbf{0.954} & \textbf{0.091} \\ \cline{2-8}
		& \multirow{4}{*}{\shortstack{$\beta_{122}$ \\ \vspace{0.1cm} \\ \footnotesize{(Class 2 Slope)}}} & Full & \multirow{4}{*}{0.500} &  0.001 & 0.001 & 0.930 & 0.096 \\ 
		& & Na\"ive &  &  -0.089 & 0.013 & 0.720 & 0.266 \\ 
		& & MAR &  &  -0.041 & 0.007 & 0.850 & 0.238 \\ 
		& & MNAR &  &  \textbf{-0.001} & \textbf{0.003} & \textbf{0.948} & \textbf{0.215} \\ \hline
		\multirow{16}{*}{$y_{2}$} & \multirow{4}{*}{\shortstack{$\beta_{211}$ \\ \vspace{0.1cm} \\ \footnotesize{(Class 1 Intercept)}}} & Full & \multirow{4}{*}{0.500} &  -0.000 & 0.002 & 0.954 & 0.189 \\ 
		& & Na\"ive &  &  0.045 & 0.006 & 0.858 & 0.224 \\ 
		& & MAR &  &  0.036 & 0.005 & 0.886 & 0.221 \\ 
		& & MNAR &  &  \textbf{0.005} & \textbf{0.003} & \textbf{0.938} & \textbf{0.210} \\   \cline{2-8}
		& \multirow{4}{*}{\shortstack{$\beta_{221}$ \\ \vspace{0.1cm} \\ \footnotesize{(Class 2 Intercept)}}} & Full & \multirow{4}{*}{-0.500} &  -0.003 & 0.003 & 0.940 & 0.196 \\ 
		& & Na\"ive &  &  0.048 & 0.015 & 0.896 & 0.379 \\ 
		& & MAR &  &  0.026 & 0.011 & 0.922 & 0.366 \\ 
		& & MNAR &  &  \textbf{0.000} & \textbf{0.007} & \textbf{0.956} & \textbf{0.310} \\  \cline{2-8}
		& \multirow{4}{*}{\shortstack{$\beta_{212}$ \\ \vspace{0.1cm} \\ \footnotesize{(Class 1 Slope)}}} & Full & \multirow{4}{*}{0.200} &  -0.001 & 0.000 & 0.918 & 0.048 \\ 
		& & Na\"ive &  &  -0.015 & \textbf{0.001} & 0.904 & 0.098 \\ 
		& & MAR &   & -0.014 & \textbf{0.001} & 0.880 & 0.096 \\ 
		& & MNAR &  &  \textbf{-0.000} & \textbf{0.001} & \textbf{0.950} & \textbf{0.093} \\  \cline{2-8}
		& \multirow{4}{*}{\shortstack{$\beta_{222}$ \\ \vspace{0.1cm} \\ \footnotesize{(Class 2 Slope)}}} & Full & \multirow{4}{*}{0.750} & 0.001 & 0.001 & 0.934 & 0.097 \\ 
		& & Na\"ive &  &  -0.102 & 0.017 & 0.646 & 0.270 \\ 
		& & MAR &  &  -0.075 & 0.012 & 0.738 & 0.262 \\ 
		& & MNAR &  &  \textbf{-0.003} & \textbf{0.004} & \textbf{0.944} & \textbf{0.237} \\ \hline
	\multirow{8}{*}{$y_1$} & \multirow{4}{*}{\shortstack{$\tilde{\beta}_{11}$ \\ \vspace{0.1cm} \\ \footnotesize{(Marginal Intercept)}}} & Full & \multirow{4}{*}{-0.582} & -0.001 & 0.001 & 0.950 & 0.147 \\ 
   &  & Na\"ive &  & 0.082 & 0.009 & 0.574 & 0.182 \\ 
   &  & MAR &  & 0.053 & 0.005 & 0.774 & \textbf{0.176} \\ 
   &  & MNAR &  & \textbf{0.017} & \textbf{0.003} & \textbf{0.896} & \textbf{0.176} \\ 
   \cline{2-8} & \multirow{4}{*}{\shortstack{$\tilde{\beta}_{12}$ \\ \vspace{0.1cm} \\ \footnotesize{(Marginal Slope)}}} & Full & \multirow{4}{*}{0.277} & 0.000 & 0.000 & 0.952 & 0.051 \\ 
   &  & Na\"ive &  & -0.065 & 0.005 & 0.364 & 0.109 \\ 
   &  & MAR &  & -0.043 & 0.003 & 0.610 & \textbf{0.101} \\ 
   &  & MNAR &  & \textbf{-0.008} & \textbf{0.001} & \textbf{0.926} & 0.104 \\ 
   \cline{2-8}\multirow{8}{*}{$y_2$} & \multirow{4}{*}{\shortstack{$\tilde{\beta}_{21}$ \\ \vspace{0.1cm} \\ \footnotesize{(Marginal Intercept)}}} & Full & \multirow{4}{*}{0.057} & -0.001 & 0.001 & 0.954 & 0.138 \\ 
   &  & Na\"ive &  & 0.107 & 0.014 & 0.364 & \textbf{0.176} \\ 
   &  & MAR &  & 0.085 & 0.010 & 0.492 & \textbf{0.176} \\ 
   &  & MNAR &  & \textbf{0.021} & \textbf{0.003} & \textbf{0.886} & 0.177 \\ 
   \cline{2-8} & \multirow{4}{*}{\shortstack{$\tilde{\beta}_{22}$ \\ \vspace{0.1cm} \\ \footnotesize{(Marginal Slope)}}} & Full & \multirow{4}{*}{0.444} & -0.000 & 0.000 & 0.944 & 0.053 \\ 
   &  & Na\"ive &  & -0.082 & 0.008 & 0.188 & 0.109 \\ 
   &  & MAR &  & -0.067 & 0.006 & 0.350 & \textbf{0.108} \\ 
   &  & MNAR &  & \textbf{-0.011} & \textbf{0.001} & \textbf{0.918} & 0.113 \\ 
   \hline
    \end{tabular}
\end{table}

\begin{figure}[htbp]
	\centering
	\centerline{\includegraphics[scale = 0.75]{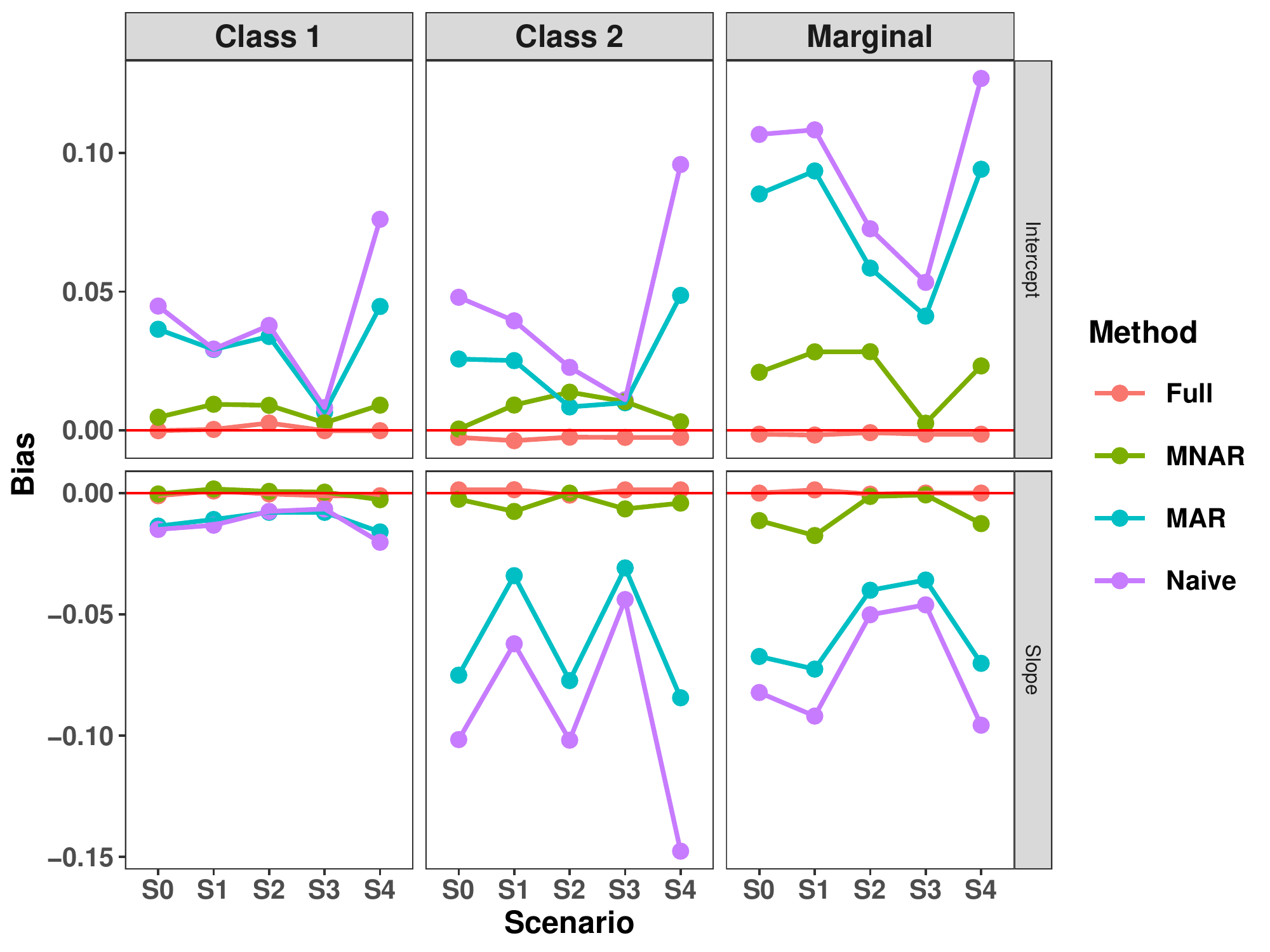}}
	\caption{Comparison of bias in parameter estimation for $y_{2}$ across data generation scenarios, by estimation method.}
	\label{f:sim_results_y2}
\end{figure}

Full simulation results for S1--S4 are provided in SM Tables C.1--C.4. The performance of the \textbf{Full} and \textbf{MNAR} methods is robust to these different data generation scenarios. Figure \ref{f:sim_results_y2} highlights how bias changes by each data generation scenario and estimation method for $y_{2}$. Overall, the \textbf{MNAR} method outperforms the \textbf{Na\"ive} and \textbf{MAR} methods. In terms of the latent class-specific parameters, while the \textbf{MNAR} method performs on par with the \textbf{Full} method, using the \textbf{Na\"ive} and \textbf{MAR} methods, the degree of bias is contingent on the specific scenario and parameter. For example, for the intercept in class 1 ($\beta_{211}$) and the slope in class 2 ($\beta_{222}$), bias under the \textbf{Na\"ive} and \textbf{MAR} methods decreases when the slopes are more different (S1) versus less different (S2). For all class-specific parameters, bias decreases when visit process missingness is reduced (S3), and bias increases when response process missingness given a clinic visit is increased (S4). With respect to the marginal effects, bias under the \textbf{Na\"ive} and \textbf{MAR} methods is smaller in S2 and S3 compared to the other scenarios. The corresponding bias comparison for $y_{1}$ in SM Figure C.3 shows similar patterns of results.

In SM Table C.5, summary statistics of subject misclassification, including the $25^{th}$, median, and $75^{th}$ percentiles, and the minimum and maximum, invariably show the advantage of the \textbf{MNAR} method compared to the \textbf{Na\"ive} and \textbf{MAR} methods across the data generation scenarios. For example, in S0, the \textbf{Full} and \textbf{MNAR} methods demonstrated a median proportion of subjects who are misclassified of 0.02 and 0.03, respectively, whereas the misclassification proportion was 0.15 using the \textbf{Na\"ive} method and 0.14 using the  \textbf{MAR} method.

\section{Discussion}

In this study, we developed a Bayesian shared parameter model for multiple longitudinal health outcomes in EHRs to account for a nonignorable visit process and response process given a clinic visit. Our proposed model targets multiple longitudinal health outcomes collected according to a clinically prescribed visit schedule. To account for underlying heterogeneity in EHR patient populations, we used a discrete latent class variable to link GMMs of the longitudinal health outcomes, the visit process, and the response process of individual health outcomes. The use of the discrete latent class variable allowed us to relax the assumption of a single, homogeneous patient population while tractably summarizing innumerable patterns of missingness from the visit and response processes into a small number of latent classes. Particularly important to EHR-based clinical research, we can easily modify our proposed Bayesian shared parameter model in order to conduct a sensitivity analysis about MAR versus MNAR missing data mechanisms for either or both the visit process and the response process given a clinic visit. Example code for model fitting, selection, and checking with our user-friendly \texttt{R} package \texttt{EHRMiss} is in Section D of the SM.

Our proposed Bayesian shared parameter model used a discrete latent class variable, targeted multiple longitudinal health outcomes, and distinguished between the visit process and the response process of individual health outcomes. In contrast, in a large clinical database, \citet{McCulloch2016} proposed a shared parameter model for a univariate longitudinal health outcome. The authors defined a single missing data process -- which they call the visit process -- as a binary indicator for whether a response on the longitudinal health outcome was observed at given time (corresponding to our definition of the response process given a clinic visit). Patient-specific random effects are used as the shared parameter. Notwithstanding their different modeling framework, \citet{McCulloch2016} show analytically that in the absence of accounting for an informative visit process, estimators of regression coefficients associated with the random effects can be badly biased. In our data simulations, we show that failure to account for a nonignorable visit process and response process given a clinic visit may result in biased estimation of latent class-specific average health trajectories, depending on whether the latent classes are well-identified. Identification of the latent classes depends on the extent to which the latent class-specific average health trajectories are different, and the extent to which the degree of missingness permits correctly classifying patients based on their observed longitudinal health outcomes. Even when estimated latent class-specific average health trajectories are largely unbiased, the marginal regression coefficients, which depend on both the class-specific trajectories and the latent class membership probabilities $\pi_{ik}$, may be poorly estimated.  

In our data application, the assumption of latent heterogeneity in weight and height z-scores, the visit process, and the response process for height z-scores given a clinic visit appeared warranted. Through the discrete latent class variable for a child’s latent class membership, the visit process and the response process for height z-scores informed parameter estimation in the longitudinal model of weight and height z-scores. The role of the visit and response processes was especially evident in the two scenarios depicted by our data application. In the first scenario (Figure \ref{f:NormaltoLow_nonLBW_K2_MAR_MARMNAR_MNAR}), a child did not have observed height z-scores during early follow-up when the latent class-specific average height z-score trajectories were easily distinguishable. In the second scenario, a child had observed z-scores during this period of follow-up, but they did not suggest a clear latent classification despite the well-separated latent class-specific average trajectories (Figure \ref{f:LowtoNormal_nonLBW_K2_MAR_MARMNAR_MNAR}). In both scenarios, the proposed Bayesian shared parameter model used a child’s patterns of observed visits and responses for height z-scores to help predict latent class membership.

We are primarily interested in two areas for future research. In this work, we were motivated by longitudinal health outcomes in EHRs with a clinically prescribed visit schedule, which we used to discretize time into observation windows during which to measure the visit process and response process given a clinic visit. However, when a prescribed visit schedule is unavailable, measuring the visit process in continuous time is consistent with the data generation in EHRs, since a patient can show up for a clinic visit at any time. We are currently modifying the proposed model to handle continuous time. Second, Bayesian methods can be especially time intensive as the number of observations grows. To enhance the practicality of our proposed model for EHR-based research, we are interested in pursuing strategies for scaling MCMC algorithms to large datasets.

EHRs are increasingly used for applied biomedical research. Rigorous treatment of the two patient-led missing data processes in EHRs, namely, the visit process and the response process of individual health outcomes given a clinic visit, may help to validate clinical findings and to stratify patient risk profiles. The proposed Bayesian shared parameter model for EHRs can be used to evaluate missing data assumptions in scientific inquiries about discovering clinically meaningful subpopulations or population-averaged associations of longitudinal health outcomes with an exposure of interest. Information contained in each patient’s visit and response processes may be valuable for allocating resources towards at-risk patient subgroups that would benefit from increased monitoring in a health care setting. Our proposed model may be applicable to other routinely collected data sources, like medical claims data.

\section{Software}
To conduct model fitting, selection, and checking, we built the user-friendly \texttt{R} package \texttt{EHRMiss} available at \url{https://github.com/anthopolos/EHRMiss}. Example code for analysis with  \texttt{EHRMiss} is in Section D in the SM.

\section{Supplementary Material}

The reader is referred to the on-line Supplementary Materials for explication of the MCMC algorithm for the proposed Bayesian shared parameter model; an addendum to the data application in early childhood weight and height measurements; explication of the simulation study design with additional results; and demonstration of the \texttt{R} package \texttt{EHRMiss}.

\section*{Acknowledgments}
Ying Wei was supported by NIH grant R01HG008980 and NSF grant DMS-1953527. Qixuan Chen was supported by NIH grant R21ES029668.  
{\it Conflict of Interest}: None declared.

\bibliographystyle{plainnat}
\bibliography{ms_bibliography}

\end{document}

\maketitle

\renewcommand{\thesection}{\Alph{section}}
\renewcommand{\thetable}{\Alph{section}.\arabic{table}}
\renewcommand{\thefigure}{\Alph{section}.\arabic{figure}}

\section{MCMC Algorithm}
\label{a:MCMCAlgorithm}
\setcounter{table}{0}    
\setcounter{figure}{0}

We explicate the MCMC algorithm for fitting the proposed shared parameter model to EHRs. In the main text, for ease of model explication, we presented a non-centered parametrization by specifying the random effects $\mathbf{b}_{ri}$ in the longitudinal health outcome model with mean zero. However, in our \verb'R' package \verb'EHRMiss' available at \url{https://github.com/anthopolos/EHRMiss}, we used a parametrization based on hierarchical centering in the longitudinal health outcomes model in order to improve chain mixing and speed model convergence \citep{Gelfand1995,Gelfand1996}. In the hierarchically centered parametrization, we centered $\mathbf{b}_{ri}$ around the average health trajectory represented by $\mathbf{\beta}_{rk}$. The MCMC algorithm herein is based on this hierarchically centered parametrization given by  
\begin{gather}
\left[
\begin{array}{c|c}
\mathbf{y}_{1i} &  \\
\vdots &  c_i = k \\
\mathbf{y}_{Ri} &  \\
\end{array}
\right]  \sim  MVN_{RJ} \begin{pmatrix}
\begin{bmatrix}
\mathbf{\beta}_{1k} \mathbf{x}_{i}^{h,T} + \mathbf{b}_{1i} \mathbf{z}_{i}^{T} \\
\vdots \\
\mathbf{\beta}_{Rk} \mathbf{x}_{i}^{h,T} + \mathbf{b}_{Ri} \mathbf{z}_{i}^{T}  
\end{bmatrix}, \,
diag(\mathbf{\Sigma}_k)
\end{pmatrix}  \label{modelyhc} \\[3\jot] 
\left[
\begin{array}{c|c}
\mathbf{b}_{1i} &  \\
\vdots &  c_i = k \\
\mathbf{b}_{Ri} &  \\
\end{array}
\right]  \sim  MVN_{Rq} \begin{pmatrix}
\begin{bmatrix}
\mathbf{\mathbf{u}_{i} \eta}_{1k}^T \\
\vdots \\
\mathbf{\mathbf{u}_{i} \eta}_{Rk}^T  
\end{bmatrix}, \,
\mathbf{\Psi}_k
\end{pmatrix}  \label{modelbhc}
\end{gather} where we use a superscript $h$ for the fixed effects design matrix $\mathbf{x}_{i}^{h}$ ($J \times p^{h}$) to indicate the change in parameterization. 
Unlike the main text, in (\ref{modelyhc}), the columns in the random effects design matrix $\mathbf{z}_{i}$ are no longer a subset of the columns in $\mathbf{x}_{i}^{h}$. For example, in a random intercept model, only $\mathbf{z}_{i}$ will include a column of ones for an intercept. In (\ref{modelbhc}), the random effects $\mathbf{b}_{1i},\dots,\mathbf{b}_{Ri}$ are distributed with mean as a function of patient-level risk factors in $\mathbf{u}_i$ ($1 \times e$) and corresponding regression coefficients in $\mathbf{\eta}_{1k},\dots,\mathbf{\eta}_{Rk}$ ($q \times e$). $diag(\mathbf{\Sigma}_k)$ is an $RJ \times RJ$ block diagonal matrix with elements $\mathbf{\Sigma}_k$ for the variance-covariance among $y_{1ij},\dots,y_{Rij}$ in each time window $j$ ($j=1,\dots,J$).

%Each y_{ri} has been defined as a J-length row vector, such that the left hand side of D.1 is R \times J, where J is the number clinical windows
%\beta_{rk} is 1 \times p^{h}
%x_{i} is J \times p^{h}, such that x_i^T is p^{h} \times J
%b_{ri} is 1 \times q
%z_i is J \times q, such that z_i^T is q \times J
%\eta_{rk} is q \times e, where e is the number of subject level covariates, and q is the number of random effects
%u_i is 1 \times e, where e is the number of subject level covariates

\subsection{Update parameters in the latent class membership model}

The Gibbs steps are given for the latent class membership model.

\begin{enumerate}
	\item Update $\xi_{ik}$. Let $\mathbf{\xi}_i^T = (\xi_{i1},\dots,\xi_{iK-1})$ be a $(K-1)$-length column vector. Per \citet{McCulloch1994}, for $i = 1,\dots,n$, the distribution of $\mathbf{\xi}_{i} \ | \ \mathbf{\delta}, c_{i}$ is a $(K - 1)$-variate normal distribution truncated over the appropriate cone in $\mathbf{R}^{K-1}$. Let $\mathbf{c}^{*}_{i}$ be a multinomial vector with entries $\mathbf{c}_{i}^{*} = (c^{*}_{i1},\dots,c^{*}_{iK})$ equal to 1 if the $i^{th}$ subject is in latent class $k$ and 0 otherwise. If $c^{*}_{ik} = 1$, then $\xi_{ik} > max(\mathbf{\xi}_{i,-k}, 0)$. If $c^{*}_{ik} = 0$, then $\xi_{ik} < max(\mathbf{\xi}_{i,-k}, 0)$. $\mathbf{\xi}_{i,-k}$ is a $K - 2$ dimensional vector of all components of $\mathbf{\xi}_{i}$ excluding $\xi_{ik}$. This algorithm avoids the problem of drawing from a truncated multivariate normal. Instead each draw is a truncated univariate normal because we are using the conditional distribution $\xi_{ik} \ | \ \mathbf{\xi}_{i,-k}, \mathbf{\delta}_k, c_i$, where $c_i = K$ if $max(\mathbf{\xi}_{i}) < 0$, or else $c_i = \ \text{index of} \ max(\mathbf{\xi}_{i})$ for $k = 1,\dots,K-1$.
	
	\item Update $\mathbf{\delta}_k$. For $k =1,\dots,K-1$, we assume the prior $\mathbf{\delta}_{k} \sim MVN_s(0, \mathbf{\Sigma}_\delta)$. The full conditional is $MVN_s(\mathbf{\mu}_{\delta_k}, \mathbf{V}_{\delta})$, where \begin{gather}
	\mathbf{V}_{\delta} = \left(\sum_{i = 1}^{n} \mathbf{w}_{i}^T \mathbf{w}_{i} + \mathbf{\Sigma}_\delta^{-1} \right)^{-1} \nonumber \\
	\mathbf{\mu}_{\delta_k} = \mathbf{V}_{\delta} \times \left(\sum_{i = 1}^{n} \mathbf{w}_{i}^T \xi_{ik} \right), \nonumber
	\end{gather} with $\mathbf{w}_{i}$ being an $s$-length row vector of patient-level risk factors, including a column of ones for an intercept. 
	
\end{enumerate}

\subsection{Update parameters in the longitudinal outcomes model}

\begin{enumerate}
	\item Update $\mathbf{\beta}_{rk}$.
	
	To update $\mathbf{\beta}_{rk}$, based on the properties of the multivariate normal distribution, we use the conditional distribution of longitudinal health outcome $r$ given health outcomes $r^{'}$ for all $r^{'} \neq r$. Let $\mathbf{y}_{ri}^{*} = (y_{riA(1)},\dots,y_{riA(n_i)})^T$. Let $\mathbf{Q}$ be a matrix of conditional coefficients defined as $\mathbf{Q} = \mathbf{I} - [diag(\mathbf{\Sigma}_k^{-1})]^{-1} \mathbf{\Sigma}_k^{-1}$, with elements $q_{rr^{'}}$ ($r = 1,\dots,R$, $r^{'} = 1,\dots,R$) \citep{gelman2014bayesian}. %Gelman BDA page 580	
	For longitudinal health outcome $r$ of patient $i$ in window $j$, the conditional distribution of $\mathbf{y}_{ri}^{*}$ given $\mathbf{y}_{r^{'}i}^{*}$ for all $r^{'} \neq r$ and latent class $c_i$ is 		
	\begin{eqnarray}
	\lefteqn{[\mathbf{y}_{ri}^{*} \, | \, \mathbf{y}_{r^{'}i}^{*} \text{ all } r^{'} \neq r, \, c_i = k] \sim} \label{conditionaly} \\
	&  & MVN_{n_i}\left(\mathbf{\beta}_{rk} \mathbf{x}_{i}^{h*,T} + \mathbf{b}_{ri} \mathbf{z}_{i}^{*,T} + \sum_{r^{'} \neq r} q_{rr^{'}} (\mathbf{y}_{r^{'}i}^{*} - \mathbf{\beta}_{r^{'}k} \mathbf{x}_{i}^{h*,T} - \mathbf{b}_{r^{'}i} \mathbf{z}_{i}^{*,T}), \, diag\left([\mathbf{\Sigma}_{krr}^{-1}]^{-1}\right) \right), \nonumber 
	\end{eqnarray} where $\mathbf{x}_i^{h*}$ ($n_i \times p^h$) is the fixed effects design matrix for time windows $A(l)$ for $l = 1,\dots,n_i$. $\mathbf{z}_i^{*}$ is the corresponding random effects design matrix.
	
	For latent classes $k = 1,\dots,K$, assuming the prior distribution $\mathbf{\beta}_{rk} \sim MVN_{p^{h}}(\mathbf{0}, \mathbf{\Sigma}_\beta)$, the full conditional is $MVN_{p^{h}}(\mathbf{\mu}_{\beta_{rk}}, \mathbf{V}_{\beta_{rk}})$, where 
	\begin{eqnarray*}
	\mathbf{V}_{\beta_{rk}} & = & \left(\sum_{i = 1}^n \mathbf{1}_{c_i = k} \times \frac{ \mathbf{x}_{i}^{h*,T} \mathbf{x}_{i}^{h*} }{[\mathbf{\Sigma}_{krr}^{-1}]^{-1}} +  \mathbf{\Sigma}_\beta^{-1}\right)^{-1} 
	\end{eqnarray*}
	\begin{eqnarray*}
	\lefteqn{\mathbf{\mu}_{\beta_{rk}}} \\
	& = & \mathbf{V}_{\beta_{rk}} \\
	& \times & \left(\sum_{i = 1}^n \mathbf{1}_{c_i = k} \times	\frac{\mathbf{x}_{i}^{h*,T} \left(\mathbf{y}_{ri}^{*,T} - \mathbf{z}_{i}^{*} \mathbf{b}_{ri}^T - (\sum_{r^{'} \neq r} q_{rr^{'}} (\mathbf{y}_{r^{'}i}^{*} - \mathbf{\beta}_{r^{'}k} \mathbf{x}_{i}^{h*,T} - \mathbf{b}_{r^{'}i} \mathbf{z}_{i}^{*,T}))^T \right)}{[\mathbf{\Sigma}_{krr}^{-1}]^{-1}} \right) 
	\end{eqnarray*}
	
	\item Update $\mathbf{b}_{ri}$. Using the conditional distribution in (\ref{conditionaly}), the full conditional is \\ $MVN_q(\mathbf{\mu}_{b_{ri}}, \mathbf{V}_{b_{ri}})$, where \begin{equation*}
	\mathbf{V}_{b_{ri}} = \sum_{k = 1}^K \mathbf{1}_{c_i = k} \left(\frac{\mathbf{z}_{i}^{*,T} \mathbf{z}_{i}^{*}}{[\mathbf{\Sigma}_{krr}^{-1}]^{-1}} + \mathbf{\Psi}_{kr}^{-1} \right)^{-1}  
   	\end{equation*}	\begin{eqnarray*}
   	\lefteqn{\mathbf{\mu}_{b_{ri}}} \\
   	&=& \mathbf{V}_{b_{ri}} \\
   	&\times& 
   	\sum_{k = 1}^K \mathbf{1}_{c_i = k} \\
   	 &\times& \left( \frac{\mathbf{z}_i^{*,T} \left(\mathbf{y}_{ri}^{*,T} - \mathbf{x}_{i}^{h*} \mathbf{\beta}_{rk}^T - (\sum_{r^{'} \neq r} q_{rr^{'}} (\mathbf{y}_{r^{'}i}^{*} - \mathbf{\beta}_{r^{'}k} \mathbf{x}_{i}^{h*,T} - \mathbf{b}_{r^{'}i} \mathbf{z}_{i}^{*,T}))^T \right)}{[\mathbf{\Sigma}_{krr}^{-1}]^{-1}} + \mathbf{\Psi}^{-1}_{kr} \mathbf{\eta}_{rk} \mathbf{u}_i^T \right) 
   	\end{eqnarray*}
   	
	%Check with Qixuan and Ying
	\item Update $\mathbf{\eta}_{rk}$. Let the elements of $\mathbf{b}_{ri}$ be indexed as $b_{rig}$ for $g = 1,\dots,q$. For the $g^{th}$ random effect, let $\mathbf{\eta}_{rkg} = (\eta_{rkg1},\dots,\eta_{rkge})^T$ ($1 \times e$). Then, $b_{rig} \sim N(\mathbf{u}_i \mathbf{\eta}_{rkg}^T, \psi_{krgg})$. Assuming the prior distribution $MVN_e(\mathbf{0}, \mathbf{\Sigma}_\eta)$, the full conditional of $\mathbf{\eta}_{rkg}$ is $MVN_e(\mathbf{\mu}_{\eta_{rkg}}, \mathbf{V}_{\eta_{rkg}})$, where
	\begin{gather}
	\mathbf{V}_{\eta_{rkg}} = \left(\sum_{i = 1}^n \mathbf{1}_{c_i = k} \times \frac{\mathbf{u}_i^T \mathbf{u}_i}{\psi_{krgg}} +  \mathbf{\Sigma}_\eta^{-1}\right)^{-1} \nonumber \\
	\mathbf{\mu}_{\eta_{rkg}} = \mathbf{V}_{\eta_{rkg}}  \times \left(\sum_{i = 1}^n \mathbf{1}_{c_i = k} \times \frac{\mathbf{u}_i^T b_{rig}}{\psi_{krgg}} \right) \nonumber
	\end{gather}
	% If prior does not have mean 0: + \mathbf{\Sigma}_\eta^{-1} \mathbf{\mu}_{\eta}^T
	
	 \item Update $\mathbf{\Sigma}_k$. Recall the $R$-length row vectors  $\mathbf{y}_{iA(l)} = (y_{1iA(l)},\dots,y_{RiA(l)})^T$, and $\mathbf{\mu}_{iA(l)} = \mathbf{x}_{iA(l)} \mathbf{\beta}_k^T  +  \mathbf{z}_{iA(l)} \mathbf{b}_i^T$.  Assuming an inverse-Wishart prior distribution $\mathbf{\Sigma}_k \sim IW(\nu_\Sigma, S_\Sigma^{-1})$, the full conditional is $IW(a_{\Sigma_k}, b_{\Sigma_k})$, where \begin{gather}
	a_{\Sigma_k} = \nu_\Sigma + \sum_{i = 1}^n \mathbf{1}_{c_i = k} \times n_i \nonumber \\
	b_{\Sigma_k} = S_\Sigma + \sum_{i = 1}^n \mathbf{1}_{c_i = k} \sum_{l = 1}^{n_i} (\mathbf{y}_{iA(l)} -  \mathbf{\mu}_{iA(l)})^T (\mathbf{y}_{iA(l)} -  \mathbf{\mu}_{iA(l)}) \nonumber
	\end{gather}

	\item Update $\mathbf{\Psi}_{k}$. The block diagonal matrix $\mathbf{\Psi}_{k}$ ($Rq \times Rq$) contains elements  $\mathbf{\Psi}_{kr}$ ($q \times q$). Assuming $\mathbf{\Psi}_{kr} \sim IW(\nu_\Psi, S_\Psi^{-1})$, the full conditional is $IW(a_{\Psi_{kr}}, b_{\Psi_{kr}})$, where \begin{gather}
	a_{\Psi_{kr}} = \nu_\Psi + \sum_{i = 1}^n \mathbf{1}_{c_i = k} \nonumber \\
	b_{\Psi_{kr}} = S_\Psi + \sum_{i = 1}^n \mathbf{1}_{c_i = k} \times (\mathbf{b}_{ri} -  \mathbf{u}_{i} \mathbf{\eta}_{rk}^T)^T (\mathbf{b}_{ri} -  \mathbf{u}_{i} \mathbf{\eta}_{rk}^T) \nonumber
	\end{gather}
	
\end{enumerate}

\subsection{Update parameters in the visit process model}
Following \citet{Albert1993}, we use a data augmentation approach \citep{Tanner1987} to model the probability of a clinic visit using Bayesian probit regression. Corresponding to the visit process for patient $i$ in clinical window $j$, we introduce latent variables $\xi_{ij}^{d}$ ($i = 1,\dots,n$, $j = 1,\dots, J$). The latent variables $\xi_{ij}^{d}$ are assumed to be distributed as $N(\mathbf{x}_{ij} \mathbf{\phi}^T_k + \mathbf{z}_{ij} \mathbf{\tau}_i^T, 1)$, where the observation-level error variance is fixed to 1. To connect latent $\xi_{ij}^{d}$ to the visit process $d_{ij}$, define $d_{ij} = 1$ if $\xi_{ij}^{d} > 0$ and $d_{ij} = 0$ if $\xi_{ij}^{d} \leq 0$. With the introduction of the latent variables, the Gibbs sampling steps are as follows.

   \begin{enumerate}
		\item Update $\xi_{ij}^{d}$. The full conditional is $\xi_{ij}^{d} \, | \, d_{ij}, \mathbf{\phi}_k, \mathbf{\tau}_i, c_i = k \sim N(\sum_{k = 1}^K \mathbf{1}_{c_i = k} \times (\mathbf{x}_{ij} \mathbf{\phi}^T_k + \mathbf{z}_{ij} \mathbf{\tau}_i^T), 1)$, truncated at the left by 0 if $d_{ij} = 1$. Otherwise, $\xi_{ij}^{d} \, | \, d_{ij}, \mathbf{\phi}_k, \mathbf{\tau}_i, c_i = k \sim N(\sum_{k = 1}^K \mathbf{1}_{c_i = k} \times (\mathbf{x}_{ij} \mathbf{\phi}^T_k + \mathbf{z}_{ij} \mathbf{\tau}_i^T), 1)$, truncated at the right by 0 if $d_{ij} = 0$. 
		
		\item Update $\mathbf{\phi}_k$. For latent classes $k = 1,\dots,K$, assuming the prior distribution $\mathbf{\phi}_{k} \sim MVN_p(\mathbf{\mu}_\phi, \mathbf{\Sigma}_\phi)$, the full conditional is $MVN_p(\mathbf{\mu}_{\phi_{k}}, \mathbf{V}_{\phi_{k}})$, where 	\begin{gather}
		\mathbf{V}_{\phi_{k}} = \left(\sum_{i = 1}^n \mathbf{1}_{c_i = k} \times \mathbf{x}_i^T \mathbf{x}_i +  \mathbf{\Sigma}_\phi^{-1}\right)^{-1} \nonumber \\
		\mathbf{\mu}_{\phi_{k}} = \mathbf{V}_{\phi_{k}} \times \left(\sum_{i = 1}^n \mathbf{1}_{c_i = k} \times	\mathbf{x}_i^T \left(\mathbf{\xi}_{i}^{{d,T}} - \mathbf{z}_i \mathbf{\tau}_{i}^T \right)  + \mathbf{\Sigma}_\phi^{-1} \mathbf{\mu}_\phi^T \right), \nonumber
		\end{gather} where the random effects design matrix $\mathbf{z}_i$ ($J \times q$) contains a subset of the columns in the fixed effects design matrix $\mathbf{x}_i$ ($J \times p$). 
				
		\item Update $\mathbf{\tau}_i$. The full conditional is $MVN_q(\mathbf{\mu}_{\tau_{i}}, \mathbf{V}_{\tau_{i}})$, where \begin{gather*}
		\mathbf{V}_{\tau_{i}} = \sum_{k = 1}^K \mathbf{1}_{c_i = k} \times \left(\mathbf{z}_i^T \mathbf{z}_i + \mathbf{\Omega}_{k}^{-1} \right)^{-1}  \\
		\mathbf{\mu}_{\tau_{i}} = \mathbf{V}_{\tau_{i}} \times \sum_{k = 1}^K \mathbf{1}_{c_i = k} \times \left(\mathbf{z}_i^T \left(\mathbf{\xi}_{i}^{d,T} - \mathbf{x}_i \mathbf{\phi}_{k}^T \right) \right)
		\end{gather*}

		\item Update $\mathbf{\Omega}_k$. Assuming an inverse-Wishart prior distribution $\mathbf{\Omega}_k \sim IW(\nu_\Omega, S_\Omega^{-1})$, the full conditional is $IW(a_{\Omega_k}, b_{\Omega_k})$, where \begin{gather}
		a_{\Omega_k} = \nu_\Omega + \sum_{i = 1}^n \mathbf{1}_{c_i = k}  \nonumber \\
		b_{\Omega_k} = S_\Omega + \sum_{i = 1}^n \mathbf{1}_{c_i = k} \times \mathbf{\tau}_{i}^T \mathbf{\tau}_{i} \nonumber
		\end{gather}
	\end{enumerate}

\subsection{Update parameters in the response process given a clinic visit model}

The Gibbs steps to update the parameters in the model for the response process given a clinic visit are analogous to the steps in the visit process model, except that we use observed clinic visits. 

For patient $i$ in clinical window $l$ with an observed visit ($l = 1,\dots,n_i$), we introduce latent variables $\xi_{riA(l)}^{m}$ ($i = 1,\dots,n$, $l = 1,\dots, n_i$). The latent variables $\xi_{riA(l)}^{m}$ are assumed to be distributed as $N(\mathbf{x}_{iA(l)} \mathbf{\lambda}^T_{rk} + \mathbf{z}_{iA(l)} \mathbf{\kappa}_{ri}^T, 1)$, where the observation-level error variance is fixed to 1. To connect latent $\xi_{riA(l)}^{m}$ to the response process $m_{riA(l)}$, define $m_{riA(l)} = 1$ if $\xi_{riA(l)}^{m} > 0$ and $m_{riA(l)} = 0$ if $\xi_{riA(l)}^{m} \leq 0$. Upon introducing the latent variables, the Gibbs sampling steps for $\mathbf{\lambda}_{rk}$, $\mathbf{\kappa}_{ri}$, and $\mathbf{\Theta}_{rk}$ proceed as in the visit process model.

\subsection{Update latent class membership}

Sample latent class indicators $c_i$ for $i = 1,\dots,n$ from $Multinomial(1; \, p_{i1},\dots,p_{iK})$, where $p_{i1},\dots,p_{iK}$ are the posterior probabilities of latent class assignment. For $k = 1,\dots,K$, 	\begin{eqnarray}
	\lefteqn{p_{ik}} \nonumber \\
	& = & Pr(c_i = k \, | \, \pi_{ik}; \, \mathbf{y}^{*}_{i}, \mathbf{b}_i; \, \mathbf{d}_{i}, \mathbf{\tau}_{i}; \, \mathbf{m}_{1i},\dots, \mathbf{m}_{Ri}, \mathbf{\kappa}_{1i},\dots,\mathbf{\kappa}_{Ri}; \, rest) \nonumber \\
	& \propto & \pi_{ik} \, f(\mathbf{y}^{*}_{i} \, | \, \mathbf{b}_i, \mathbf{\beta}_k, \mathbf{\Sigma}_k^{*}) \, f(\mathbf{b}_{i} \, | \, \mathbf{\Psi}_k) \nonumber \\
	& \times & f(\mathbf{d}_{i} \, | \, \mathbf{\tau}_{i}, \mathbf{\phi}_k) \, f(\mathbf{\tau}_{i} \, | \, \mathbf{\Omega}_k) \nonumber \\
	& \times & \prod_{r = 1}^R f(\mathbf{m}_{ri} \, | \, \mathbf{\kappa}_{ri}, \mathbf{\lambda}_{rk}) \, f(\mathbf{\kappa}_{ri} \, | \, \mathbf{\Theta}_{rk}), \nonumber
	\end{eqnarray} where $\mathbf{y}^{*}_{i} = (\mathbf{y}_{iA(1)}^T,\dots, \mathbf{y}_{iA(n_{i})}^T)$, and $\mathbf{\Sigma}_k^{*}$ is an $n_i R \times n_i R$ block diagonal matrix with elements $\mathbf{\Sigma}_k$ ($R \times R$) for each $\mathbf{y}_{iA(l)}$ ($l=1,\dots,n_i$).

\section{Analysis of Early Childhood Weight and Height Measurements Addendum}

\setcounter{table}{0}    
\setcounter{figure}{0}    

\begin{figure}[htbp]
	\centering
	\includegraphics[scale=0.75]{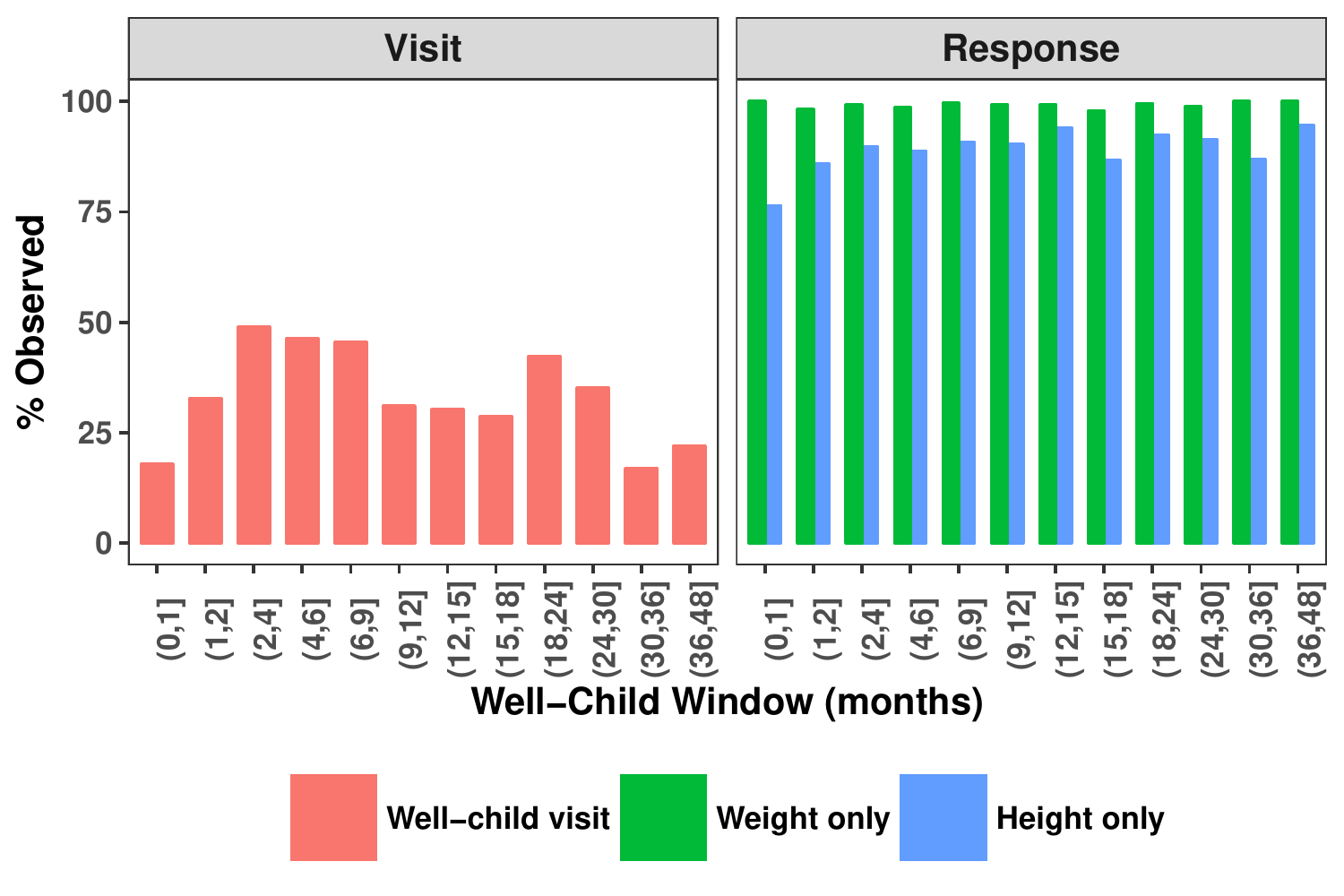}
	\caption{Patterns of observed visits and observed responses in weight and height z-scores given a clinic visit.}
	\label{f:EHR_weight_height_missing_3}
\end{figure}

Based on the well-child windows, Figure \ref{f:EHR_weight_height_missing_3} presents the patterns of observed visits and responses for weight and height given a clinic visit. Of 5,988 well-child windows (499 children $\times$ 12 well-child windows), 67\% correspond to missed visits. Among 1,983 observed visits, only 17 weight measurements are missing ($<$ 1\%), whereas 207 height measurements (10\%) are missing.

\subsection{Longitudinal trajectories of weight and height z-scores using the \textbf{MNAR} method}

\subsubsection{Model selection}

In GMMs, the number of latent classes $K$ is assumed to be fixed and known. A challenge in data applications with GMMs is to select among models with varying numbers of latent classes. We compared $K=2,3,4$-class models based on the \textbf{MNAR} method using model information criteria, including the Bayesian Information Criterion (BIC) \citep{Schwarz1978} and a modified version of the Deviance Information Criterion (DIC) \citep{Spiegelhalter2002} known as the DIC3 recommended for latent variable models \citep{Celeux2006}; the log-pseudo marginal likelihood (LPML) \citep{Geisser1979,Gelfand1994,Ibrahim2001}; and a graphical technique known as latent class identifiability displays (LCIDs) \citep{Garrett2000}. To select among models with varying numbers of latent classes, we use two model information criteria, including the Bayesian Information Criterion (BIC) \citep{Schwarz1978} and a modified version of the Deviance Information Criterion (DIC) \citep{Spiegelhalter2002} known as the DIC3 \citep{Celeux2006}, the log-pseudo marginal likelihood \citep{Geisser1979,Gelfand1994,Ibrahim2001}, and a graphical technique known as latent class identifiability displays (LCIDs) \citep{Garrett2000}. With the objective of clinical interpretability, we prioritized a solution to the number of latent classes that represented distinctive clusters of children. 

We calculate the BIC using the marginal density of $\mathbf{y}^{*}_i$, $\mathbf{d}_i$, $\mathbf{m}_{1i},\dots,\mathbf{m}_{Ri}$ after integrating out $c_i$ and random effects $\mathbf{b}_i$, $\mathbf{\tau}_i$, and $\mathbf{\kappa}_{1i},\dots,\mathbf{\kappa}_{Ri}$, given by \begin{eqnarray}
\lefteqn{f(\mathbf{y}^{*}_{i}, \mathbf{d}_{i}, \mathbf{m}_{1i}, \dots, \mathbf{m}_{Ri} \, | \, \mathbf{\pi}; \, \mathbf{\beta}, \mathbf{\Sigma}, \mathbf{\Psi}; \, \mathbf{\phi}, \mathbf{\Omega}; \, \mathbf{\lambda}, \mathbf{\Theta})} \label{e:marginaldensity} \\
& = & \sum_{k=1}^K \pi_{ik} \,\, \left( \int_{b_i} f(\mathbf{y}^{*}_i \, | \, \mathbf{b}_i, \mathbf{\beta}_{k}, \mathbf{\Sigma}_k^{*}) \, f(\mathbf{b}_i \, | \, \mathbf{\Psi}_k) \, \partial  \mathbf{b}_i \right) \nonumber \\
& \times & \left( \int_{\tau_i} f(\mathbf{d}_i \, | \, \mathbf{\tau}_i, \mathbf{\phi}_k) \, f(\mathbf{\tau}_i \, | \, \mathbf{\Omega}_k) \,  \partial \mathbf{\tau}_i \right)  \nonumber \\
& \times & \left( \int_{\kappa_{Ri}} \dots \int_{\kappa_{1i}} f(\mathbf{m}_{1i} \, | \, \mathbf{\kappa}_{1i}, \mathbf{\lambda}_{1k}) \, f(\mathbf{\kappa}_{1i} \, | \, \mathbf{\Theta}_{1k}) \dots f(\mathbf{m}_{Ri} \, | \, \mathbf{\kappa}_{Ri}, \mathbf{\lambda}_{Rk}) \, f(\mathbf{\kappa}_{Ri} \, | \, \mathbf{\Theta}_{Rk}) \, \partial \mathbf{\kappa}_{1i}, \dots, \partial \mathbf{\kappa}_{Ri} \right), \nonumber  
\end{eqnarray} where we can analytically compute only the integral for $\mathbf{y}^{*}_i$, and we estimate the integrals for $\mathbf{d}_i$ and  $\mathbf{m}_{1i},\dots,\mathbf{m}_{Ri}$ using numerical integration. We then define the BIC as \begin{eqnarray}
\text{BIC} & = & \sum_{i = 1}^n \text{log} f(\mathbf{y}^{*}_{i}, \mathbf{d}_{i}, \mathbf{m}_{1i}, \dots, \mathbf{m}_{Ri} \, | \, \hat{\mathbf{\pi}}; \, \hat{\mathbf{\beta}}, \hat{\mathbf{\Sigma}}, \hat{\mathbf{\Psi}}; \, \hat{\mathbf{\phi}}, \hat{\mathbf{\Omega}}; \, \hat{\mathbf{\lambda}}, \hat{\mathbf{\Theta})} + d \, \text{log} N_{\text{eff}}, \nonumber
\end{eqnarray} where $\hat{\mathbf{\pi}}$, $\hat{\mathbf{\beta}}$, $\hat{\mathbf{\Sigma}}$, $\hat{\mathbf{\Psi}}$, $\hat{\mathbf{\phi}}$, $\hat{\mathbf{\Omega}}$, $\hat{\mathbf{\lambda}}$, $\hat{\mathbf{\Theta}}$ are the Bayesian estimators of the unknown parameters; $d$ is the number of free parameters; and $N_{\text{eff}}$ is the effective sample size from the model of $\mathbf{y}^{*}_i$, which we estimate by accounting for correlations among longitudinal measurements from same patient \citep{Jones2011}. The first term is a measure of goodness of fit, and the second term provides a penalty for model complexity.

Since the number of free parameters may be unclear in Bayesian hierarchical models, the DIC was proposed \citep{Spiegelhalter2002} in which the number of effective parameters is estimated as the difference between the posterior mean deviance and a point estimate of the deviance commonly computed with the posterior mean estimator of unknown parameters. However, because in finite mixture modeling, the posterior mean estimator often leads to a negative effective number of parameters, the DIC3 is instead  based on the estimator of the marginal density in (\ref{e:marginaldensity}) \citep{Celeux2006}. Like the BIC, the DIC3 penalizes goodness of fit by model complexity. Smaller values of BIC and DIC3 indicate a preferred model.

Based on leave-one-out cross-validation, the LPML is a summary measure of a model’s predictive utility, and is calculated as the log of the product of the conditional predictive ordinate (CPO) for each patient under a given model \citep{Geisser1979,Gelfand1994,Ibrahim2001}. For patient $i$, the CPO is the marginal posterior predictive density given that patient $i$ is excluded from the dataset. Using   (\ref{e:marginaldensity}), a Monte Carlo estimate of the CPO for patient $i$ is given by  
\begin{eqnarray*}
 \widehat{\text{CPO}}_i & = & \frac{\text{G}}{\sum_{g = 1}^G (1 / f(\mathbf{y}^{*}_{i}, \mathbf{d}_{i}, \mathbf{m}_{1i}, \dots, \mathbf{m}_{Ri} \, | \, \mathbf{\pi}^{g}; \, \mathbf{\beta}^{g}, \mathbf{\Sigma}^{g}, \mathbf{\Psi}^{g}; \, \mathbf{\phi}^{g}, \mathbf{\Omega^{g}}; \, \mathbf{\lambda}^{g}, \mathbf{\Theta}^{g}))},  \end{eqnarray*} where superscript $g$ indexes parameters sampled over MCMC iterations $g=1,\dots,G$ following a burn-in period. Then, the LPML is estimated by $\widehat{\text{LPML}} = \sum_{i=1}^n \text{log} \widehat{\text{CPO}}_i$. Larger values of the LPML indicate a preferred model.

Latent class identifiability displays (LCIDs) \citep{Garrett2000} overlay the prior versus posterior distributions for the regression coefficients $\mathbf{\delta}_k$ in the latent class membership model: When the prior and posterior distributions are largely overlapping, the number of latent classes may be too large given the data. 

\begin{figure}[htbp]
	\centering
	\includegraphics[scale=0.75]{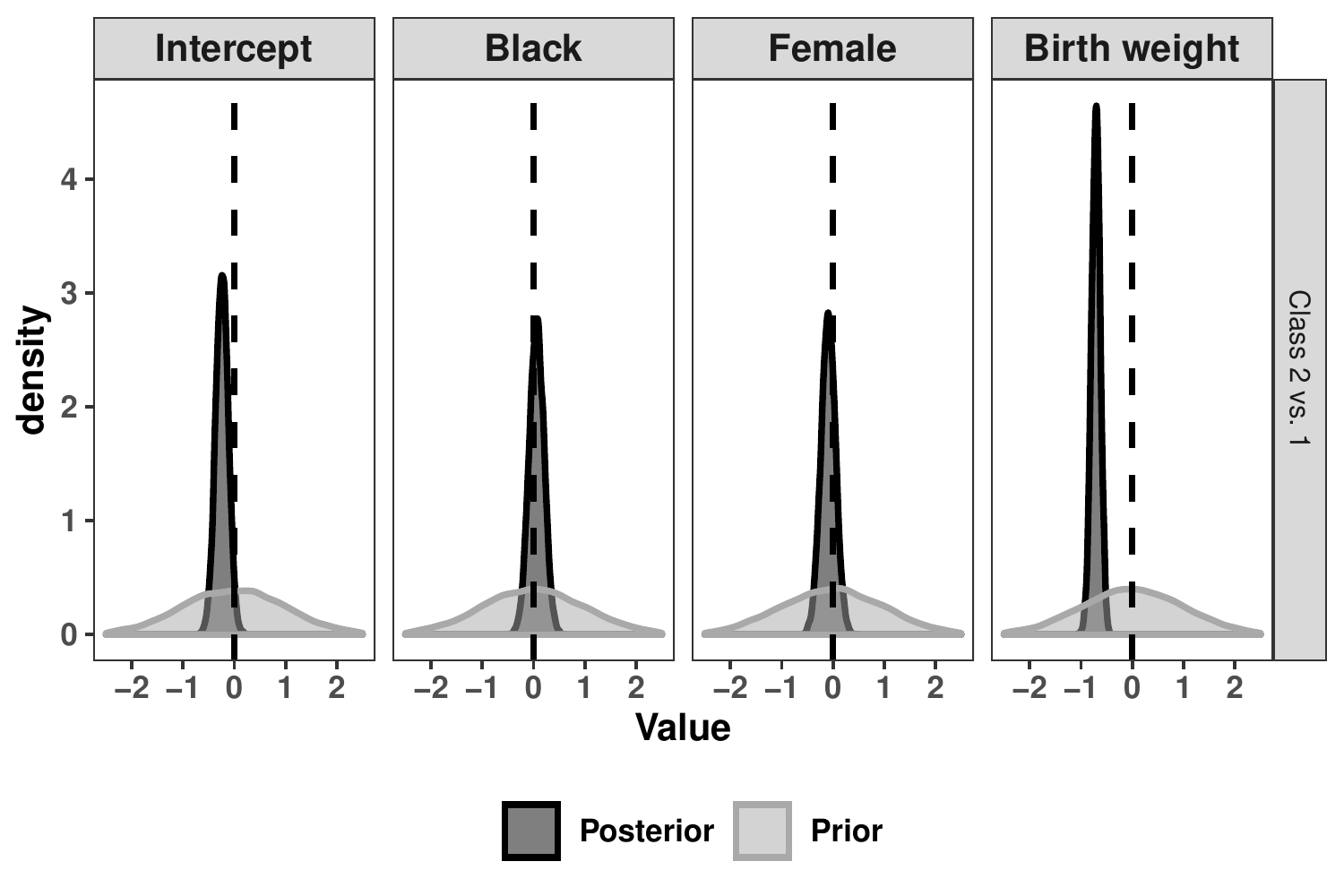}
	\caption{Posterior versus prior distributions for the covariates in the multinomial probit model of latent class membership using the \textbf{MNAR} method, $K=2$.}
	\label{f:K2_MNAR_MNAR_Prior_Posterior}
\end{figure}

\begin{figure}[htbp]
	\centering
	\includegraphics[scale=0.75]{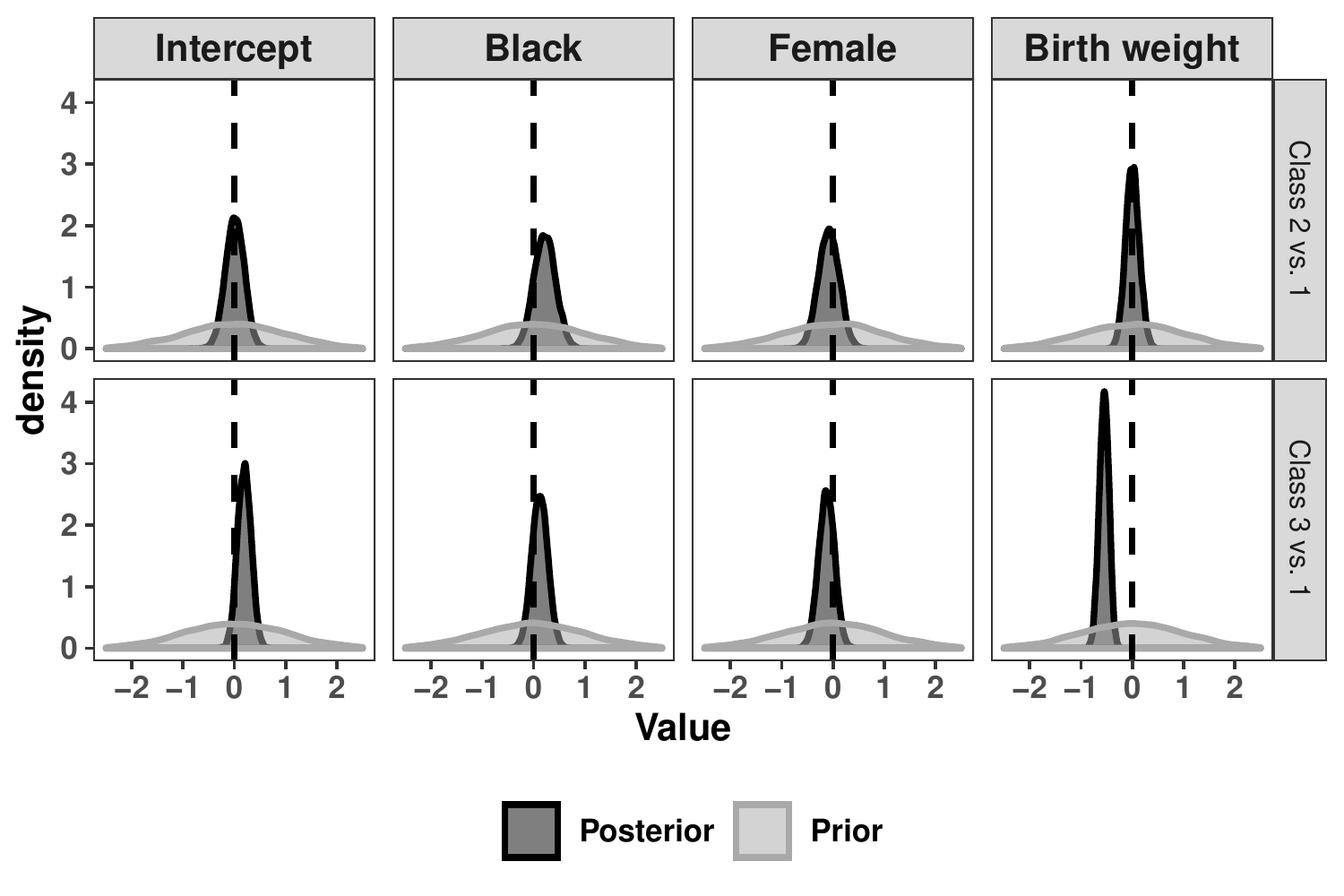}
	\caption{Posterior versus prior distributions for the covariates in the multinomial probit model of latent class membership using the \textbf{MNAR} method, $K=3$.}
	\label{f:K3_MNAR_MNAR_Prior_Posterior}
\end{figure}

\begin{figure}[htbp]
	\centering
	\includegraphics[scale=0.75]{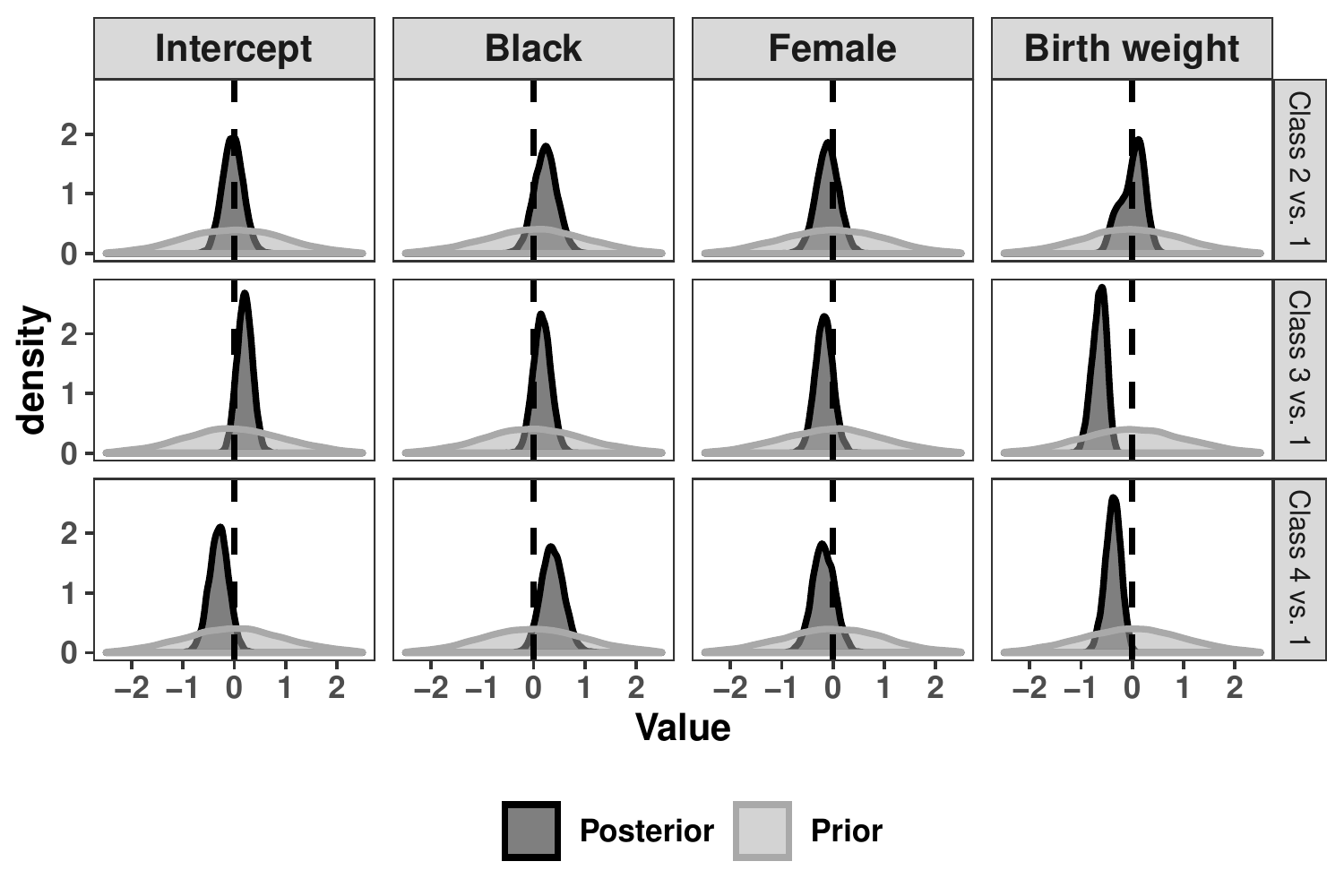}
	\caption{Posterior versus prior distributions for the covariates in the multinomial probit model of latent class membership using the \textbf{MNAR} method, $K=4$.}
	\label{f:K4_MNAR_MNAR_Prior_Posterior}
\end{figure}

In Table \ref{t:table_model_comparison}, the DIC3 and LPML both chose the 3-class model, whereas the BIC selected the 4-class model. For each of the $K$-class models, Figures \ref{f:K2_MNAR_MNAR_Prior_Posterior}--\ref{f:K4_MNAR_MNAR_Prior_Posterior} show the posterior versus prior distributions for the covariates in the latent class membership model. Compared to the 4-class model, the 3-class model appears to identify distinctive population subgroups according to race, sex, and birth weight. We therefore selected the 3-class model. 
\begin{table}[htbp]
	\caption{Comparison of models with $K=2,3,4$ latent class using the \textbf{MNAR} method. A preferred model is indicated by smaller values of the BIC and DIC3 and larger values of the LPML.}
	\begin{tabular}{lrrr}
		\hline
		&           \multicolumn{3}{c}{$K$} \\ 
		\multicolumn{1}{c}{Criterion} & \multicolumn{1}{c}{2} &  \multicolumn{1}{c}{3}  & \multicolumn{1}{c}{4}  \\ \hline
		BIC          &  21625   & 21226     &    \textbf{20939}  \\
		DIC3         &  23263   & \textbf{22716}     &    22741  \\ 
		LPML         &  -12770  & \textbf{-12230}    &    -13500  \\ \hline
	\end{tabular}
	\label{t:table_model_comparison}
\end{table}

\clearpage

\subsubsection{The 3-class model using the \textbf{MNAR} method}

Based on the maximum of a child's mean posterior probabilities of belonging to each latent class, we assigned approximately one-third of children to each subgroup, with subgroup mean (median) probability ranging from 0.81 to 0.84 (0.84 to 0.93) (Table \ref{t:table_K3_latent_class_assignment}).

\begin{table}[htbp]
	\caption{Posterior latent class assignment in the 3-class model using the \textbf{MNAR} method. Children were assigned to the Normal, increasing; Normal, decreasing; or Low trajectory subgroups according to the maximum of the mean posterior probabilities of class assignment.}
	\begin{tabular}{lccc}
		\hline
		& \multicolumn{1}{c}{Normal,} & \multicolumn{1}{c}{Normal,} & \multirow{2}{*}{Low} \\ 
		& \multicolumn{1}{c}{increasing} & \multicolumn{1}{c}{decreasing} &  \\ \cline{2-4}
		Predicted class size (\%) & 162 (32) & 160 (32) & 177 (35)\\ 
		Mean probability &  0.81 & 0.84 & 0.84 \\ 
		Median probability & 0.84 & 0.93 & 0.93 \\ 
		\hline
	\end{tabular}
	\label{t:table_K3_latent_class_assignment}
\end{table}

\begin{figure}[htbp]
	\centering
	\includegraphics[scale=0.75]{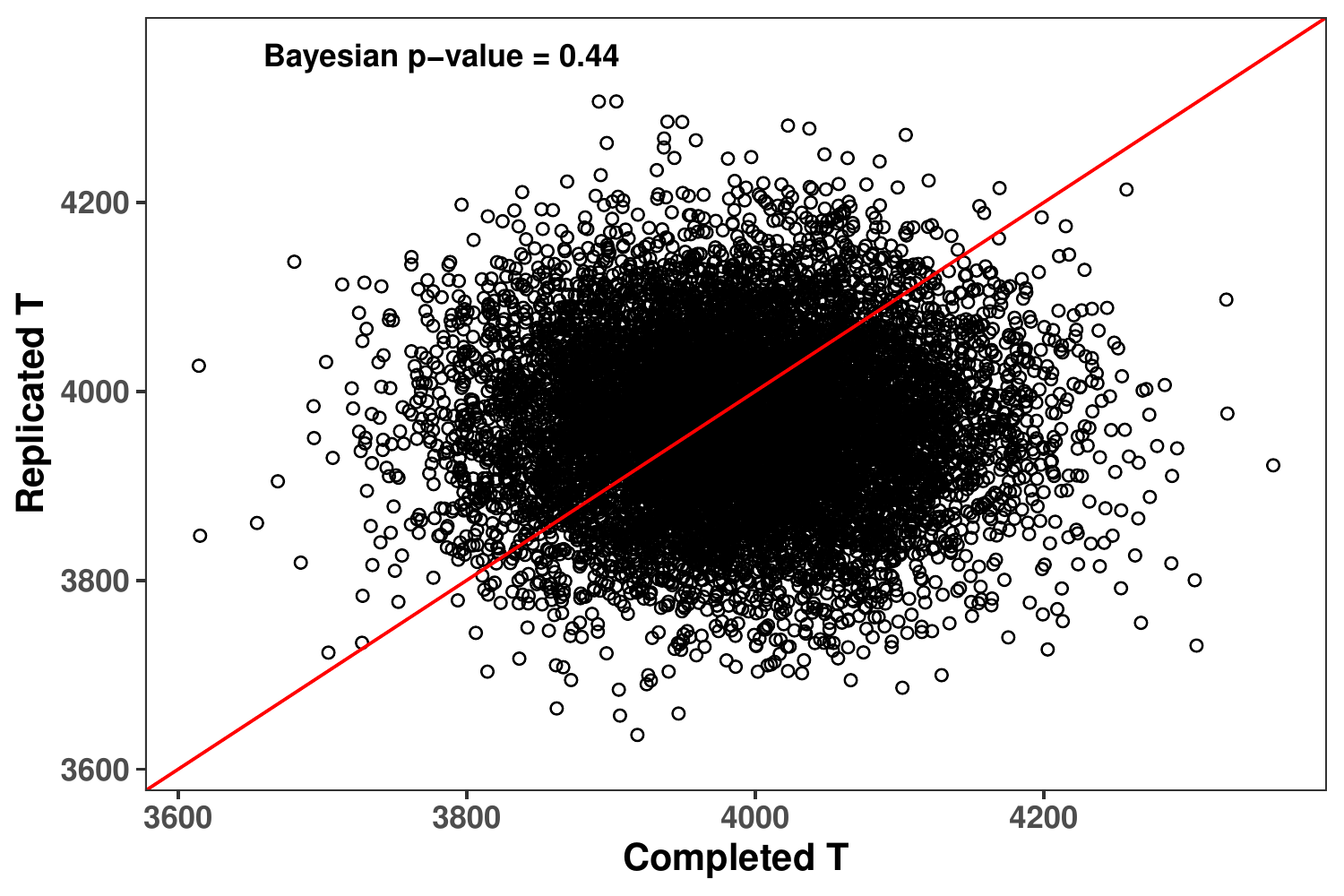}
	\caption{Posterior predictive checking for the 3-class model estimated using the \textbf{MNAR} method. Completed $T$ is computed using the completed data. Replicated $T$ is computed using the replicated completed datasets from the posterior predictive distribution.}
	\label{f:Figure_completed_discrepancy_K3_MNAR_MNAR_Y2}
\end{figure}

\begin{figure}[htbp]
	\centering
	\includegraphics[scale=0.75]{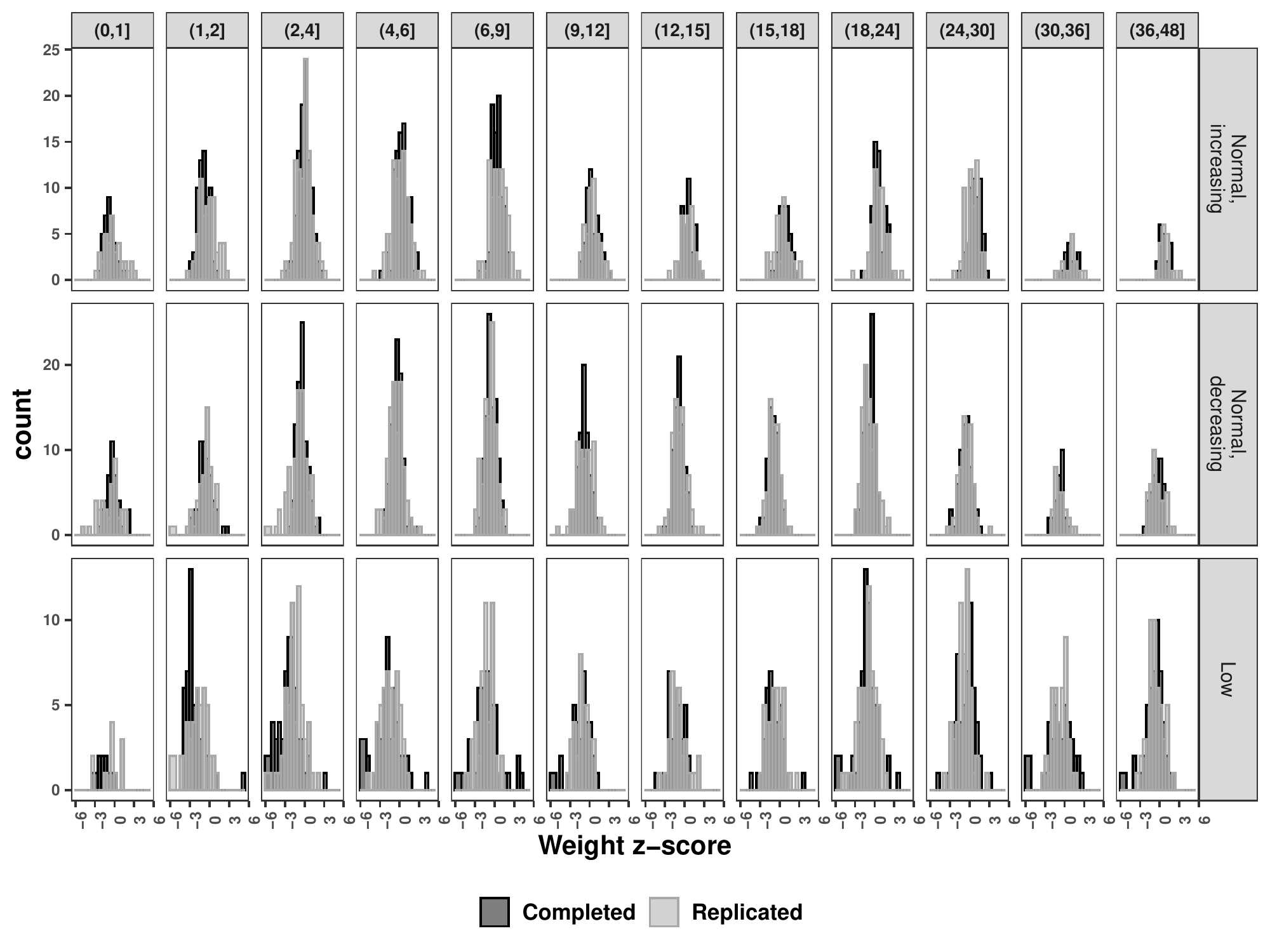}
	\caption{Histograms of completed and replicated completed weight z-scores from the posterior predictive distribution, by subgroup and well-child window, assuming 3 latent classes and using the \textbf{MNAR} method.}
	\label{f:Figure_K3_PPDHistWeight_MNAR_MNAR_Y2_2}
\end{figure}

\begin{figure}[htbp]
	\centering
	\includegraphics[scale=0.75]{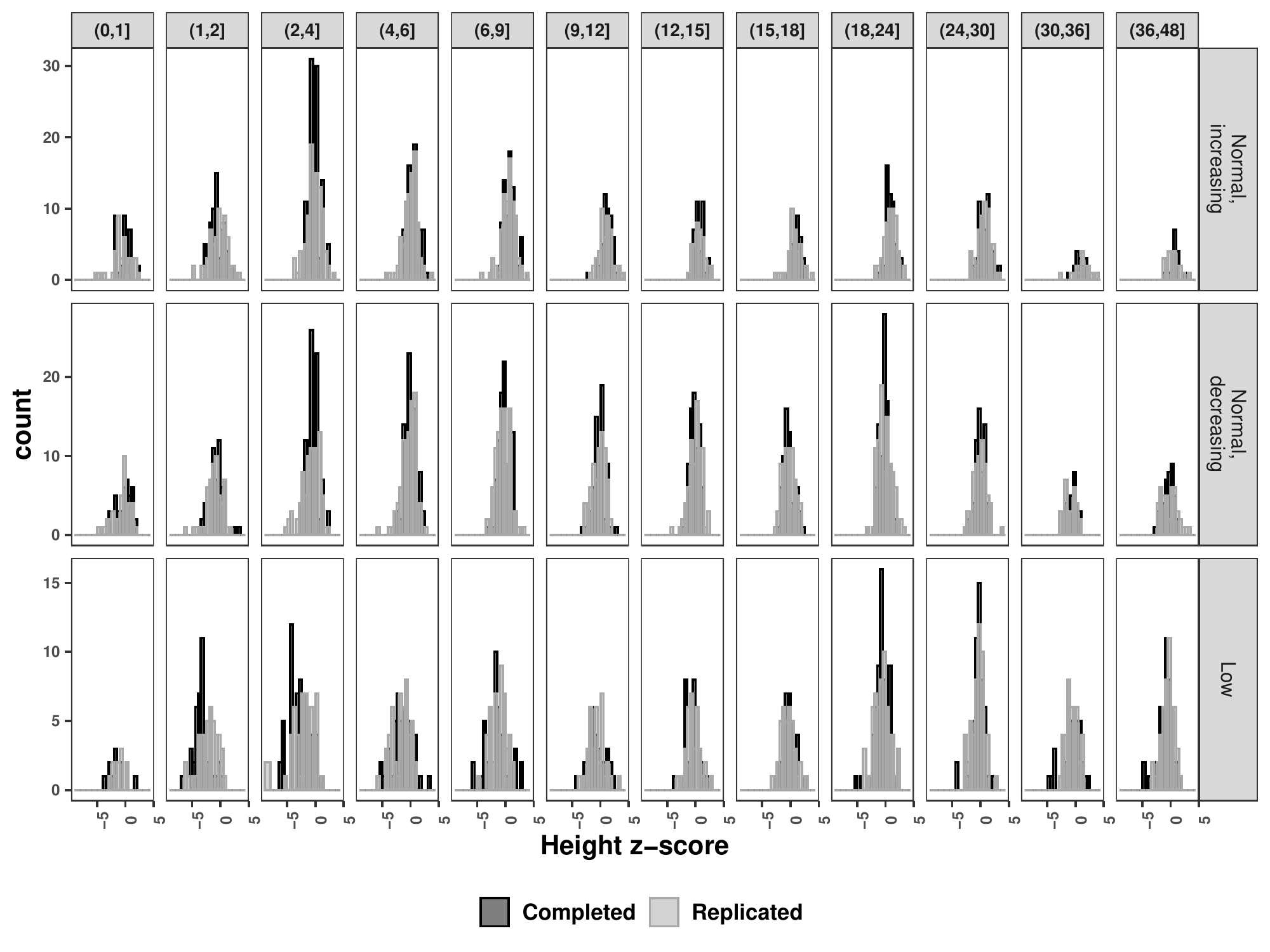}
	\caption{Histograms of completed and replicated completed height z-scores from the posterior predictive distribution, by subgroup and well-child window, assuming 3 latent classes and using the \textbf{MNAR} method.}
	\label{f:Figure_K3_PPDHistHeight_MNAR_MNAR_Y2_2}
\end{figure}

Comparison of the completed datasets that include observed and imputed weight and height z-scores in each well-child window, and replicates of the completed datasets drawn from the posterior predictive distribution demonstrated adequate model fit. Figure \ref{f:Figure_completed_discrepancy_K3_MNAR_MNAR_Y2} presents a scatter plot of the discrepancy measure $T$ across MCMC samples, with the horizontal and vertical axes being $T$ based on the completed and replicated datasets, respectively. Comparing completed and replicated $T$, the Bayesian predictive p-value of 0.44 suggests adequate overall model fit. In addition, we compared histograms of randomly selected completed and replicated datasets of weight and height z-scores. In Figures \ref{f:Figure_K3_PPDHistWeight_MNAR_MNAR_Y2_2} and \ref{f:Figure_K3_PPDHistHeight_MNAR_MNAR_Y2_2}, the distribution of z-scores by subgroup and well-child window appears largely consistent between the completed and replicated datasets.

\clearpage

\subsection{Child classification using the different estimation methods in 2-class models}

\begin{figure}[htbp]
	\centering
	\includegraphics[scale = 0.75]{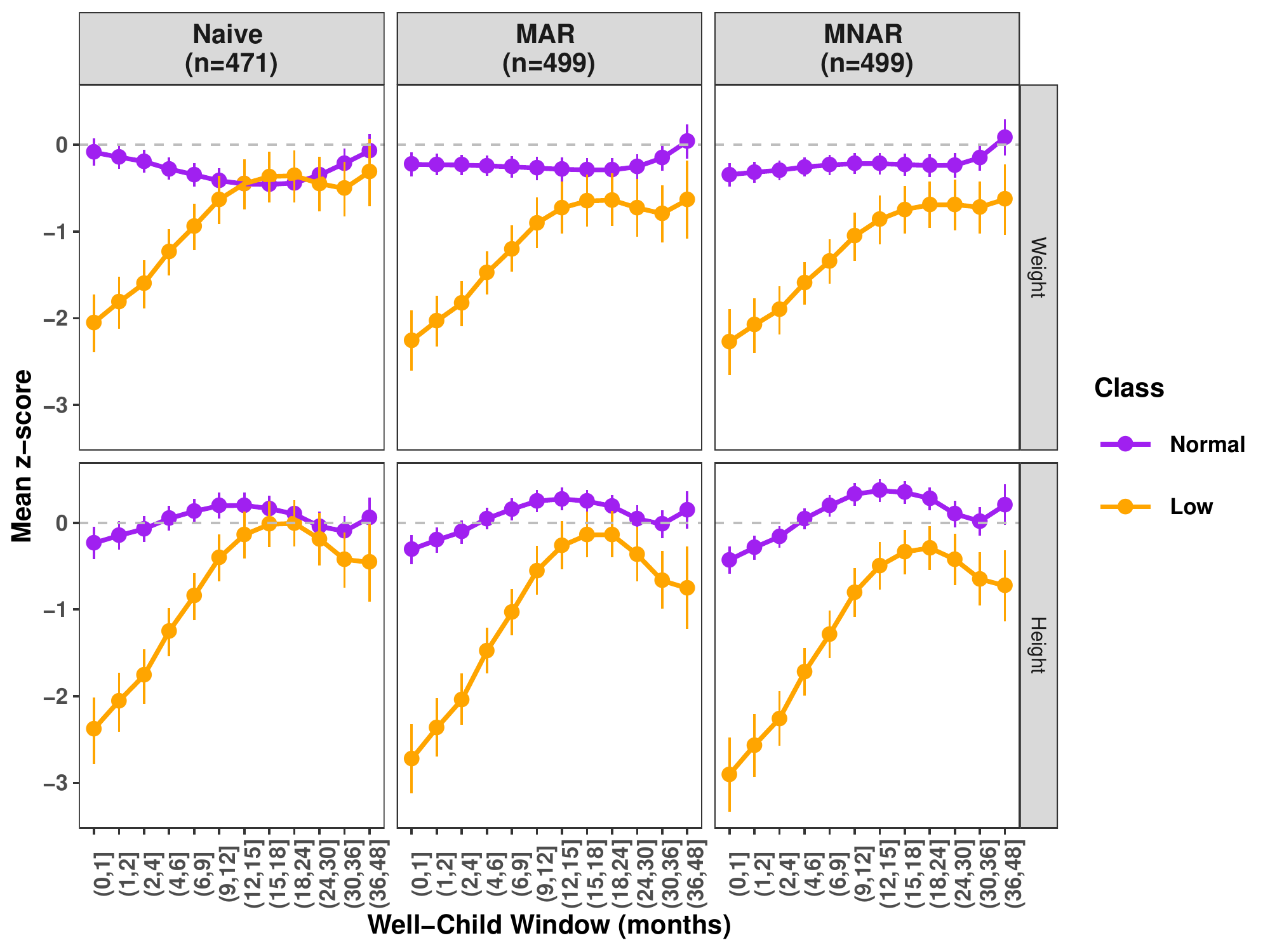}
	\caption{Latent class-specific average trajectories of weight and height z-scores estimated by the \textbf{Na\"ive}, \textbf{MAR}, and \textbf{MNAR} methods, assuming 2 latent classes. $n$ refers to the number of children included by each method.}
	\label{f:K2_LCSpecific_Y_Mechanism}
\end{figure}

\begin{figure}[htbp]
	\centering
	\includegraphics[scale=0.75]{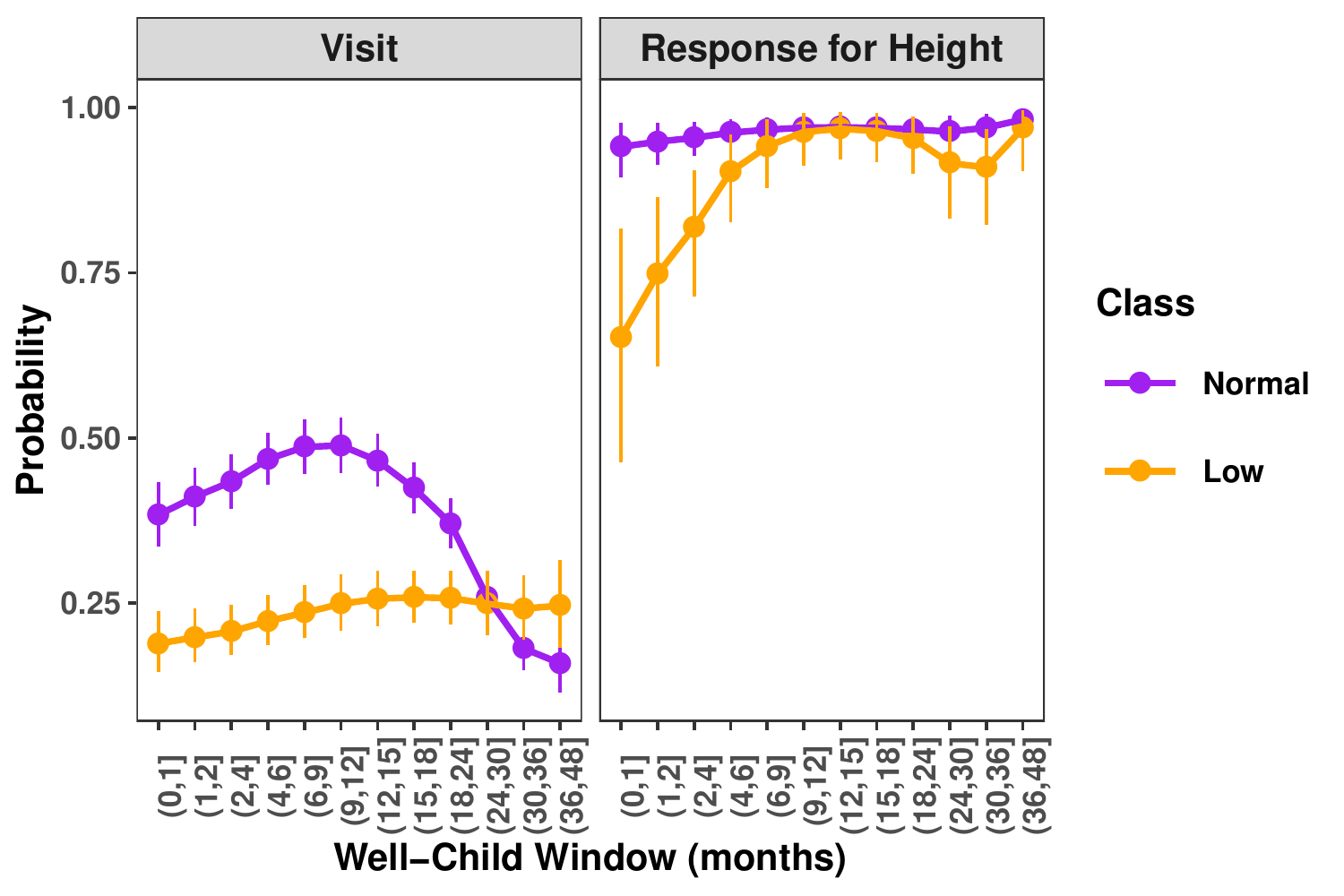}
	\caption{Latent class-specific trajectories of the probability of a clinic visit and the probability of a response for height z-scores using the \textbf{MNAR} method, assuming 2 latent classes.}
	\label{f:K2_LCSpecific_DM_Mechanism}
\end{figure}

The \textbf{Na\"ive}, \textbf{MAR}, and \textbf{MNAR} methods each detected a Normal trajectory subgroup (purple) and a Low trajectory subgroup (orange) (Figures \ref{f:K2_LCSpecific_Y_Mechanism} and \ref{f:K2_LCSpecific_DM_Mechanism}).

\clearpage

Table \ref{t:table_K2_latent_class_assignment} presents a summary of posterior latent class assignment under the three estimation methods. The \textbf{MNAR} method assigned about 8\% fewer children to the Normal subgroup than the other methods. The mean (median) probability of latent class assignment in each subgroup ranged from 0.87 to 0.93 (0.92 to 0.99). 

\begin{table}[htbp]
	\caption{Posterior latent class assignment in the 2-class models based on assigning children to the Normal or Low trajectory subgroup according to the maximum of the mean posterior probabilities of class assignment. The \textbf{Na\"ive}, \textbf{MAR}, and \textbf{MNAR} methods are shown.}
	\begin{tabular}{lcc}
	\hline
	& \multicolumn{1}{c}{Normal} & \multicolumn{1}{c}{Low} \\ \cline{2-3}
	\textbf{Na\"ive} ($n=471$): &  &  \\ 
	Predicted class size (\%) & 304 (65) & 167 (35) \\ 
	Mean probability & 0.87 & 0.88 \\ 
	Median probability & 0.92 & 0.97 \\ \hline
	\textbf{MAR} ($n=499$): & & \\ 
	Predicted class size (\%) & 335 (67) & 164 (33) \\ 
	Mean probability & 0.90 & 0.91 \\ 
	Median probability & 0.96 & 0.99 \\ \hline
	\textbf{MNAR} ($n=499$): & & \\ 
	Predicted class size (\%) & 292 (59) & 207 (41) \\ 
	Mean probability & 0.93 & 0.87 \\ 
	Median probability & 0.99 & 0.95 \\ 
	\hline
	\end{tabular}
	\label{t:table_K2_latent_class_assignment}
\end{table}

Table \ref{t:table_K2_cross_classification} cross-classifies the 499 children by their latent class assignment from the \textbf{MAR} and \textbf{MNAR} methods, and the birth weight variable from the latent class membership model.

\begin{table}[htbp]
	\centering
	\caption{Cross-classification of 499 children assigned to the Normal and Low trajectory subgroups by the \textbf{MAR} and \textbf{MNAR} methods, according to low birth weight (LBW) status.} 
	\begin{tabular}{lcccccc}
	\hline
	&  \multicolumn{5}{c}{\textbf{MNAR}} \\ \cline{2-6}
	&  \multicolumn{2}{c}{Non-LBW}  & & \multicolumn{2}{c}{LBW}\\
	&  \multicolumn{1}{c}{Normal} & \multicolumn{1}{c}{Low} & & \multicolumn{1}{c}{Normal} & \multicolumn{1}{c}{Low} \\ \cline{2-3} \cline{5-6}
	\multicolumn{1}{c}{\textbf{MAR}}  &  & & & & \\ \cline{1-1}
	Normal & 256 &  52 & & 16 &  11\\ 
	Low    &  17 &  58 & &  3 &  86 \\ 
	\hline
	\end{tabular}
\label{t:table_K2_cross_classification}
\end{table}

\clearpage

\section{Simulation Study Addendum}
\label{a:AddendumSimulation}

\setcounter{table}{0}    
\setcounter{figure}{0}

\subsection{Design}

We designed the study based on the real data analysis with 2 latent classes estimated with the \textbf{MNAR} method. For $i=1,\dots,500$ subjects over $j=1,\dots,12$ time windows, we generated longitudinal outcomes of interest $y_{1ij}$ and $y_{2ij}$, with about 60\% and 40\% of subjects in latent classes 1 and 2, respectively. We assumed the missing data mechanisms for the visit process and response process for $y_{2}$ are MNAR, while $y_{1}$ is fully observed given a clinic visit. In this setting, we considered the following five specific scenarios (S0-S4): 

\begin{enumerate}
    \item Under S0, we mimicked the latent class-specific trajectories and missingness proportions in the real data analysis. True parameter values for variance components were selected according to the real data analysis. 
    
    First, we generated the latent class membership of subject $i$  as 
    \begin{gather}
    \left[
    \begin{array}{c|c}
    c_{i} &  w_{i} \\
    \end{array}
    \right]  \sim  Bernoulli \begin{pmatrix}
    \pi_{i2}
    \end{pmatrix}, \nonumber
    \end{gather} where using a probit link function, $\pi_{i2} = \Phi\{-0.25 -  w_{i} \}$. $\pi_{i2}$ is the probability that subject $i$ belongs to latent class 2, and $w_{i}$ is a scaled and centered simulated variable for a subject's birth weight. 
    
    Then, we generated the longitudinal outcomes, visit process, and response process given a clinic visit conditional on a subject's latent class membership as \begin{gather}
\left[
\begin{array}{c|c}
y_{1ij} &  \multirow{2}{*}{$c_i = k$} \\
y_{2ij} &  
\end{array}
\right]  \sim  MVN_{2} \begin{pmatrix}
\begin{bmatrix}
\beta_{1k1} + \beta_{1k2} x_{ij} + b_{1i1} \\
\beta_{2k1} + \beta_{2k2} x_{ij} + b_{2i1}  
\end{bmatrix}, \,
\mathbf{\Sigma}_k
\end{pmatrix} \label{e:s1modely} \\[3\jot] 
\left[
\begin{array}{c|c}
b_{1i1} &  \multirow{2}{*}{$c_i = k$}  \\
b_{2i1} &  
\end{array}
\right]  \sim  MVN_{2} \begin{pmatrix}
\begin{bmatrix}
0 \\ 
0  
\end{bmatrix}, \,
\begin{bmatrix}
\Psi_{k1} & 0 \\
0 & \Psi_{k2} 
\end{bmatrix}
\end{pmatrix} \label{e:s1modelb} \\[3\jot]
\left[
\begin{array}{c|c}
d_{ij} &  c_i = k   \\
\end{array}
\right]  \sim  Bernoulli \begin{pmatrix} \Phi\{
\phi_{k1} + \phi_{k2} x_{ij} + \tau_{i1} \}
\end{pmatrix} \label{e:s1modeld} \\[3\jot] 
\left[
\begin{array}{c|c}
\tau_{i1} &  c_i = k
\end{array}
\right]  \sim  Normal \begin{pmatrix}
0, \,
0.25
\end{pmatrix} \nonumber \\[3\jot]
\left[
\begin{array}{c|c}
m_{2i,A(l)} &  c_i = k  \\
\end{array}
\right]  \sim  Bernoulli \begin{pmatrix} \Phi\{
\lambda_{2k1} + \lambda_{2k2} x_{iA(l)} + \kappa_{2i1} \}
\end{pmatrix}  \label{e:s1modelm} \\[3\jot] 
\left[
\begin{array}{c|c}
\kappa_{2i1} &  c_i = k
\end{array}
\right]  \sim  Normal \begin{pmatrix}
0, \,
\Theta_{2k}
\end{pmatrix} \label{e:s1modelkappa}
\end{gather} In (\ref{e:s1modely}), for $y_{1ij}$, in latent class 1, $\mathbf{\beta}_{11} = (\beta_{111}, \, \beta_{112})^T = (-0.25, \, 0.1)$, and in latent class 2, $\mathbf{\beta}_{12} = (\beta_{121}, \, \beta_{122})^T = (-1, \, 0.5)$. For $y_{2ij}$, $\mathbf{\beta}_{21} = (\beta_{211}, \, \beta_{212})^T = (0.5, \, 0.2)$, and $\mathbf{\beta}_{22} = (\beta_{221}, \, \beta_{222})^T = (-0.5, \,  0.75)$. The latent class-specific variance-covariances of $y_{1ij}$, $y_{2ij}$ are $\mathbf{\Sigma}_1 = \begin{bsmallmatrix} 0.5 & 0.2 \\ 0.2 & 0.5  \end{bsmallmatrix}$ and $\mathbf{\Sigma}_2 = \begin{bsmallmatrix} 1.5 & 1 \\ 1 & 1.5  \end{bsmallmatrix}$. In (\ref{e:s1modelb}), for the random intercept of $y_{1ij}$, the latent class-specific variances are $\Psi_{11} = \Psi_{21} = 0.6$. For $y_{2ij}$, $\Psi_{12} = 0.6$ and $\Psi_{22} = 0.4$. 

For the visit process, in (\ref{e:s1modeld}), $\mathbf{\phi}_k = (\phi_{k1}, \, \phi_{k2})^T$, with $\mathbf{\phi}_1 = (-0.2, \, -0.8)$ and $\mathbf{\phi}_2 = (-0.8, \, 0.2)$. 

For the response process of $y_{2ij}$ given a clinic visit, in (\ref{e:s1modelm}), $\mathbf{\lambda}_{2k} = (\lambda_{2k1}, \, \lambda_{2k2})^T$, with $\mathbf{\lambda}_{21} = (1.9, \, 0.1)$ and $\mathbf{\lambda}_{22} = (1.1, \, 0.25)$. The latent class-specific random intercept variances (\ref{e:s1modelkappa}) are $\Theta_{21} = 0.5$ and $\Theta_{22} = 1.5$.

Depicting $y_{1}$ and $y_{2}$ in S0, Figure \ref{f:Figure_data_gen_S0} shows that in early follow-up when the latent class-specific average trajectories are better separated, missingness in $y_{2}$ is high in class 2, while in later follow-up, missingness in $y_{2}$ is high in class 1.

\begin{figure}[htbp]
	\centering
	\includegraphics[scale = 0.75]{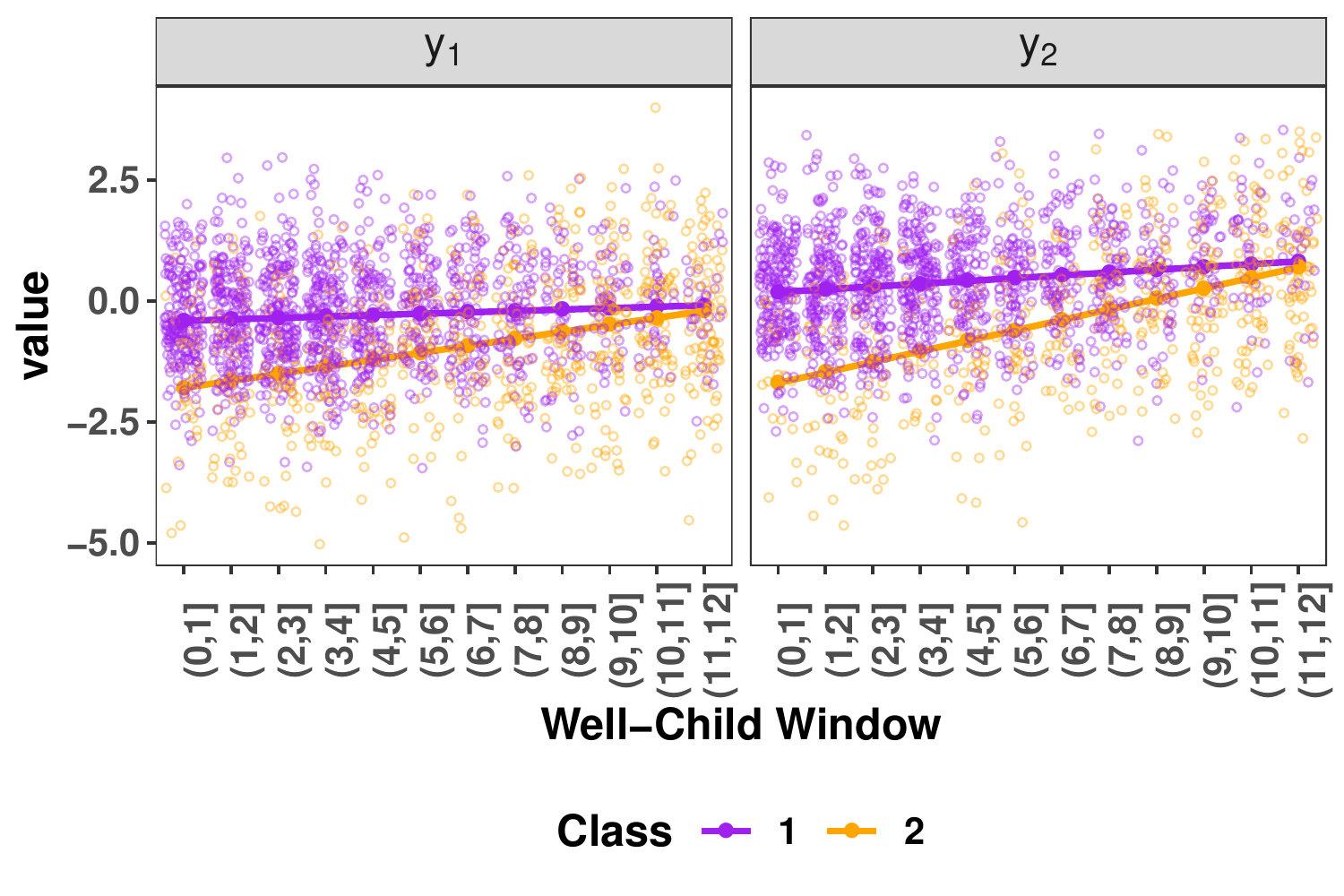}
	\caption{Average latent class-specific trajectories for $y_{1}$ and $y_{2}$ overlaid by points for observed measurements, under S0.}
	\label{f:Figure_data_gen_S0}
\end{figure}

\clearpage

\item In S1, we modified S0 by increasing the difference in the slopes for the latent class-specific trajectories of $y_{2}$: We made the slope in latent class 2 steeper. Specifically, in (\ref{e:s1modely}), $\beta_{222} = 1$. No other changes to S0 were made.

\item In S2, we modified S0 by decreasing the difference in the slopes for the latent class-specific trajectories of $y_{2}$: We made the slope in latent class 2 nearly parallel to the latent class 1 slope. Specifically, in (\ref{e:s1modely}), $\beta_{222} = 0.3$. No other changes to S0 were made.

\item In S3, we altered the visit process of S0 to reduce the percent of missed clinic visits in each latent class whilst maintaining the general visit process trajectory. In (\ref{e:s1modeld}), we set $\mathbf{\phi}_1 = (0.4, \, -0.2)$ and $\mathbf{\phi}_2 = (-0.1, \, 0.9)$. These changes resulted in 35\% missed clinic visits in latent class 1, and 55\% missed clinic visits in latent class 2. No other changes to S0 were made.

\item In S4, we modified S0 by increasing the percent of missed responses of $y_{2}$ in each latent class whilst maintaining the general response process trajectory. In (\ref{e:s1modelm}), we set $\mathbf{\lambda}_{21} = (0.8, \, 0.1)$ and $\mathbf{\lambda}_{22} = (0.5, \, 0.2)$. These changes resulted in 25\% missed responses in $y_{2}$ in latent class 1, and 35\% missed responses in $y_{2}$ in latent class 2. No other changes to S0 were made.
\end{enumerate}

\begin{figure}[htbp]
	\centering
	\includegraphics[scale = 0.75]{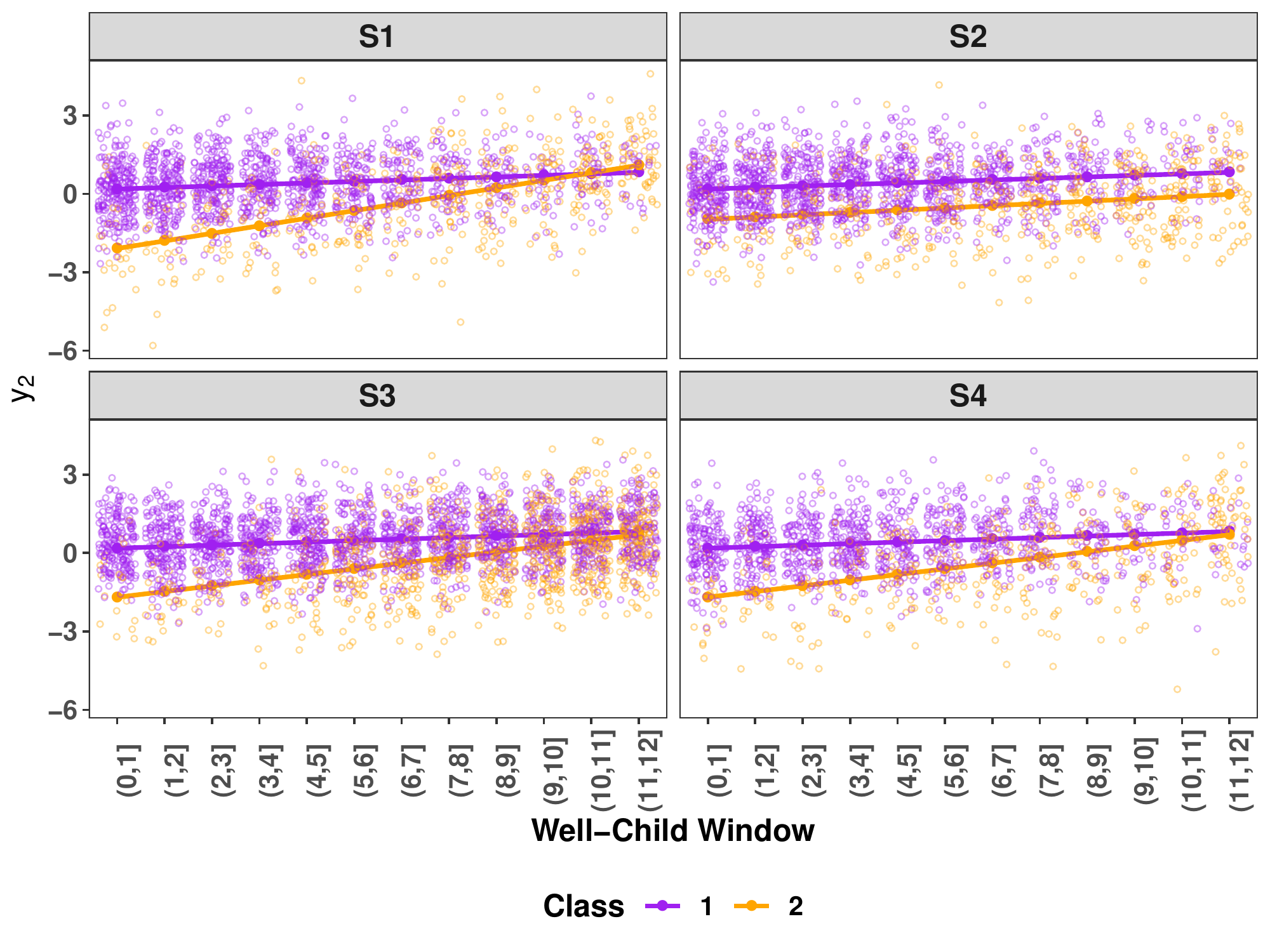}
	\caption{Average latent class-specific trajectories for $y_{2}$ overlaid by points for observed measurements, under S1 to S4. Compared to S0, in S1, the latent class-specific slopes are more different. In S2, they are more similar. In S3, the percent of missed clinic visits is reduced compared to S0. In S4, the percent of missed responses given a clinic visit is increased compared to S0. }
	\label{f:Figure_data_gen_scenarios1to4}
\end{figure}

Figure \ref{f:Figure_data_gen_scenarios1to4} portrays S1 -- S4 for $y_{2}$.

\clearpage

\subsection{Results}

% latex table generated in R 3.5.1 by xtable 1.8-3 package
% Tue Jul 30 08:52:51 2019
\begin{table}[ht]
\centering
\caption{Simulation results of S1 for parameter estimation of intercept $\beta_{rk1}$ and slope $\beta_{rk2}$ for longitudinal outcome $r$ in latent class $k$, along with the corresponding marginal intercept and slope, $\tilde{\beta}_{r1}$ and $\tilde{\beta}_{r2}$, respectively, under the \textbf{Full}, \textbf{Na\"ive}, \textbf{MAR}, and \textbf{MNAR} methods. The \textbf{Full} method is the benchmark. The best performing method among the \textbf{Na\"ive}, \textbf{MAR}, and \textbf{MNAR} methods is in bold.} 
\label{t:param_DG2}
\footnotesize
\begin{tabular}{cclcrrrr}
  \hline
\multicolumn{1}{c}{Outcome} &\multicolumn{1}{c}{Parameter} & \multicolumn{1}{c}{Method} & \multicolumn{1}{c}{Truth} &  \multicolumn{1}{c}{Bias} & \multicolumn{1}{c}{MSE} & \multicolumn{1}{c}{Coverage} & \multicolumn{1}{c}{Length} \\ 
  \hline
\multirow{16}{*}{$y_{1}$} & \multirow{4}{*}{\shortstack{$\beta_{111}$ \\ \vspace{0.1cm} \\ \footnotesize{(Class 1 Intercept)}}} & Full & \multirow{4}{*}{-0.250}  &  0.002 & 0.002 & 0.960 & 0.188 \\ 
  & & Na\"ive &  &  0.018 & 0.004 & 0.930 & 0.217 \\ 
  & & MAR &  &  0.006 & \textbf{0.003} & \textbf{0.948} & 0.212 \\ 
  & & MNAR &  &  \textbf{0.005} & \textbf{0.003} & 0.946 & \textbf{0.209} \\ \cline{2-8}
  & \multirow{4}{*}{\shortstack{$\beta_{121}$ \\ \vspace{0.1cm} \\ \footnotesize{(Class 2 Intercept)}}} & Full & \multirow{4}{*}{-1.000} & -0.003 & 0.003 & 0.952 & 0.228 \\ 
  & & Na\"ive &  & \textbf{0.005} & 0.014 & 0.922 & 0.404 \\ 
  & & MAR &  &  -0.010 & 0.009 & \textbf{0.946} & 0.369 \\ 
  & & MNAR &  & 0.006 & \textbf{0.007} & 0.932 & \textbf{0.314} \\ \cline{2-8}
 	& \multirow{4}{*}{\shortstack{$\beta_{112}$ \\ \vspace{0.1cm} \\ \footnotesize{(Class 1 Slope)}}} & Full & \multirow{4}{*}{0.100} &  -0.000 & 0.000 & 0.940 & 0.048 \\ 
  & & Na\"ive &  &  -0.008 & \textbf{0.001} & 0.936 & 0.096 \\ 
  & & MAR &  &  -0.006 & \textbf{0.001} & \textbf{0.950} & 0.092 \\ 
  & & MNAR &  &  \textbf{0.000} & \textbf{0.001} & 0.946 & \textbf{0.091} \\ \cline{2-8}
 	& \multirow{4}{*}{\shortstack{$\beta_{122}$ \\ \vspace{0.1cm} \\ \footnotesize{(Class 2 Slope)}}}  & Full & \multirow{4}{*}{0.500} &  -0.001 & 0.001 & 0.946 & 0.096 \\ 
  & & Na\"ive &  &  -0.055 & 0.009 & 0.854 & 0.273 \\ 
  & & MAR &  &  -0.020 & 0.005 & 0.928 & 0.242 \\ 
  & & MNAR &  &  \textbf{-0.005} & \textbf{0.003} & \textbf{0.940} & \textbf{0.214} \\ \hline
 	\multirow{16}{*}{$y_{2}$} & \multirow{4}{*}{\shortstack{$\beta_{211}$ \\ \vspace{0.1cm} \\ \footnotesize{(Class 1 Intercept)}}} & Full & \multirow{4}{*}{0.500} &  0.000 & 0.002 & 0.938 & 0.188 \\ 
  & & Na\"ive &  &  0.029 & 0.004 & 0.894 & 0.214 \\ 
  & & MAR &   & 0.029 & 0.004 & 0.914 & 0.212 \\ 
  & & MNAR &  &  \textbf{0.009} & \textbf{0.003} & \textbf{0.932} & \textbf{0.209} \\ \cline{2-8}
  	& \multirow{4}{*}{\shortstack{$\beta_{221}$ \\ \vspace{0.1cm} \\ \footnotesize{(Class 2 Intercept)}}}  & Full & \multirow{4}{*}{-0.500} & -0.004 & 0.003 & 0.942 & 0.194 \\ 
  & & Na\"ive &  &  0.039 & 0.014 & 0.890 & 0.380 \\ 
  & & MAR &  &  0.025 & 0.011 & 0.916 & 0.364 \\ 
  & & MNAR &  &  \textbf{0.009} & \textbf{0.008} & \textbf{0.934} & \textbf{0.311} \\ \cline{2-8}
 	& \multirow{4}{*}{\shortstack{$\beta_{212}$ \\ \vspace{0.1cm} \\ \footnotesize{(Class 1 Slope)}}} & Full & \multirow{4}{*}{0.200} &  0.001 & 0.000 & 0.946 & 0.048 \\ 
  & & Na\"ive &  & -0.013 & \textbf{0.001} & 0.904 & 0.095 \\ 
  & & MAR &  &  -0.011 & \textbf{0.001} & 0.902 & 0.094 \\ 
  & & MNAR &  &  \textbf{0.002} & \textbf{0.001} & \textbf{0.934} & \textbf{0.093} \\ \cline{2-8}
 	& \multirow{4}{*}{\shortstack{$\beta_{222}$ \\ \vspace{0.1cm} \\ \footnotesize{(Class 2 Slope)}}}  & Full & \multirow{4}{*}{1.000}  & 0.001 & 0.001 & 0.944 & 0.097 \\ 
  & & Na\"ive &  &  -0.062 & 0.012 & 0.814 & 0.287 \\ 
 & & MAR &  &  -0.034 & 0.007 & 0.896 & 0.275 \\ 
 & & MNAR &  &  \textbf{-0.008} & \textbf{0.004} & \textbf{0.954} & \textbf{0.239} \\ \hline
\multirow{8}{*}{$y_1$} & \multirow{4}{*}{\shortstack{$\tilde{\beta}_{11}$ \\ \vspace{0.1cm} \\ \footnotesize{(Marginal Intercept)}}} & Full & \multirow{4}{*}{-0.582} & -0.001 & 0.001 & 0.950 & 0.147 \\ 
   &  & Na\"ive &  & 0.070 & 0.008 & 0.653 & 0.183 \\ 
   &  & MAR &  & 0.049 & 0.004 & 0.807 & \textbf{0.176} \\ 
   &  & MNAR &  & \textbf{0.019} & \textbf{0.003} & \textbf{0.915} & 0.177 \\ 
   \cline{2-8} & \multirow{4}{*}{\shortstack{$\tilde{\beta}_{12}$ \\ \vspace{0.1cm} \\ \footnotesize{(Marginal Slope)}}} & Full & \multirow{4}{*}{0.277} & -0.000 & 0.000 & 0.962 & 0.051 \\ 
   &  & Na\"ive &  & -0.056 & 0.004 & 0.501 & 0.110 \\ 
   &  & MAR &  & -0.037 & 0.002 & 0.695 & \textbf{0.101} \\ 
   &  & MNAR &  & \textbf{-0.009} & \textbf{0.001} & \textbf{0.909} & 0.104 \\ 
   \cline{2-8}\multirow{8}{*}{$y_2$} & \multirow{4}{*}{\shortstack{$\tilde{\beta}_{21}$ \\ \vspace{0.1cm} \\ \footnotesize{(Marginal Intercept)}}} & Full & \multirow{4}{*}{0.058} & -0.002 & 0.001 & 0.948 & 0.138 \\ 
   &  & Na\"ive &  & 0.108 & 0.014 & 0.359 & 0.178 \\ 
   &  & MAR &  & 0.094 & 0.011 & 0.450 & \textbf{0.177} \\ 
   &  & MNAR &  & \textbf{0.028} & \textbf{0.003} & \textbf{0.857} & \textbf{0.177} \\ 
   \cline{2-8} & \multirow{4}{*}{\shortstack{$\tilde{\beta}_{22}$ \\ \vspace{0.1cm} \\ \footnotesize{(Marginal Slope)}}} & Full & \multirow{4}{*}{0.554} & 0.001 & 0.000 & 0.932 & 0.056 \\ 
   &  & Na\"ive &  & -0.092 & 0.010 & 0.144 & 0.113 \\ 
   &  & MAR &  & -0.073 & 0.006 & 0.279 & \textbf{0.112} \\ 
   &  & MNAR &  & \textbf{-0.018} & \textbf{0.001} & \textbf{0.865} & 0.115 \\ 
   \hline
\end{tabular}
\end{table}

\clearpage

% latex table generated in R 3.5.1 by xtable 1.8-3 package
% Thu Aug 01 00:37:29 2019
\begin{table}[ht]
\centering
\caption{Simulation results of S2 for parameter estimation of intercept $\beta_{rk1}$ and slope $\beta_{rk2}$ for longitudinal outcome $r$ in latent class $k$, along with the corresponding marginal intercept and slope, $\tilde{\beta}_{r1}$ and $\tilde{\beta}_{r2}$, respectively, under the \textbf{Full}, \textbf{Na\"ive}, \textbf{MAR}, and \textbf{MNAR} methods. The \textbf{Full} method is the benchmark. The best performing method among the \textbf{Na\"ive}, \textbf{MAR}, and \textbf{MNAR} methods is in bold.} 
\label{t:param_DG3}
\footnotesize
\begin{tabular}{cclcrrrr}
  \hline
\multicolumn{1}{c}{Outcome} & \multicolumn{1}{c}{Parameter} & \multicolumn{1}{c}{Method} & \multicolumn{1}{c}{Truth} &  \multicolumn{1}{c}{Bias} & \multicolumn{1}{c}{MSE} & \multicolumn{1}{c}{Coverage} & \multicolumn{1}{c}{Length} \\ 
  \hline
	\multirow{16}{*}{$y_{1}$} & \multirow{4}{*}{\shortstack{$\beta_{111}$ \\ \vspace{0.1cm} \\ \footnotesize{(Class 1 Intercept)}}}  & Full & \multirow{4}{*}{-0.250}  &  0.003 & 0.002 & 0.942 & 0.190 \\ 
   && Na\"ive &  & 0.054 & 0.007 & 0.832 & 0.231 \\ 
   && MAR &  &  0.036 & 0.005 & 0.890 & 0.225 \\ 
   && MNAR &  &  \textbf{0.001} & \textbf{0.003} & \textbf{0.956} & \textbf{0.209} \\ \cline{2-8}
  	& \multirow{4}{*}{\shortstack{$\beta_{121}$ \\ \vspace{0.1cm} \\ \footnotesize{(Class 2 Intercept)}}}  & Full & \multirow{4}{*}{-1.000} &  -0.003 & 0.003 & 0.950 & 0.231 \\ 
  && Na\"ive &  &  0.057 & 0.016 & 0.885 & 0.397 \\ 
  & & MAR &  &  0.025 & 0.010 & 0.928 & 0.366 \\ 
  & & MNAR &  &  \textbf{0.008} & \textbf{0.006} & \textbf{0.962} & \textbf{0.314} \\ \cline{2-8}
 	& \multirow{4}{*}{\shortstack{$\beta_{112}$ \\ \vspace{0.1cm} \\ \footnotesize{(Class 1 Slope)}}} & Full & \multirow{4}{*}{0.100} &  -0.000 & 0.000 & 0.948 & 0.048 \\ 
  & & Na\"ive &  &  -0.014 & \textbf{0.001} & 0.913 & 0.101 \\ 
  & & MAR &   & -0.012 & \textbf{0.001} & 0.934 & 0.096 \\ 
  & & MNAR &  &  \textbf{-0.002} & \textbf{0.001} & \textbf{0.960} & \textbf{0.091} \\ \cline{2-8}
 	& \multirow{4}{*}{\shortstack{$\beta_{122}$ \\ \vspace{0.1cm} \\ \footnotesize{(Class 2 Slope)}}}  & Full & \multirow{4}{*}{0.500} &  0.000 & 0.001 & 0.938 & 0.097 \\ 
  & & Na\"ive &  &  -0.107 & 0.017 & 0.593 & 0.252 \\ 
  & & MAR &   & -0.062 & 0.008 & 0.790 & 0.227 \\ 
  & & MNAR &  &  \textbf{-0.003} & \textbf{0.003} & \textbf{0.958} & \textbf{0.216} \\ \hline
  \multirow{16}{*}{$y_{2}$} & \multirow{4}{*}{\shortstack{$\beta_{211}$ \\ \vspace{0.1cm} \\ \footnotesize{(Class 1 Intercept)}}}  & Full & \multirow{4}{*}{0.500} &  0.003 & 0.003 & 0.926 & 0.191 \\ 
  && Na\"ive &  &  0.038 & 0.006 & 0.866 & 0.236 \\ 
  && MAR &   & 0.034 & 0.005 & 0.866 & 0.231 \\ 
  && MNAR &  &  \textbf{0.009} & \textbf{0.003} & \textbf{0.936} & \textbf{0.210} \\ \cline{2-8}
  	& \multirow{4}{*}{\shortstack{$\beta_{221}$ \\ \vspace{0.1cm} \\ \footnotesize{(Class 2 Intercept)}}} & Full & \multirow{4}{*}{-0.500} &  -0.003 & 0.002 & 0.950 & 0.197 \\ 
   && Na\"ive &  &  0.023 & 0.010 & 0.929 & 0.363 \\ 
   && MAR &   & \textbf{0.008} & 0.009 & \textbf{0.932} & 0.349 \\ 
   && MNAR &  &  0.014 & \textbf{0.007} & \textbf{0.932} & \textbf{0.310} \\ \cline{2-8}
  	& \multirow{4}{*}{\shortstack{$\beta_{212}$ \\ \vspace{0.1cm} \\ \footnotesize{(Class 1 Slope)}}}  & Full & \multirow{4}{*}{0.200} &  -0.000 & 0.000 & 0.940 & 0.048 \\ 
  && Na\"ive &  &  -0.008 & \textbf{0.001} & 0.917 & 0.099 \\ 
  & & MAR &   & -0.008 & \textbf{0.001} & \textbf{0.942} & 0.098 \\ 
  & & MNAR &  &  \textbf{0.001} & \textbf{0.001} & 0.932 & \textbf{0.093} \\ \cline{2-8}
  	& \multirow{4}{*}{\shortstack{$\beta_{222}$ \\ \vspace{0.1cm} \\ \footnotesize{(Class 2 Slope)}}}  & Full & \multirow{4}{*}{0.300} &  -0.001 & 0.001 & 0.942 & 0.096 \\ 
  & & Na\"ive &  &  -0.102 & 0.014 & 0.597 & 0.236 \\ 
  & & MAR &   & -0.077 & 0.010 & 0.726 & \textbf{0.233} \\ 
  & & MNAR &  &  \textbf{-0.000} & \textbf{0.003} & \textbf{0.944} & 0.235 \\ \hline
\multirow{8}{*}{$y_1$} & \multirow{4}{*}{\shortstack{$\tilde{\beta}_{11}$ \\ \vspace{0.1cm} \\ \footnotesize{(Marginal Intercept)}}} & Full & \multirow{4}{*}{-0.582} & -0.000 & 0.001 & 0.952 & 0.147 \\ 
   &  & Na\"ive &  & 0.086 & 0.010 & 0.524 & 0.182 \\ 
   &  & MAR &  & 0.058 & 0.006 & 0.742 & \textbf{0.175} \\ 
   &  & MNAR &  & \textbf{0.017} & \textbf{0.002} & \textbf{0.946} & 0.176 \\ 
   \cline{2-8} & \multirow{4}{*}{\shortstack{$\tilde{\beta}_{12}$ \\ \vspace{0.1cm} \\ \footnotesize{(Marginal Slope)}}} & Full & \multirow{4}{*}{0.277} & 0.000 & 0.000 & 0.956 & 0.051 \\ 
   &  & Na\"ive &  & -0.068 & 0.006 & 0.318 & 0.108 \\ 
   &  & MAR &  & -0.047 & 0.003 & 0.542 & \textbf{0.100} \\ 
   &  & MNAR &  & \textbf{-0.009} & \textbf{0.001} & \textbf{0.934} & 0.104 \\ 
   \cline{2-8}\multirow{8}{*}{$y_2$} & \multirow{4}{*}{\shortstack{$\tilde{\beta}_{21}$ \\ \vspace{0.1cm} \\ \footnotesize{(Marginal Intercept)}}} & Full & \multirow{4}{*}{0.057} & -0.001 & 0.001 & 0.946 & 0.139 \\ 
   &  & Na\"ive &  & 0.073 & 0.008 & 0.595 & \textbf{0.173} \\ 
   &  & MAR &  & 0.058 & 0.006 & 0.718 & \textbf{0.173} \\ 
   &  & MNAR &  & \textbf{0.028} & \textbf{0.003} & \textbf{0.888} & 0.176 \\ 
   \cline{2-8} & \multirow{4}{*}{\shortstack{$\tilde{\beta}_{22}$ \\ \vspace{0.1cm} \\ \footnotesize{(Marginal Slope)}}} & Full & \multirow{4}{*}{0.244} & -0.001 & 0.000 & 0.930 & 0.050 \\ 
   &  & Na\"ive &  & -0.050 & 0.003 & 0.518 & \textbf{0.104} \\ 
   &  & MAR &  & -0.040 & 0.002 & 0.670 & \textbf{0.104} \\ 
   &  & MNAR &  & \textbf{-0.001} & \textbf{0.001} & \textbf{0.948} & 0.111 \\ 
   \hline
\end{tabular}
\end{table}

\clearpage

% latex table generated in R 3.5.1 by xtable 1.8-3 package
% Mon Aug 05 08:13:46 2019
\begin{table}[ht]
\centering
\caption{Simulation results of S3 for parameter estimation of intercept $\beta_{rk1}$ and slope $\beta_{rk2}$ for longitudinal outcome $r$ in latent class $k$, along with the corresponding marginal intercept and slope, $\tilde{\beta}_{r1}$ and $\tilde{\beta}_{r2}$, respectively, under the \textbf{Full}, \textbf{Na\"ive}, \textbf{MAR}, and \textbf{MNAR} methods. The \textbf{Full} method is the benchmark. The best performing method among the \textbf{Na\"ive}, \textbf{MAR}, and \textbf{MNAR} methods is in bold.} 
\label{t:param_DG5}
\footnotesize
\begin{tabular}{cclcrrrr}
  \hline
		\multicolumn{1}{c}{Outcome} & \multicolumn{1}{c}{Parameter} & \multicolumn{1}{c}{Method} & \multicolumn{1}{c}{Truth} &  \multicolumn{1}{c}{Bias} & \multicolumn{1}{c}{MSE} & \multicolumn{1}{c}{Coverage} & \multicolumn{1}{c}{Length} \\ 
  \hline
	\multirow{16}{*}{$y_{1}$} & \multirow{4}{*}{\shortstack{$\beta_{111}$ \\ \vspace{0.1cm} \\ \footnotesize{(Class 1 Intercept)}}} & Full & \multirow{4}{*}{-0.250} &  -0.002 & 0.002 & 0.950 & 0.190 \\ 
  && Na\"ive &  &  0.008 & 0.003 & \textbf{0.954} & 0.199 \\ 
  && MAR &  &  \textbf{-0.001} & 0.003 & \textbf{0.946} & 0.197 \\ 
  && MNAR &  &  \textbf{-0.001} & \textbf{0.002} & 0.958 & \textbf{0.195} \\ \cline{2-8}
  	& \multirow{4}{*}{\shortstack{$\beta_{121}$ \\ \vspace{0.1cm} \\ \footnotesize{(Class 2 Intercept)}}}  & Full & \multirow{4}{*}{-1.000} &  0.000 & 0.003 & 0.956 & 0.230 \\ 
  && Na\"ive &  &  0.014 & 0.009 & 0.922 & 0.338 \\ 
  && MAR &  &  \textbf{0.001} & 0.007 & \textbf{0.944} & 0.307 \\ 
  && MNAR &  &  0.008 & \textbf{0.005} & 0.940 & \textbf{0.274} \\ \cline{2-8}
 	& \multirow{4}{*}{\shortstack{$\beta_{112}$ \\ \vspace{0.1cm} \\ \footnotesize{(Class 1 Slope)}}} & Full & \multirow{4}{*}{0.100} &  -0.000 & 0.000 & 0.930 & 0.048 \\ 
  && Na\"ive &  &  -0.004 & \textbf{0.000} & \textbf{0.944} & 0.066 \\ 
  && MAR &   & -0.004 & \textbf{0.000} & 0.930 & 0.064 \\ 
  && MNAR &  & \textbf{0.001} & \textbf{0.000} & 0.938 & \textbf{0.062} \\ \cline{2-8}
 	& \multirow{4}{*}{\shortstack{$\beta_{122}$ \\ \vspace{0.1cm} \\ \footnotesize{(Class 2 Slope)}}} & Full & \multirow{4}{*}{0.500} &  0.001 & 0.001 & 0.930 & 0.096 \\ 
  && Na\"ive &  &  -0.040 & 0.005 & 0.878 & 0.222 \\ 
  && MAR &   & -0.017 & 0.003 & 0.924 & 0.194 \\ 
  && MNAR &   & \textbf{-0.004} & \textbf{0.002} & \textbf{0.968} & \textbf{0.176} \\ \hline
	\multirow{16}{*}{$y_{2}$} & \multirow{4}{*}{\shortstack{$\beta_{211}$ \\ \vspace{0.1cm} \\ \footnotesize{(Class 1 Intercept)}}}  & Full & \multirow{4}{*}{0.500} &  -0.000 & 0.002 & 0.954 & 0.189 \\ 
  && Na\"ive &  &  0.008 & \textbf{0.003} & 0.946 & 0.199 \\ 
  && MAR &   & 0.006 & \textbf{0.003} & 0.936 & 0.198 \\ 
  && MNAR &  & \textbf{0.003} & \textbf{0.003} & \textbf{0.952} & \textbf{0.195} \\ \cline{2-8}
 	& \multirow{4}{*}{\shortstack{$\beta_{221}$ \\ \vspace{0.1cm} \\ \footnotesize{(Class 2 Intercept)}}}  & Full & \multirow{4}{*}{-0.500} &  -0.003 & 0.003 & 0.940 & 0.196 \\ 
  && Na\"ive &  &  0.011 & 0.007 & \textbf{0.938} & 0.311 \\ 
  && MAR &   & \textbf{0.010} & 0.007 & 0.918 & 0.296 \\ 
  && MNAR &  &  \textbf{0.010} & \textbf{0.005} & 0.924 & \textbf{0.263} \\ \cline{2-8}
 & \multirow{4}{*}{\shortstack{$\beta_{212}$ \\ \vspace{0.1cm} \\ \footnotesize{(Class 1 Slope)}}} & Full & \multirow{4}{*}{0.200} &  -0.001 & 0.000 & 0.918 & 0.048 \\ 
  && Na\"ive &  &  -0.007 & \textbf{0.000} & 0.916 & 0.066 \\ 
  && MAR &  &  -0.008 & \textbf{0.000} & 0.918 & 0.065 \\ 
  && MNAR &  &  \textbf{0.000} & \textbf{0.000} & \textbf{0.944} & \textbf{0.064} \\ \cline{2-8}
  	& \multirow{4}{*}{\shortstack{$\beta_{222}$ \\ \vspace{0.1cm} \\ \footnotesize{(Class 2 Slope)}}} & Full & \multirow{4}{*}{0.750} &  0.001 & 0.001 & 0.934 & 0.097 \\ 
  && Na\"ive &  &  -0.044 & 0.006 & 0.872 & 0.223 \\ 
  && MAR &   & -0.031 & 0.004 & 0.885 & 0.212 \\ 
  && MNAR &   & \textbf{-0.007} & \textbf{0.003} & \textbf{0.928} & \textbf{0.190} \\    \hline
\multirow{8}{*}{$y_1$} & \multirow{4}{*}{\shortstack{$\tilde{\beta}_{11}$ \\ \vspace{0.1cm} \\ \footnotesize{(Marginal Intercept)}}} & Full & \multirow{4}{*}{-0.582} & -0.001 & 0.001 & 0.950 & 0.147 \\ 
   &  & Na\"ive &  & 0.043 & 0.004 & 0.794 & 0.168 \\ 
   &  & MAR &  & 0.025 & 0.003 & 0.887 & \textbf{0.161} \\ 
   &  & MNAR &  & \textbf{0.001} & \textbf{0.002} & \textbf{0.934} & 0.162 \\ 
   \cline{2-8} & \multirow{4}{*}{\shortstack{$\tilde{\beta}_{12}$ \\ \vspace{0.1cm} \\ \footnotesize{(Marginal Slope)}}} & Full & \multirow{4}{*}{0.277} & 0.000 & 0.000 & 0.952 & 0.051 \\ 
   &  & Na\"ive &  & -0.036 & 0.002 & 0.646 & 0.092 \\ 
   &  & MAR &  & -0.023 & 0.001 & 0.805 & \textbf{0.084} \\ 
   &  & MNAR &  & \textbf{0.000} & \textbf{0.000} & \textbf{0.954} & 0.086 \\ 
   \cline{2-8}\multirow{8}{*}{$y_2$} & \multirow{4}{*}{\shortstack{$\tilde{\beta}_{21}$ \\ \vspace{0.1cm} \\ \footnotesize{(Marginal Intercept)}}} & Full & \multirow{4}{*}{0.057} & -0.001 & 0.001 & 0.954 & 0.138 \\ 
   &  & Na\"ive &  & 0.053 & 0.005 & 0.742 & 0.161 \\ 
   &  & MAR &  & 0.041 & 0.004 & 0.785 & \textbf{0.160} \\ 
   &  & MNAR &  & \textbf{0.003} & \textbf{0.002} & \textbf{0.909} & \textbf{0.160} \\ 
   \cline{2-8} & \multirow{4}{*}{\shortstack{$\tilde{\beta}_{22}$ \\ \vspace{0.1cm} \\ \footnotesize{(Marginal Slope)}}} & Full & \multirow{4}{*}{0.444} & -0.000 & 0.000 & 0.944 & 0.053 \\ 
   &  & Na\"ive &  & -0.046 & 0.003 & 0.476 & 0.092 \\ 
   &  & MAR &  & -0.036 & 0.002 & 0.652 & \textbf{0.091} \\ 
   &  & MNAR &  & \textbf{-0.001} & \textbf{0.001} & \textbf{0.909} & 0.092 \\ 
   \hline
\end{tabular}
\end{table}

\clearpage

\begin{table}[ht]
\centering
\caption{Simulation results of S4 for parameter estimation of intercept $\beta_{rk1}$ and slope $\beta_{rk2}$ for longitudinal outcome $r$ in latent class $k$, along with the corresponding marginal intercept and slope, $\tilde{\beta}_{r1}$ and $\tilde{\beta}_{r2}$, respectively, under the \textbf{Full}, \textbf{Na\"ive}, \textbf{MAR}, and \textbf{MNAR} methods. The \textbf{Full} method is the benchmark. The best performing method among the \textbf{Na\"ive}, \textbf{MAR}, and \textbf{MNAR} methods is in bold.} 
\label{t:param_DG6}
\footnotesize
\begin{tabular}{cclcrrrr}
  \hline
\multicolumn{1}{c}{Outcome} & \multicolumn{1}{c}{Parameter} & \multicolumn{1}{c}{Method} & \multicolumn{1}{c}{Truth} & \multicolumn{1}{c}{Bias} & \multicolumn{1}{c}{MSE} & \multicolumn{1}{c}{Coverage} & \multicolumn{1}{c}{Length} \\ 
  \hline
	\multirow{16}{*}{$y_{1}$} & \multirow{4}{*}{\shortstack{$\beta_{111}$ \\ \vspace{0.1cm} \\ \footnotesize{(Class 1 Intercept)}}} & Full & \multirow{4}{*}{-0.250} &  -0.002 & 0.002 & 0.950 & 0.190 \\ 
  & & Na\"ive &  & 0.046 & 0.007 & 0.870 & 0.249 \\ 
  & & MAR &   & 0.034 & 0.004 & 0.900 & 0.222 \\ 
  & & MNAR &  &  \textbf{0.005} & \textbf{0.003} & \textbf{0.964} & \textbf{0.210} \\ \cline{2-8}
 	& \multirow{4}{*}{\shortstack{$\beta_{121}$ \\ \vspace{0.1cm} \\ \footnotesize{(Class 2 Intercept)}}} & Full & \multirow{4}{*}{-1.000}  &  0.000 & 0.003 & 0.956 & 0.230 \\ 
  && Na\"ive &  &  0.091 & 0.024 & 0.830 & 0.424 \\ 
  && MAR &  &  0.015 & 0.011 & 0.928 & 0.374 \\ 
  && MNAR &  &  \textbf{0.010} & \textbf{0.006} & \textbf{0.942} & \textbf{0.316} \\ \cline{2-8}
 	& \multirow{4}{*}{\shortstack{$\beta_{112}$ \\ \vspace{0.1cm} \\ \footnotesize{(Class 1 Slope)}}}  & Full & \multirow{4}{*}{0.100} &  -0.000 & 0.000 & 0.930 & 0.048 \\ 
  && Na\"ive &  &  -0.012 & \textbf{0.001} & 0.920 & 0.117 \\ 
  && MAR &  &  -0.009 & \textbf{0.001} & 0.930 & 0.095 \\ 
  && MNAR &  &  \textbf{-0.002} & \textbf{0.001} & \textbf{0.950} & \textbf{0.091} \\ \cline{2-8}
 & \multirow{4}{*}{\shortstack{$\beta_{122}$ \\ \vspace{0.1cm} \\ \footnotesize{(Class 2 Slope)}}} & Full & \multirow{4}{*}{0.500} &  0.001 & 0.001 & 0.930 & 0.096 \\ 
  & & Na\"ive &  & -0.122 & 0.022 & 0.590 & 0.282 \\ 
  && MAR &  &  -0.052 & 0.007 & 0.846 & 0.235 \\ 
  && MNAR &  & \textbf{-0.003} & \textbf{0.003} & \textbf{0.952} & \textbf{0.216} \\ \hline
 	\multirow{16}{*}{$y_{2}$} & \multirow{4}{*}{\shortstack{$\beta_{211}$ \\ \vspace{0.1cm} \\ \footnotesize{(Class 1 Intercept)}}} & Full & \multirow{4}{*}{0.500} &  -0.000 & 0.002 & 0.954 & 0.189 \\ 
  & & Na\"ive &  &  0.076 & 0.011 & 0.754 & 0.250 \\ 
  && MAR &  &  0.045 & 0.006 & 0.888 & 0.233 \\ 
  & & MNAR &  &  \textbf{0.009} & \textbf{0.004} & \textbf{0.922} & \textbf{0.220} \\ \cline{2-8}
 	& \multirow{4}{*}{\shortstack{$\beta_{221}$ \\ \vspace{0.1cm} \\ \footnotesize{(Class 2 Intercept)}}} & Full & \multirow{4}{*}{-0.500} &  -0.003 & 0.003 & 0.940 & 0.196 \\ 
  & & Na\"ive &  &  0.096 & 0.025 & 0.778 & 0.405 \\ 
  & & MAR &  &  0.049 & 0.015 & 0.874 & 0.388 \\ 
  && MNAR &  &  \textbf{0.003} & \textbf{0.007} & \textbf{0.950} & \textbf{0.334} \\ \cline{2-8}
  	& \multirow{4}{*}{\shortstack{$\beta_{212}$ \\ \vspace{0.1cm} \\ \footnotesize{(Class 1 Slope)}}}  & Full & \multirow{4}{*}{0.200} &  -0.001 & 0.000 & 0.918 & 0.048 \\ 
  && Na\"ive &  &  -0.020 & \textbf{0.001} & 0.882 & 0.116 \\ 
  & & MAR &  &  -0.016 & \textbf{0.001} & 0.898 & 0.109 \\ 
  && MNAR &  &  \textbf{-0.003} & \textbf{0.001} & \textbf{0.930} & \textbf{0.105} \\ \cline{2-8}
 	& \multirow{4}{*}{\shortstack{$\beta_{222}$ \\ \vspace{0.1cm} \\ \footnotesize{(Class 2 Slope)}}}  & Full & \multirow{4}{*}{0.750} &  0.001 & 0.001 & 0.934 & 0.097 \\ 
   && Na\"ive &  &  -0.148 & 0.030 & 0.478 & 0.285 \\ 
   && MAR &  &  -0.084 & 0.014 & 0.740 & 0.280 \\ 
   && MNAR &  & \textbf{-0.004} & \textbf{0.004} & \textbf{0.976} & \textbf{0.258} \\ \hline
   \multirow{8}{*}{$y_1$} & \multirow{4}{*}{\shortstack{$\tilde{\beta}_{11}$ \\ \vspace{0.1cm} \\ \footnotesize{(Marginal Intercept)}}} & Full & \multirow{4}{*}{-0.582} & -0.001 & 0.001 & 0.950 & 0.147 \\ 
   &  & Na\"ive &  & 0.097 & 0.012 & 0.512 & 0.191 \\ 
   &  & MAR &  & 0.063 & 0.006 & 0.696 & \textbf{0.175} \\ 
   &  & MNAR &  & \textbf{0.020} & \textbf{0.003} & \textbf{0.903} & 0.177 \\ 
   \cline{2-8} & \multirow{4}{*}{\shortstack{$\tilde{\beta}_{12}$ \\ \vspace{0.1cm} \\ \footnotesize{(Marginal Slope)}}} & Full & \multirow{4}{*}{0.277} & 0.000 & 0.000 & 0.952 & 0.051 \\ 
   &  & Na\"ive &  & -0.074 & 0.007 & 0.360 & 0.120 \\ 
   &  & MAR &  & -0.046 & 0.003 & 0.574 & \textbf{0.100} \\ 
   &  & MNAR &  & \textbf{-0.009} & \textbf{0.001} & \textbf{0.918} & 0.105 \\ 
   \cline{2-8}\multirow{8}{*}{$y_2$} & \multirow{4}{*}{\shortstack{$\tilde{\beta}_{21}$ \\ \vspace{0.1cm} \\ \footnotesize{(Marginal Intercept)}}} & Full & \multirow{4}{*}{0.057} & -0.001 & 0.001 & 0.954 & 0.138 \\ 
   &  & Na\"ive &  & 0.127 & 0.019 & 0.236 & 0.185 \\ 
   &  & MAR &  & 0.094 & 0.012 & 0.494 & \textbf{0.184} \\ 
   &  & MNAR &  & \textbf{0.023} & \textbf{0.003} & \textbf{0.901} & 0.187 \\ 
   \cline{2-8} & \multirow{4}{*}{\shortstack{$\tilde{\beta}_{22}$ \\ \vspace{0.1cm} \\ \footnotesize{(Marginal Slope)}}} & Full & \multirow{4}{*}{0.444} & -0.000 & 0.000 & 0.944 & 0.053 \\ 
   &  & Na\"ive &  & -0.096 & 0.010 & 0.174 & 0.120 \\ 
   &  & MAR &  & -0.070 & 0.006 & 0.374 & \textbf{0.118} \\ 
   &  & MNAR &  & \textbf{-0.013} & \textbf{0.001} & \textbf{0.940} & 0.124 \\ 
   \hline
\end{tabular}
\end{table}
\clearpage

\begin{figure}[htbp]
	\centering
	\centerline{\includegraphics[scale = 0.75]{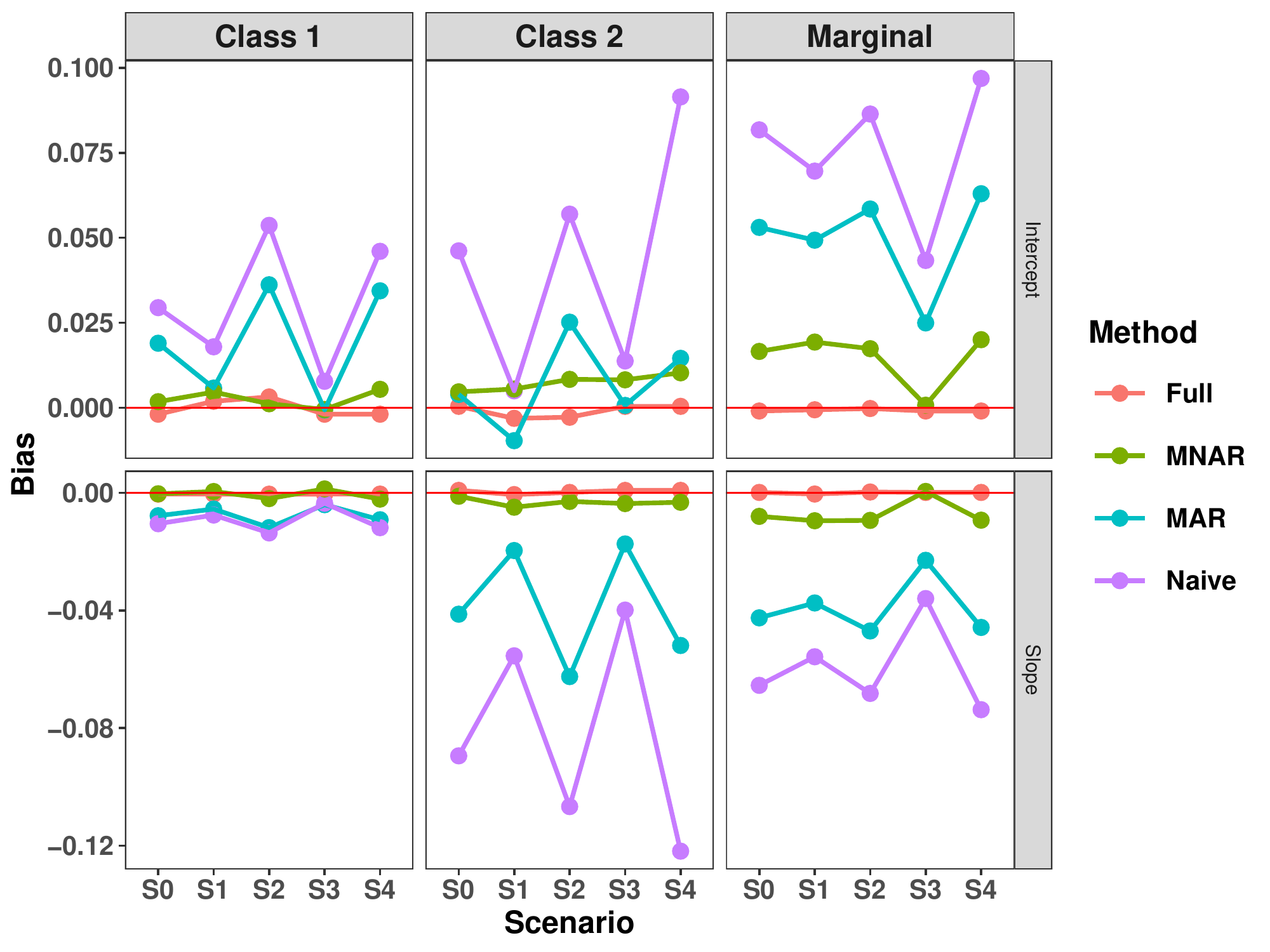}}
	\caption{Comparison of bias in parameter estimation for $y_{1}$ across data generation scenarios, by estimation method.}
	\label{f:sim_results_y1}
\end{figure}

\clearpage

\begin{table}[ht]
	\centering
	\caption{Simulation results for subject misclassification by data generation scenario under the \textbf{Full}, \textbf{Na\"ive}, \textbf{MAR}, and \textbf{MNAR} methods. The \textbf{Full} method is the benchmark. The best performing method among the \textbf{Na\"ive}, \textbf{MAR}, and \textbf{MNAR} methods is in bold.} 
	\begin{tabular}{cllrrrr}
		\hline
	&	&  & \multicolumn{3}{c}{Percentile} &  \\ \cline{4-6}
\multicolumn{1}{c}{Scenario}	&	\multicolumn{1}{c}{Method} & \multicolumn{1}{c}{Min} & \multicolumn{1}{c}{25} & \multicolumn{1}{c}{50} & \multicolumn{1}{c}{75} & \multicolumn{1}{c}{Max} \\ 
		\hline
\multirow{4}{*}{S0} & Full & 0.00 & 0.01 & 0.02 & 0.02 & 0.04 \\ 
	 &	Na\"ive & 0.09 & 0.14 & 0.15 & 0.16 & 0.20 \\ 
	 &	MAR & 0.09 & 0.13 & 0.14 & 0.16 & 0.20 \\ 
	 &	MNAR & \textbf{0.01} & \textbf{0.03} & \textbf{0.03} & \textbf{0.04} & \textbf{0.06} \\ 
		\hline
\multirow{4}{*}{S1} &	Full & 0.00 & 0.01 & 0.01 & 0.01 & 0.03 \\ 
  & Na\"ive & 0.09 & 0.12 & 0.13 & 0.14 & 0.19 \\ 
  & MAR & 0.08 & 0.12 & 0.13 & 0.14 & 0.18 \\ 
  & MNAR & \textbf{0.01} & \textbf{0.02} & \textbf{0.03} & \textbf{0.03} & \textbf{0.06} \\ 
  \hline
\multirow{4}{*}{S2} &  	Full & 0.00 & 0.02 & 0.02 & 0.03 & 0.05 \\ 
	&	Na\"ive & 0.08 & 0.14 & 0.15 & 0.17 & 0.22 \\ 
	&	MAR & 0.10 & 0.13 & 0.14 & 0.16 & 0.20 \\ 
	&	MNAR & \textbf{0.01} & \textbf{0.03} & \textbf{0.03} & \textbf{0.04} & \textbf{0.06} \\ 
	\hline
\multirow{4}{*}{S3} & 		Full & 0.00 & 0.01 & 0.02 & 0.02 & 0.04 \\ 
	&	Na\"ive & 0.07 & 0.10 & 0.11 & 0.12 & 0.16 \\ 
	&	MAR & 0.06 & 0.09 & 0.10 & 0.11 & 0.14 \\ 
	&	MNAR & \textbf{0.00} & \textbf{0.02} & \textbf{0.02} & \textbf{0.02} & \textbf{0.04} \\ 
	\hline
\multirow{4}{*}{S4} & 			Full & 0.00 & 0.01 & 0.02 & 0.02 & 0.04 \\ 
	&	Na\"ive & 0.11 & 0.15 & 0.17 & 0.18 & 0.25 \\ 
	&	MAR & 0.11 & 0.14 & 0.15 & 0.17 & 0.21 \\ 
	&	MNAR & \textbf{0.01} & \textbf{0.03} & \textbf{0.04} & \textbf{0.04} & \textbf{0.07} \\ 
	\hline
	\end{tabular}
	\label{t:misClass}
\end{table}

\clearpage

\section{\texttt{R} package \texttt{EHRMiss}}
\label{a:RPackage}

\setcounter{table}{0}    
\setcounter{figure}{0}

Our \verb'R' package \verb'EHRMiss' is available for download at \url{https://github.com/anthopolos/EHRMiss}. We provide an example of how to use the package to implement the proposed model for longitudinal health outcomes in EHRs with a nonignorable visit process and response process given a clinic visit. 

\vspace{20pt}

\begin{lstlisting}[language=R]
#------ Load data from the EHRMiss package
data(data)
names(data)
dim(data)


#------- Model details
### Number of outcomes
J <- 2
### Number of latent classes
K <- 2


#------- Specify the formulas for the design matrices and put the formulas in regf
regf <- list(LatentClass = ~ 1 + birthweight,
             YRe = ~ 1,
             YObs = ~ -1 + time,
             YSub = ~ 1,
             DObs = ~ 1 + time,
             DRe = ~ 1,
             MObs = ~ 1 + time,
             MRe = ~ 1)

#-------- MCMC preparation
m <- length(all.vars(regf[["LatentClass"]])) + 1
p <- length(all.vars(regf[["YSub"]])) + 1
s <- length(all.vars(regf[["YObs"]]))
f <- length(all.vars(regf[["DObs"]])) + 1
e <- length(all.vars(regf[["MObs"]])) + 1

### Number of random effects, assumed the same for all models
q <- length(all.vars(regf[["YRe"]])) + 1

### Prior distributions
priors <- list(list(rep(0, m), diag(1, m)), list(rep(0, s), diag(100, s)),
               list(rep(0, p), diag(10000, p)),
               list(1, 1),
               list(diag(c(0.5, 0.6), J), (J + 2)),
               list(rep(0, f), diag(100, f)),
               list(1, 1),
               list(rep(0, e), diag(100, e)),
               list(scale = 1, df = 1))

### Initial values
inits <- list(matrix(rep(0, m * (K - 1)), nrow = m, ncol = (K - 1)),
              list(matrix(rnorm(s*K), ncol = K, nrow = s), matrix(rnorm(s*K), ncol = K, nrow = s)),
              list(array(rnorm(p*q*K), dim = c(p, q, K)), array(rnorm(p*q*K), dim = c(p, q, K))),
              list(array(rep(0.4, K), dim = c(q, q, K)), array(rep(0.4, K), dim = c(q, q, K))),
              array(c(1, 0, 0, 1, 0.5, 0, 0, 0.5), dim = c(J, J, K)),
              matrix(rnorm(f*K), ncol = K),
              array(rep(0.5, K), dim = c(q, q, K)),
              list(matrix(rnorm(e*K), ncol = K)),
              list(array(rep(0.5, K), dim = c(q, q, K))))


#------- Fit model with MNAR visit process, MNAR response process for Y2, Y1 fully observed given a clinic visit
n.samples <- 2000
burn <- 1000
update <- 10
monitor <- TRUE


#MNAR Visit process
#MNAR response process for Y2
#See ?MVNYBinaryMiss
res <- MVNYBinaryMiss(K = K, J = J, data = data, regf = regf, imputeResponse = TRUE, Mvec = 2, modelVisit = TRUE, modelResponse = TRUE, priors = priors, inits = inits, n.samples = n.samples, burn = burn, monitor = monitor, update = update, modelComparison = TRUE, sims = FALSE)

#Printed to console: posterior summaries of model parameters, posterior latent class membership, model comparison statistics, label switching diagnostic

#Get Bayesian posterior predictive p-value, see ?get_discrepancy_plot
#Set working directory to where discrepancy samples are written
store_T_completed <- read.table("store_T_completed.txt", header = FALSE, sep = ",")
get_discrepancy_plot(store_T_completed)

\end{lstlisting}

\bibliography{sm_bibliography}